\documentclass[11pt, tightenlines, prd, english, superscriptaddress, nofootinbib, preprintnumbers, aps, notitlepage]{revtex4-1}
\usepackage{color}
\usepackage{babel}
\usepackage{graphicx}
\usepackage{amsfonts}
\usepackage{amsmath, amsthm, amssymb, mathrsfs}
\usepackage{float}
\usepackage[caption=false]{subfig}
\usepackage{subfiles}
\usepackage{bm}
\usepackage{hyperref}
\usepackage{enumerate}
\usepackage{calc}
\usepackage{placeins}

\usepackage{colordvi}

\makeatletter
\@ifundefined{textcolor}{}
{%
 \definecolor{BLACK}{gray}{0}
 \definecolor{WHITE}{gray}{1}
 \definecolor{RED}{rgb}{1,0,0}
 \definecolor{GREEN}{rgb}{0,1,0}
 \definecolor{BLUE}{rgb}{0,0,1}
 \definecolor{CYAN}{cmyk}{1,0,0,0}
 \definecolor{MAGENTA}{cmyk}{0,1,0,0}
 \definecolor{YELLOW}{cmyk}{0,0,1,0}
 }

\newcommand{\rd}{{\rm d}}
\newcommand{\re}{\mathbb{R}}
\newcommand{\Hil}{\mathcal{H}}
\newcommand{\id}{\mathbb{I}}
\newcommand{\lat}{\mathcal{L}}
\newcommand{\sgn}{{\rm sgn}}
\newcommand{\dom}{\mathcal{D}}
\newcommand{\ub}[1]{\underline{#1}}

\begin{document}


\title{Emergent de Sitter epoch of the Loop Quantum Cosmos: a detailed analysis}

\author{Mehdi Assanioussi}
  \email{mehdi.assanioussi@fuw.edu.pl}
  \affiliation{II. Institute for Theoretical Physics, University of Hamburg,\\ Luruper Chaussee 149, 22761 Hamburg, Germany.}

\author{Andrea Dapor}
  \email{andrea.dapor@gravity.fau.de}
  \affiliation{Institute for Quantum Gravity, Friedrich-Alexander University Erlangen-N\"urnberg, Staudstra\ss e 7, 91058 Erlangen, Germany.}
  \affiliation{Department of Physics and Astronomy, Louisiana State University, Baton Rouge, LA 70803, USA.}

\author{Klaus Liegener}
  \email{liegener1@lsu.edu}
  \affiliation{Institute for Quantum Gravity, Friedrich-Alexander University Erlangen-N\"urnberg, Staudstra\ss e 7, 91058 Erlangen, Germany.}
  \affiliation{Department of Physics and Astronomy, Louisiana State University, Baton Rouge, LA 70803, USA.}

\author{Tomasz Paw{\l}owski}
	\email{tomasz.pawlowski@uwr.edu.pl}
	\affiliation{Institute for Theoretical Physics, Faculty of Physics and Astronomy, University of Wroc{\l}aw,
	pl. M. Borna 9, 50-204  Wroc{\l}aw, Poland}

\date{\today{}}

\begin{abstract}
	\noindent We present a detailed analysis of a quantum model for Loop Quantum Cosmology based on strict application of the Thiemann regularization algorithm for the Hamiltonian in Loop Quantum Gravity, extending the results presented previously in our brief report. This construction leads to a qualitative modification of the bounce paradigm. Quantum gravity effects still lead to a quantum bounce connecting deterministically large classical Universes. However, the evolution features a large epoch of de Sitter Universe, with emergent cosmological constant of Planckian order, smoothly transiting into a spatially flat expanding Universe. Moreover, we present an effective Hamiltonian describing the quantum evolution to high accuracy and for which the dynamics can be solved analytically.
\end{abstract}

\maketitle

\documentclass[../main.tex]{subfiles}

\section{Introduction}

Modern experiments and precise cosmological observations constantly expand the frontiers of our knowledge of the Universe and its evolution at largest scales. The influx on high precision CMB measurements and the birth of gravitational wave astronomy \cite{GW1,GW2} give hope for making the models describing the very early Universe dynamics -- where the quantum nature of gravity is expected to play an important role -- experimentally testable. It is therefore particularly important to bring the available models/theories of the interaction between geometry and matter at highest energy scales to the level where concrete physical predictions can be made in unambiguous manner. One of the most popular initiatives to bring relativity and quantum theory to a common footing is Loop Quantum Gravity (LQG) \cite{Rov04,AL04,Thi07}. LQG exploits the fact that general relativity (GR) in its background-independent Hamiltonian formulation is equivalent to a Yang-Mills gauge theory \cite{YM54,KS75,Cre85} and it is therefore possible to proceed with its quantization in a well-known and mathematically rigorous manner. Despite LQG reaching the level of maturity, where the physical Hilbert space and the analog of the Schr\"odinger evolution equation generating the dynamics could be constructed \cite{LOST06,Thi98a,Thi98b}, attempts of applying it in its full form to study the implications for cosmology have not been successful so far. 
Yet, in the last two decades the subfield of Loop Quantum Cosmology (LQC) emerged. Here, one imports regularization techniques from LQG directly to symmetry reduced (usually cosmological) spacetimes \cite{Boyo-book,Boj99a,Boj99b,Boj05,Ash09_Cos,AP11}. Due to this symmetry-reduction, the phase space of the theory becomes coordinatized by quasi-global degrees of freedom (in case of inhomogeneous spacetimes, for example, by Fourier modes of the inhomogeneities) becoming finite dimensional for homogeneous cosmology models.
This allows to proceed by investigating effects of quantum geometry in the Planck regime \cite{AAN12}. In particular, the LQC model of a Friedman-Lema\^itre-Robertson-Walker (FLRW) Universe led to the replacement of the big bang initial singularity by a bounce, connecting two (semi-)classical FLRW spacetimes \cite{APS06a,APS06b,APS06c,BP08,PA11}. This was achieved by dynamically evolving semiclassical states (in the sense of small relative uncertainties) starting from a chosen moment of time corresponding to large expanding Universe. In most cases, for that purpose one selects Gaussian states in the ``energy'' representation -- the canonical momentum of a matter field serving as the internal clock that parametrizes the quantum evolution. Subsequently, the studies of the full quantum dynamics of isotropic spacetimes were generalized to nonisotropic ones \cite{MBMMP09-1,MBMMP09-2,P-Loops17}, including in particular the Kantowski-Sachs chart of the interior of the Schwarzschild black hole \cite{AOP18-1,AOP18-2}.
Interestingly, the genuine quantum trajectories defined by the time evolution of the expectation values of certain observables (volume, its momentum, energy density, Hubble rate, etc.) for these states are reproduced to accuracy well below quantum variances by the so-called {\it effective Hamiltonian}, which is constructed by replacing a set of 
``elementary'' operators (volume and $U(1)$ components of holonomies) forming the Hamiltonian constraint operator with their expectation values \cite{SV05}.\footnote{The validity of this heuristic procedure is supported by a series of works where the attempt of evaluating the correct expectation value of the Hamiltonian constraint was made. In particular the effective Hamiltonian was confirmed to reproduce the latter in the limit of low energy and low relative dispersion \cite{Tav08}. Also, the modified Friedmann equation -- one of the equations of motion generated by the Hamiltonian constraint -- has been derived explicitly on the genuine quantum level in context of isotropic cosmology with dust field as the internal clock \cite{HP11-dust}.}

In its present form, however, the construction of the framework of LQC used by the majority of the community (later referred to as the \emph{mainstream LQC} or \emph{standard LQC}) involves making particular choices between non-equivalent alternatives in certain key steps of the construction. One of such steps is known as ``regularization process'', and consists of reexpressing the Hamiltonian constraint in terms of the extended operators (i.e., holonomies and fluxes). In the pioneering works \cite{Boj99a,Boj99b,APS06a,APS06b,APS06c} part of the gravitational Hamiltonian constraint involving the extrinsic curvature (the so-called ``Lorentzian part'') has been regularized by reexpressnig it in terms of the spatial Ricci curvature. While it is possible to implement it in full LQG \cite{AAL14,AALM,ALM15}, it differs significantly from the regularization algorithm originally proposed by Thiemann. Unlike in standard quantum mechanics, in LQG it is not known whether different regularization algorithms lead to similar dynamical predictions. Indeed, the quasi-phenomenological analysis of the full LQG scalar constraint in its isotropic sector\cite{DL17a,DL17b,LR19} -- performed via evaluating the expectation values on coherent states peaked on isotropic cosmological spacetimes -- has revealed that, in the leading order in $\hbar$, the effective Hamiltonian generating the dynamics differs significantly from the Hamiltonian of effective LQC.
An alternative approach -- known as the Quantum Reduced Loop Gravity (QRLG) and based on the quantization of those spacetimes which, upon a suitable gauge-fixing, take diagonal form -- is claimed to yield yet different corrections \cite{ABCL17}.
On the other hand, if one implements in the context of studies of \cite{DL17a} the construction of the Lorentzian part of the Hamiltonian constraint proposed in \cite{ALM15}, one is left with the mainstream LQC effective constraint as the leading order approximation.
In order to track down the nature of this discrepancy, it is then important to reexamine the implementation of the original Thiemann algorithm in full (that is, including the Lorentzian part of the Hamiltonian constraint) in the LQC framework.
The LQC reduction to this regularization algorithm has already been considered in the literature \cite{YDM09}, however in those works the analysis was not developed to the level allowing for verification of the dynamical predictions. Our work \cite{ADLP18} and the detailed analysis presented in this article close this gap.  

With the ever extending reach of LQC, the dynamical consequences of different regularizations must be understood before further studies can be conducted. These studies include the several extensions beyond flat FLRW, like positive and negative curvature \cite{SKL06,APSV07,CK13,Van06}, inclusion of a cosmological constant \cite{Sin09,KP09,PA11} or extension to non-isotropic cosmologies  \cite{AWE09,AWE09b,SS17}.
Also, it is critical to extend the new construction to the context of perturbative LQC by studies similar to those of\cite{AAN12,AAN13b,BJMM18} or in context of nonperturbative inhomogeneous LQC like the studies of Gowdy cosmologies \cite{BD08a,BD08b,BOP17}. In the former case, some results have already been obtained \cite{Agu18}. To pave the way for all these constructions, we will present here a detailed analysis of the quantum model as well as its effective dynamics for the Thiemann regularization in the LQC framework.

In section \ref{Section_scalar_field} we present how the Thiemann regularization (denoted by `TR') can be implemented as an operator on the physical Hilbert space of LQC. For this purpose, we will work in the $\bar{\mu}$-scheme, also called {\it improved dynamics}. Since the Euclidean term can be treated as in mainstream LQC, we pay special attention to the Lorentzian part due to which non-trivial modifications arise. When coupled to a massless scalar field, the scalar constraint can be promoted to an evolution operator.
In section \ref{sec:evol-prop} we investigate certain properties of this evolution operator and its self-adjoint extensions.
In section \ref{sec:dynmSector} we discuss how the implementation of the scalar constraint leads to the physical Hilbert space with a suitable set of physical observables. All of this is in analogy to mainstream LQC and the numerical investigations can therefore be executed in the same way as in \cite{APS06a,APS06b,APS06c}.
In section \ref{sec:effDynm} the effective dynamics of this model is carefully investigated, and the solution to the equations of motion is found analytically.
The simulations of the quantum dynamics are presented in section \ref{sec:results} and are shown to be well approximated by the effective dynamics. This justifies the terminology.
In section \ref{sec:conclusion} we summarise our results and finish with a prospect on further research.

\documentclass[../main.tex]{subfiles}

\section{Flat FRW with scalar field}
\label{Section_scalar_field}

In this section we recall the framework behind isotropic LQC. For more details we refer to appendix \ref{ReviewA} or the several reviews in the literature (see e.g. \cite{Boj05,Ash09_Cos,AP11}). We pay special attention to different regularizations of the Hamiltonian operator and derive in detail the regularization from \cite{YDM09}, which is inspired by the Thiemann regularization of the Lorentzian part.

\subsection{Review of LQC kinematics}

The starting point of LQC is the Hamiltonian formulation of GR in terms of Ashtekar-Barbero variables \cite{Ash86,Ash87,Ash88,Bar94,Bar95}. The phase space of GR is coordinatized by the Ashtekar connection $A^i_a(x)$ and the inverse densitized triad $E^b_i(x)$ which, for isotropic flat FLRW spacetime, read ($a,b,...=1,2,3$ are spatial indices and $i,j,..=,1,2,4$ are internal $SU(2)$ indices)
\begin{align}\label{frw-metric}
A^i_a(x)= V_o^{-1/3} c\delta^i_a,\hspace{30pt}E^a_i(x)= V_o^{-2/3} p \delta^i_a
\end{align}
where $V_0$ is the coordinate volume of a chosen spatial cell. Upon reducing to the symmetric sector, their Poisson bracket on the reduced phase space becomes
\begin{align}
\{A^i_a(x),E^b_j(y)\}=8\pi G \gamma \delta^i_j\delta^b_a\delta^{(3)}(x,y)\hspace{30pt}\to \hspace{30pt}\{c,p\}=\frac{8\pi G \gamma}{3}
\end{align}
where $G$ is the gravitational coupling constant and $\gamma\in \mathbb{R}-\{0\}$ is a free choice and called the {\it Barbero-Immirzi parameter} \cite{RT97}.
Mimicking the quantization procedure in the full theory, one wants to regularize the classical constraints via holonomies of the connection.

As outlined in the appendix \ref{ReviewA}, we will work throughout this article with a different choice of variables. These are a rescaled connection and the physical volume of the chosen cell:
\begin{align}\label{bV_def}
b:=c\bar{\mu},\hspace{30pt} V:=p^{3/2},\hspace{30pt}\{b,V\}=\frac{2\alpha}{\hbar}
\end{align}
with $\alpha=2\pi G\hbar \gamma\sqrt{\Delta}$ and $\bar{\mu}$ the regularization parameter, used in what is known as the $\bar{\mu}$-scheme or {\it improved dynamics} \cite{APS06c}
\begin{align} \label{mu-bar-def}
\bar{\mu}:=\frac{\sqrt{\Delta}}{\sqrt{|p|}}, \hspace{30pt}\Delta:= 2\pi \sqrt{3}\gamma G \hbar \approx 2.61 \ell^2_{\rm Pl}
\end{align}
where $\ell_{\rm Pl}$ is the Planck length and $\Delta$ is the smallest non-vanishing area eigenvalue from the full theory.

The volume is promoted to a multiplication operator $\hat{V}$ on the kinematical Hilbert space $\mathcal{H}_{\rm gr}$, which is the subspace of symmetric states of $L_2(\bar{\mathbb{R}},d\mu_{\rm Bohr}(v))$. And the exponential  $\mathcal{N}:=e^{ib/2}$ is represented by a shift operator:
\begin{align}
\hat{V}|v\rangle=\alpha|v|\; |v\rangle,\hspace{30pt}    \hat{\mathcal{N}}|v\rangle=|v+1\rangle
\end{align}
where volume-eigenstates $|v\rangle$ are normalized with respect to the Kronecker delta
\begin{align}
\langle v | v' \rangle = \delta_{vv'}
\end{align}
This finishes the kinematical set up of LQC. Now, one has to turn towards quantization of the scalar constraint, which in terms of Ashtekar-Barbero variables reads
\begin{align}
C=C_E+C_L
\end{align}
where Euclidean and Lorentzian parts are respectively (details in the appendix \ref{ReviewA})
\begin{align}
C_E=\frac{1}{16\pi G}\frac{\epsilon_{ijk}E^a_jE^b_k}{\sqrt{\det(q)}}F^i_{ab},\hspace{30pt}C_L=-(1+\gamma^2)\frac{1}{16\pi G}\frac{\epsilon_{ijk}E^a_jE^b_k}{\sqrt{\det(q)}}\epsilon_{imn}K^m_a K^n_b
\end{align}
This will be focus of the next subsection.

\subsection{Scalar constraint with the new (Thiemann) regularization}
The regularization of the Euclidean part $C_E$ is explained in the appendix \ref{ReviewA}, and its quantization reads
\begin{align}
\hat{C}_E^{\bar\mu}[N] |v\rangle = \dfrac{3N\alpha}{4(16\pi G)\Delta}\left(F(v+2) \hat{\mathcal{N}}^4 - F_0(v) {\rm id} + F(v-2) \hat{\mathcal{N}}^{-4} \right)|v\rangle
\end{align}
where the functions $F_0$ and $F$ are given in (\ref{F-0-F}). The regularization of $C_L$ used in mainstream LQC is based on relations which are only true in cosmology:
\begin{align} \label{MAIN_true-only-in-cosmo}
\gamma K^i_a |_{\rm cos}= A^i_a|_{\rm cos}\hspace{20pt}{\rm and}\hspace{20pt}2\gamma^2 K^i_{[a}K^j_{b]}|_{\rm cos}=\epsilon_{ijk} F^k_{ab}|_{\rm cos}
\end{align}
Using these relations, one finds that in classical cosmology the Lorentzian part is proportional to the Euclidean part. It can therefore be regularized in the same way. Hence, we can say that the philosophy of mainstream LQC is ``first reduce, then regularize''. On the other hand, one can propose a new regularization scheme for $C_L$, which follows the opposite philosophy: ``first regularize, then reduce''. In other words, we first consider a regularization of $C_L$ which is valid in full GR -- incidentally, the one due to Thiemann \cite{Thi98a,Thi98b} and currently used in LQG -- and where the Lorentzian part is {\it not} proportional to the Euclidean part. Afterwards, we reduce to the sector of flat cosmology and promote the resulting expression to a quantum operator in LQC.\footnote{The philosophy behind this procedure is the same which led to the quantum operators for the Euclidean part, which was based on cosmological expressions {\it after} implementing the regularization (\ref{ThiemannI}).}
\\
\\
Let us start by pointing out the second Thiemann identity, which is true in full GR, and can be regularized using a regularization parameter $\epsilon > 0$ independent of the phase space variables:
\begin{align}\label{ThiemannII}
 \tau_j K^j_a=\frac{1}{8\pi G\gamma^3}\{\tau_j A^j_a,\{C_E[1],V\}\}=-\frac{1}{8\pi G\gamma^3 \epsilon} h_a\{h^\dagger_a,\{C_E^{\epsilon}[1],V\}\}+\mathcal{O}(\epsilon)
\end{align}
where $h_a$ is the holonomy of a path oriented along coordinate direction $a$ and of coordinate length $\epsilon$. $\tau_j:=-i\sigma_j/2$ are the generators of the Lie algebra $\mathfrak{su}(2)$, with $\sigma_j$ being the Pauli matrices. However, one has to be careful in passing from $\epsilon$ to $\bar{\mu}$, which is phase space dependent. Indeed, Thiemann identity (\ref{ThiemannII}) is only correct if $\epsilon$ is independent of the phase space point. Thus, instead of performing the replacement $\epsilon\to\bar{\mu}$ in (\ref{ThiemannII}), we make use of the following observation from \cite{YDM09}, which is true {\it only} in cosmology:
\begin{align}
\tau_j K^j_a = -\frac{4}{3\bar{\mu}(16\pi G)\gamma^3} h_a\{h^\dagger_a,\{C_E^{\bar{\mu}}[1],V\}\} + \mathcal{O}(\Delta)
\end{align}
where $\bar\mu$ is given in (\ref{mu-bar-def}). With this identity one finds (see appendix \ref{ReviewA})
\begin{align} \label{correct-new-CL}
C^{\bar{\mu}}_L[N] = -\frac{(1+\gamma^2)N}{\gamma^7(4\pi G)^4}\frac{\epsilon^{abc}  }{9\Delta^{3/2}}\text{Tr}\left(h_a\{h_a^\dagger,\{C_E^{\bar{\mu}}[1],V\}\} \sqrt V h_b\{h_b^\dagger,V\} \sqrt V h_c\{h_c^\dagger,\{C_E^{\bar{\mu}}[1],V\}\}\right)
\end{align}
\\
The quantization of (\ref{correct-new-CL}) on the Hilbert space of LQC can now be done in the standard way: promoting $h$ and $V$ to operators and recalling that $\widehat{\{.,.\}} = [.,.]/(i\hbar)$, we find
\begin{align}
\hat C^{\bar{\mu}}_L[N] = \frac{(1+\gamma^2)N}{\gamma^7(4\pi G)^4}\frac{i\;\epsilon^{abc}}{9\Delta^{3/2}\hbar^5}  \text{Tr}\left(\hat h_a[\hat h_a^\dagger,[\hat C_E^{\epsilon}[1],\hat V]] \sqrt{\hat V} \hat h_b [\hat h_b^\dagger,\hat V] \sqrt{\hat V} \hat h_c[\hat h_c^\dagger,[\hat C_E^{\epsilon}[1],\hat V]]\right)
\end{align}
Its action on $|v\rangle$ reads (details in appendix \ref{ReviewA})
\begin{align}
\hat{C}_L^{\bar{\mu}}[N]|v\rangle & = \frac{3N\alpha}{16\pi G \Delta 2^{10}}\frac{1+\gamma^2}{4\gamma^2}\left(G(v-4)\hat{\mathcal{N}}^{-8} - G_0(v) \id + G(v+4)\mathcal{\hat{N}}^8
\right)|v\rangle\label{Lorentzian_quant_op}
\end{align}
where the functions $G(v)$ and $G_0(v)$ are given in (\ref{horriblecoefficents}).

This is the new quantum operator for the Lorentzian part of the scalar constraint. The sum of the Euclidean part (\ref{Euclidean_quant_op}) and this Lorentzian part (\ref{Lorentzian_quant_op}) completes the alternative quantization of the scalar constraint for flat cosmology:
\begin{align}
\hat{C}^{\bar{\mu}}[N] := \hat{C}^{\bar{\mu}}_E[N]+\hat{C}^{\bar{\mu}}_L[N]
\end{align}

So far we discussed the gravitation degrees of freedom. In this work, we consider the matter content to be a massless, free scalar field $\phi$ that is minimally coupled to gravity. The field serves as a physical clock with respect to which we deparametrize the system. The action of matter is:
\begin{align}
S_\phi=-\frac{1}{2}\int_{\mathcal{M}}d^4 x\sqrt{-g}g^{\mu\nu} (\partial_\mu \phi)(\partial_\nu \phi)
\end{align}
Upon a Legendre transformation and in the presence of an isotropic, spatially flat metric, the above equation leads to the total scalar constraint:
\begin{align}\label{eq:Const-Def}
C_{tot}[N] = C_E[N]+C_L[N]+C_{\phi}[N],\hspace{50pt} C_{\phi}[N]= N\;|p|^{-\frac{3}{2}}p^2_\phi /2
\end{align}
where $p_\phi$ is the canonical conjugate momentum to $\phi$. We follow the strategy of \cite{ACS08}, where the lapse function is chosen to be $N=2V$. This convenient choice makes $C_{\phi}[2V]$ independent of the geometric variables.
Then, using Schr\"ondinger representation for $\phi$, the matter part of the constraint can be promoted to an operator
\begin{align}
\hat{C}_{\phi} = \id _{\mathcal{H}_{\rm gr}} \otimes (i \hbar \partial_\phi)^2
\end{align}
on the direct product Hilbert space $\Hil_{\rm kin}= \Hil_{\rm gr} \otimes \Hil_{\phi}$, with $\mathcal H_{\phi} = L_2(\mathbb R, d\phi)$.

To express the full quantum constraint equation in $\mathcal{H}_{\rm kin}$ one chooses a symmetric ordering for gravitational part of the scalar constraint with respect to the volume operator in the lapse function, i.e.
\begin{align}\label{Def_Theta}
-\hbar^2 \partial_\phi^2 =-2\sqrt{\hat{V}}(\hat{C}^{\bar{\mu}}_E[1]+\hat{C}^{\bar{\mu}}_L[1])\sqrt{\hat{V}}=:\hbar^2 \Theta_{\rm TR}
\end{align}
For the physical time evolution, one has to take the square root of (\ref{Def_Theta}) and hence we will investigate $\sqrt{|\Theta_{\rm TR}|}$ in the next chapter.\\
For the remainder of this paper we will proceed in a ``large $v$ approximation'', where the operator shall be defined only in the region $v>8$ such that the absolute values in the functions $F$ and $G$ may be dropped. In this case the expressions simplify to
\begin{align} \label{correct-Cs}
\begin{array}{c}
\hat{C}^{\bar\mu}_E[N] |v\rangle = \dfrac{3N\alpha}{2(16\pi G)\Delta}\left(-(v+2)\hat{\mathcal{N}}^4+2v \ \id-(v-2)\hat{\mathcal{N}}^{-4}\right)|v\rangle
\\
\\
\hat{C}^{\bar\mu}_L[N] |v\rangle = \dfrac{3N\alpha}{2(16\pi G)\Delta} \dfrac{1+\gamma^2}{4\gamma^2}\left((v+4)\hat{\mathcal{N}}^8-2v \ \id+(v-4)\hat{\mathcal{N}}^{-8}\right)|v\rangle
\end{array}
\end{align}
Plugging this into (\ref{Def_Theta}) we find finally:
\begin{align}\label{final_theta}
\Theta_{\rm TR}= \frac{3}{(16\pi G)\hbar^2\Delta}\sqrt{\hat{V}}\left(-s
\hat{\mathcal{N}}^4\hat{V}\hat{\mathcal{N}}^4+ \hat{\mathcal{N}}^2\hat{V}\hat{\mathcal{N}}^2+2(s-1)\hat{V}+\hat{\mathcal{N}}^{-2}\hat{V}\hat{\mathcal{N}}^{-2}- s\hat{\mathcal{N}}^{-4}\hat{V}\hat{\mathcal{N}}^{-4}\right)\sqrt{\hat{V}}
\end{align}
where $s:=(1+\gamma^2)/(4\gamma^2)$.\\
Unlike the standard LQC, where the evolution operator is a difference operator of the $2$nd order, in this case $\Theta_{\rm TR}$ is a difference operator of the $4$th order.

\documentclass[./main.tex]{subfiles}

\section{Properties of the evolution operator}
\label{sec:evol-prop}

Unlike the full LQG, the models of LQC (including the one investigated here) are usually sufficiently simple to 
allow determining explicitly the spectrum of the quantum Hamiltonian constraint and its components, as well as evaluating explicitly the physical Hilbert space basis elements defined by the spectral decomposition of these 
operators. Having that at one's disposal, it is then relatively straightforward to solve the Hamiltonian constraint 
using group averaging methods \cite{ALMMT-ga,Mar95,Mar99,Mar00}. These techniques (standard for LQC, \cite{APS06b}) will be 
employed here directly. A central step in this application is the systematic spectral analysis of the evolution 
operator $\Theta_{\rm TR}$.\\
The operator itself is well defined on the domain of finite sums of the volume eigenstates $|v\rangle$ being dense 
in $\Hil_{\rm gr}$. However, the problem is that $\Hil_{\rm gr}$ itself is nonseparable. Fortunately, the method of 
splitting $\Hil_{\rm gr}$ into superselection sectors, used in mainstream LQC \cite{ABL,APS06b}, can still be applied here:  
the sets (the 'lattices') $\lat_{\epsilon} = \epsilon+4\mathbb{Z}$, $\epsilon\in (0,4]$ are preserved by action of $\Theta_{\rm TR}$ and the set of observables used to describe the dynamics (which is the case here, as in the mainstream LQC). Hence, one can divide $\Hil_{\rm gr}$ into separable subspaces of square summable functions supported on a given 
lattice. The structure of this division allows to select just one superselection sector and work with it without loss of generality of the results. We then focus our attention on the sector corresponding 
to $\epsilon=4$.\footnote{The sector of states $|v=0\rangle$ decouples from the rest of the lattice, thus evolves independently.} \\
Furthermore, we use the fact that the matter field present in the model is parity-invariant (that is, it is invariant with respect to 
the change of sign of $v$ encoding the triad orientation), to conclude that the parity reflection is a large gauge transformation. In such situation we
can further divide the Hilbert space into the superselection sectors of symmetric and antisymmetric states, of which we choose the 
former.\footnote{Choosing the antisymmetric sector in LQC models without fermions affects only the details of the discrete spectra, thus does 
not produce significant differences in the dynamical predictions. See for example \cite{BP08}.} As a consequence, we end up with the 
sector of square summable functions supported on the semi-lattice $4\mathbb{Z}^+$.\\ 
Having selected the separable superselection sector, we can now probe the spectrum of (the self-adjoint extensions of) $\Theta_{\rm TR}$ and construct 
the basis of the physical Hilbert space composed of the ``energy'' eigenvectors. For that we need to analyze the generalized eigenvalue 
problem for this operator.

\subsection{The eigenvalue problem and representations}
\label{section:eigen-p}

Given the choice of superselection sectors discussed above, we restrict the domain of definiteness of $\Theta_{\rm TR}$ to the space $\dom$ of finite sums
\begin{equation}\label{eq:dom-def}
	\dom :=  \{ |\psi\rangle \in \Hil_{\rm gr}:\ |\psi\rangle = \sum_{n=1}^{N}c_n|4n\rangle, \ c_n\in\mathbb{C}, \ N\in \mathbb{N} \} .
\end{equation}
Consider now the generalized eigenvalue problem
\begin{align}\label{eq:eig-prob}
(\Psi_{\lambda}| \Theta^{\dagger}_{\rm TR} - \lambda^\star \id |\psi\rangle = 0,\hspace{30pt}  \forall |\psi\rangle \in \dom \; . 
\end{align}
The direct inspection of the form of $\Theta_{\rm TR}$ \eqref{final_theta} shows that the above equation can be solved recursively as follows:
\begin{itemize}
	\item The value of $\Psi_{\lambda}(v) := (\Psi_{\lambda}|v\rangle^\star$ at $v=12$ is determined by the pair 
		$\Psi_{\lambda}(v=4), \Psi_{\lambda}(v=8)$ ($v=0$ decouples, while for $v=-4$ we use the symmetry of $\Psi$).
	\item The value of $\Psi_{\lambda}(v) := (\Psi_{\lambda}|v\rangle^\star$ at $v=16$ is determined by the triple 
		$\Psi_{\lambda}(v=12), \Psi_{\lambda}(v=8), \Psi_{\lambda}(v=4)$ ($v=0$ again decouples).
	\item For each $n \in \mathbb Z^+$, the value $\Psi_{\lambda}(v=4(n+4))$ is determined by a quadruple 
		$\Psi_{\lambda}(v=4(n+3)),\Psi_{\lambda}(v=4(n+2)),\Psi_{\lambda}(v=4(n+1)),\Psi_\lambda(v=4n)$.
\end{itemize}
In consequence the whole eigenvector is uniquely determined by the first two values $\Psi_{\lambda}(v=4), \Psi_{\lambda}(v=8)$, 
thus the space of solutions has dimension $2$.
\\
A particularly interesting subset of solutions are the eigenvectors corresponding to $\lambda\in\re$ as all the physical Hilbert space 
elements will necessarily belong to this subset. Under this restriction the real and imaginary part of $\Psi_{\lambda}(v)$ 
decouple due to reality of operator $\Theta_{\rm TR}$. Thus, without loss of generality one can assume the reality of $\Psi_{\lambda}(v)$.\\
Unfortunately, even with this simplification the eigenvalue problem can only be solved numerically (see fig.~\ref{fig:eigenf}). What we can infer 
from the numerical solutions is the qualitative behavior of the eigenfunctions. Since the dynamics is generated by the operator 
$\sqrt{|\Theta_{\rm TR}|}$ we are interested in positive eigenvalues $\lambda=\omega^2$. For a given eigenfunction 
$\Psi_{\lambda=\omega^2}$, we observe two $\omega$-dependent regions for $v\in4\mathbb{Z}^+$: the exponential suppression region (for small $v$) 
and the (quasi)-oscillatory region for $v$ above a certain critical ($\omega$-dependent) value. This picture is quite characteristic 
to the cosmic bounce, however the oscillatory pattern is much more complicated than in the mainstream LQC, indicating much richer large volume (or more precisely low energy) structure. To determine it, we employ the analytic studies of the eigenvector asymptotics, using the technique originally specified in \cite{KP09}. In order to not break the reasoning flow, the details of the derivation are presented in 
Appendix \ref{app:asympt}. Here we just present the result:
\begin{equation}\label{eq:eig-asympt}
	\Psi_{\lambda=\omega^2} (v)
	= \dfrac{1}{\sqrt v} N_F(\omega) \cos(k\ln(v)+\sigma_F(\omega)) 
	+ \dfrac{1}{|v|} N_S(\omega) \cos( \Omega_S v + \kappa(\omega)/v + \sigma_S(\omega)) + O(v^{-2}) , 
\end{equation}
where $N_F$, $N_S$ are normalization constants, $k$, $\Omega_S$ and $\kappa$ are related in the following way
\begin{align}\label{eq:eig-asympt-params}
	\omega &= \sqrt{12\pi G} k , \qquad 
	\cos(4\Omega_S) = \frac{1-2s}{2s} , \qquad
	\kappa(\omega) = \frac{2s-3}{2\sqrt{4s-1}} + \frac{4 s k^2}{\sqrt{4s-1}} ,
\end{align}
and $\sigma_F$, $\sigma_S$ are phase shifts.
\begin{figure}[h!]
	\begin{centering}
		\includegraphics[width=0.7\textwidth]{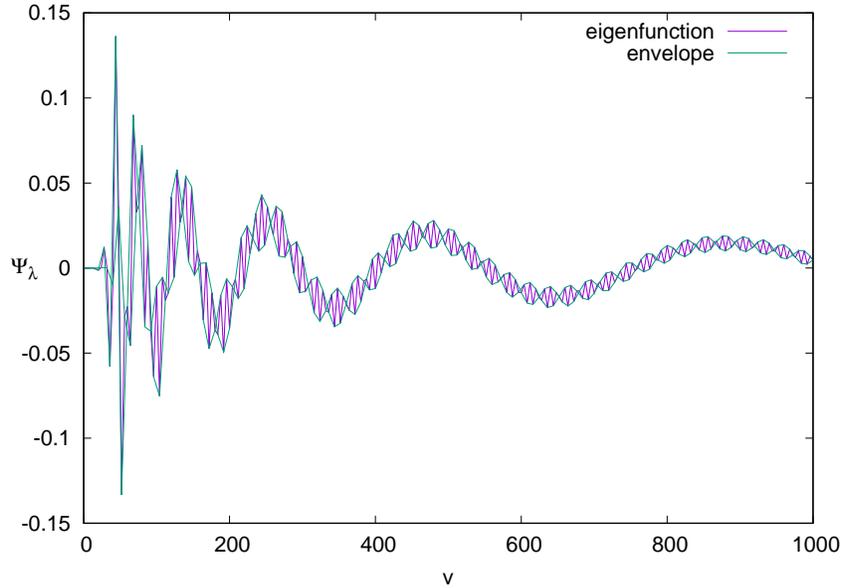}
	\end{centering}
	\caption{An example of the eigenfunction $\Psi_{\lambda}$ to the evolution operator $\Theta_{\rm TR}$ corresponding to the eigenvalue $\lambda = 12\pi G k^2$ (where $k=10$). One can observe: $(i)$ the reflected wave pattern and $(ii)$ the asymptotic approach to a combination of two asymptotic waveforms given by eq. \eqref{eq:eig-asympt}. For better visualization of the behavior an envelope (green line) compensating for rapid oscillations due to $\Omega_s>\pi$ has been added.}
	\label{fig:eigenf}
\end{figure}

The comparison with the asymptotic form of evolution operator eigenfunctions in mainstream LQC \cite{APS06b,PA11} shows that, for large 
$v$, the eigenfunctions $\Psi_{\lambda=\omega^2}$ converge to a linear combination of two terms: one coincides with the eigenfunction obtained in mainstream LQC; the other agrees with the eigenfunction of mainstream LQC with positive cosmological constant. Comparing the expressions for $\Omega_S$ and $\kappa$ (the latter up to an additive constant) with their analogs in mainstream LQC listed in eq.~(4.2) of \cite{Paw16} allows to cast the new model as mainstream LQC with a cosmological constant given by
\begin{equation}\label{barecosconst}
\Lambda = \frac{8\pi G\rho_{c}^{{\rm LQC},\Lambda=0}}{1+\gamma^2}
\end{equation}
where $\rho_{c}^{{\rm LQC},\Lambda=0} = 3/(8\pi G\gamma^2\Delta)$ is the critical energy density of matter as obtained in mainstream LQC without cosmological constant. In the following, we will denote this quantity simply by $\rho_c$.
\\
As is well known \cite{Sin09}, mainstream LQC admits a classical limit in which the cosmological constant is renormalized. The effective cosmological constant is related to the ``bare'' one, $\Lambda$, by
\begin{align}
\Lambda_{{\rm eff}} = \Lambda\left(1-\frac{\Lambda}{8\pi G\rho_c}\right) 
\end{align}
which, given (\ref{barecosconst}), in the new model reads
\begin{align} \label{lambda-eff-from-q}
\Lambda_{{\rm eff}}= \frac{3}{\Delta(1+\gamma^2)^2} .
\end{align}
Given the similarity between the new model and mainstream LQC with cosmological constant, it is convenient to use the methods already applied in the literature \cite{PA11}.\\

The crucial first step is the transformation to the momentum $b$
\begin{equation}\label{eq:vb-transform}
	\tilde \psi(b) = [\mathcal{F}\psi](b) = \sum_{v\in\lat_{4}} |v|^{-1/2} \psi(v) e^{(i/2)vb} , 
\end{equation}
where for the selected superselection sector, the domain of $b$ is a circle of radius $1/2$ and the parity reflection symmetry transforms into the symmetry
\begin{equation}
	\tilde \psi(b) = \tilde \psi(\pi-b) .
\end{equation}
In this coordinate the evolution operator takes the form
\begin{equation}\label{eq:theta-brep}
	\Theta_{\rm TR} = 12\pi G\gamma^2 \left[(\sin(b)\partial_b)^2 - s (\sin(2b)\partial_b)^2\right] .
\end{equation}
Plugging it into the Klein-Gordon form (\ref{Def_Theta}) of the Hamiltonian constraint, we observe that in the coordinates $(\phi,b)$ it becomes a partial differential 
equation of mixed signature with the boundary defined by
\begin{equation}
	\cos(b_o) = 1/\sqrt{4s} .
\end{equation}
For $b$ such that $|\cos(b)|<\cos(b_o)$ the constraint is hyperbolic, whereas for $|\cos(b)|>\cos(b_o)$ it becomes elliptic, which indicates that the latter
is a classically forbidden region. It is then sensible to introduce a coordinate $x(b)$ such that 
\begin{equation}\label{eq:Theta-x}
	\Theta_{\rm TR} = -12\pi G\,\sgn(|x|-x_o)\partial_x^2 , \qquad x_o = -x(b_o) .
\end{equation}
Unlike in \cite{PA11} the relation $x \leftrightarrow b$ can be expressed analytically and is given by
\begin{equation}\label{eq:x-func}
	x(b) = \begin{cases} 
						\frac{1}{2}\ln\left[ 1 - \frac{2\sqrt{1-(1+\gamma^2)\sin^2(b)}}{\cos(b) + \sqrt{1-(1+\gamma^2)\sin^2(b)}}\right] - \frac{\pi}{2} , & 
						0 < b < b_o ,\\
						-\arctan\left(\frac{\cos(b)}{\sqrt{(1+\gamma^2)\sin^2(b)-1}}\right) , & b_o < b < \pi - b_o , \\
						\frac{1}{2}\ln\left[ 1 - \frac{2\sqrt{1-(1+\gamma^2)\sin^2(b)}}{\cos(b) + \sqrt{1-(1+\gamma^2)\sin^2(b)}}\right] + \frac{\pi}{2} , & 
						\pi-b_o < b < \pi .\\
	       \end{cases}
\end{equation}
The new coordinate spans the entire real line, with 
\begin{equation}
	\lim_{b\to 0} x(b) = -\infty ,\quad x(b_o) = -\pi/2 ,\quad x(\pi/2) = 0 ,\quad x(\pi-b_o)=\pi/2 ,\quad \lim_{b\to \pi} x(b) = +\infty , 
\end{equation}
and is globally continuous, but not differentiable at the points $x=\pm\pi/2$. The parity reflection symmetry transforms into the symmetry with respect to the 
reflection about $x=0$, namely $\psi(x) = \psi(-x)$ (this follows from the fact that $\psi(v) = \psi(-v)$ implies $\tilde\psi(b) = \tilde\psi(\pi - b)$ and that, by direct observation of (\ref{eq:x-func}), $x(\pi-b) = -x(b)$). Due to the non-differentiability at $\pm \pi/2$, an application of the form \eqref{eq:Theta-x} to the eigenvalue problem (\ref{eq:eig-prob}) will generate nontrivial boundary terms at $x=\pm\pi/2$. The derivation, being a straightforward application of 
the solution from Sec.~IIIA of \cite{PA11}, is briefly outlined in the Appendix \ref{sec:aux-math}. Its result is that at $x=\pm\pi/2$ the eigenfunction 
$\Psi_{\lambda}(x) := \tilde \Psi_{\lambda}(b)$ corresponding to an arbitrary complex eigenvalue $\lambda$ needs to be continuous but not necessarily differentiable, thus satisfying
\begin{equation}\label{eq:eig-sol}
	\Psi_{\lambda}(x) = \zeta \begin{cases}
	                      \cos(\sqrt{\lambda/(12\pi G)}\,|x| + \varphi), & |x|>\pi/2 , \\
	                      \frac{\cos(\sqrt{\lambda/(12\pi G)}(\pi/2) + \varphi)}{\cosh(\sqrt{\lambda/(12\pi G)}(\pi/2))} 
	                      \cosh(\sqrt{\lambda/(12\pi G)}\,x), & |x|\leq\pi/2 ,
	                    \end{cases}
\end{equation}
where $\zeta$ is a free complex constant and $\varphi$ is a free phase shift.
The non-differentiability at $\pm\pi/2$ will be a crucial determinant of the structure of self-adjoint extensions of $\Theta_{\rm TR}$.

\subsection{Self-adjointness, extensions}
\label{self-adj-ext}

A crucial initial step in probing the unitary time evolution of physical states generated by $\Theta_{\rm TR}$ (more precisely $\sqrt{|\Theta_{\rm TR}|}$) is 
determining whether it admits any self-adjoint extension and whether such extension is unique. Within the mainstream LQC framework the evolution operator of the the model of flat isotropic universe with scalar field admits a unique self-adjoint extension, whereas the analogous operator in presence of positive cosmological constant admits an entire family. Since the large $v$ asymptotics of eigenvectors features the properties of the eigenvectors of both these models (see subsection \ref{section:eigen-p}), the answer to the above question is nontrivial. To answer it, we again employ the techniques from \cite{KP09,Paw16}.

The direct inspection of \eqref{final_theta} shows that it is symmetric. Also, the elements $\psi$ of the domain $\dom$ satisfy 
(due to smoothness in $b$, as they are the finite sums defined in \eqref{eq:dom-def} transformed via \eqref{eq:vb-transform}) the conditions
\begin{equation}\label{eq:deriv-at-boundary}
	\lim_{x\to\pm\infty} \partial_x \psi(x) = \lim_{x\to\pm\infty} (\partial_x b) \partial_b \psi(x(b)) = 0, \ \ \ \ \ \text{and} \ \ \ \ \ [\partial_x\psi](\pm\pi/2) = 0.
\end{equation}
due to $\partial_x b$ being zero at those points.

In order to determine the structure of self-adjoint extensions of $\Theta_{\rm TR}$ we need to investigate its deficiency spaces \cite{ReedSimon}. They can be
defined as the spaces of normalizable solutions to the eigenvalue problem \eqref{eq:eig-prob} for the eigenvalues $\pm 24\pi G i$\footnote{Precisely, 
the deficiency functions are defined as normalizable solutions to the eigenvalue problem for $\lambda=\pm i$, however one can safely rescale the 
eigenvalues by any real factor.} 
\begin{equation}\label{eq:defic-def}
	\mathcal{K}^{\pm} = \{ \psi\in \Hil_{\rm gr}:\ \forall \chi\in\Hil_{\rm gr}\ \langle\psi|\Theta^{\dagger}_{\rm TR} \mp 24\pi Gi\id|\chi\rangle = 0 \} .
\end{equation}
The form of all $\Psi^\pm\in\mathcal{K}^\pm$ can be determined by solving the eigenvalue equation of $\Theta_{TR}$ (as given in (\ref{eq:Theta-x})) for $\lambda_\pm = \pm 24 \pi G i$. Neglecting the non-decaying solutions, demanding continuity at $x = \pm \pi/2$, and using the symmetry $x \to -x$, we find
\begin{equation}\label{eq:Psi_pm}
  \Psi^\pm(x) = \zeta \begin{cases}
										(e^\pi-1) e^{(\pm i-1)|x|} , & |x|>\pi/2 , \\
										e^{(1\pm i)x} + e^{-(1\pm i)x} , & |x|\leq\pi/2 ,
                  \end{cases}
\end{equation}
where the phase $\varphi$ has been absorbed in the free complex constant $\zeta$. Therefore, both deficiency spaces are of dimension $1$. In such case, the operator admits  a family of self-adjoint extensions, each associated with a unitary transformation $U_\sigma : \mathcal{K}^+ \to \mathcal{K}^-$. For ${\rm dim}(\mathcal{K}^{\pm})=1$ all the unitary transformations are just phase rotations, that is, for chosen \emph{normalized} deficiency functions $\Psi^{\pm}_o$, $U_\sigma$ acts as
\begin{equation}
	U_{\sigma}\Psi^+_o = e^{i\sigma}\Psi^-_o .
\end{equation}
The extensions of the domain are by Theorem X.2 of \cite{ReedSimon} of the form 
\begin{equation} 
	\mathcal{D}_{\sigma} = \{ \psi + c(\Psi^+_{o} + U_{\sigma}\Psi^+_{o}), \ \psi\in\dom, \Psi^+_{o}\in\mathcal{K}^+, c\in\mathbb{C} \} .
\end{equation}
A convenient property of the extension elements is that the ratios of their left and right derivatives at the boundary $x=\pm\pi/2$ depends on the extension only. Indeed, the elements of $\dom$ do not contribute to the derivatives, which leaves only the relatively easy to evaluate contribution of the deficiency functions: for all $\psi_\sigma\in\dom_\sigma$ one has
\begin{align} \label{eq:ext-cond}
	\frac{\lim_{x\to^+\pi/2}\partial_x\psi_\sigma}{\lim_{x\to^-\pi/2} \partial_x\psi_\sigma} 
	& = \frac{\lim_{x\to^--\pi/2}\partial_x\psi_\sigma}{\lim_{x\to^+-\pi/2}\partial_x\psi_\sigma}
	= \frac{\lim_{x\to^+\pi/2}\partial_x(e^{-i\sigma/2}\Psi^+_o + e^{i\sigma/2}\Psi^-_o)}{\lim_{x\to^-\pi/2}\partial_x(e^{-i\sigma/2}\Psi^+_o + e^{i\sigma/2}\Psi^-_o)}
	= \notag
	\\
	& = \tanh(\pi/2)\frac{\cos(\sigma/2)+\sin(\sigma/2)}{\cos(\sigma/2)-\sin(\sigma/2)} =: -\tan(\beta)
\end{align}
where in the second step we used (\ref{eq:deriv-at-boundary}).
\\
By direct inspection one can check, that the relation between $\beta\in [0,\pi)$ and $U_\sigma$ is bijective, thus $\beta$ can replace $\sigma$ as the extension label. This in turn allows to associate to a choice of a self-adjoint extension a physical meaning: each extension corresponds to particular boundary conditions at $x=\pm\pi/2$. 

Each extension (now denoted as $\dom_\beta$) of the original domain $\dom$ is dense in $\Hil_{\rm gr}$. Furthermore, by self-adjointness, the spectrum of each extension $\Theta_\beta$ of $\Theta_{\rm TR}$ is real. Since in the considered physical system only the positive part of $\Theta_{\rm TR}$ is relevant (due to the solution of the constraint (\ref{Def_Theta})), its spectral decomposition will distinguish a proper\footnote
{
By choosing for example a smooth function supported on a compact interval within $|x|<\pi/2$, one can show explicitly that $\Theta_{\rm TR}$ is not positive definite.
}
subspace $\Hil_\beta$ of $\Hil_{\rm gr}$. Each subspace $\Hil_{\beta}$ is spanned by a basis composed of normalized eigenvectors \eqref{eq:eig-sol} corresponding to eigenvalues $\lambda>0$ and satisfying the condition \eqref{eq:ext-cond} (reducing the originally $2$-dimensional eigenspace to a $1$-dimensional one)
\begin{equation}\label{eq:eigenv-real}
  \Psi_{\beta,k}(x) = \zeta \begin{cases} 
                          \cos(k|x|+\varphi({\beta,} k)) , & |x|>\pi/2 , \\
                          \frac{\cos(k\pi/2+\varphi({\beta,} k))}{\cosh(k\pi/2)}\cosh(kx) , & |x|\leq \pi/2 ,
                        \end{cases}
\end{equation} 
where $\lambda=\omega^2=12\pi G k^2$ and $\varphi(\beta,k)$ is fixed by \eqref{eq:ext-cond} to
\begin{equation}
  \tan(k\pi/2+\varphi({\beta,}k)) = \tan(\beta)\tanh(k\pi/2) .
\end{equation}
Thus, the eigenspaces are non-degenerate.

Recalling the asymptotic behavior of eigenfunctions for large $v$ \eqref{eq:eig-asympt}, we observe that the considered eigenfunctions are Dirac delta normalizable (i.e. their norm is proportional to $\delta(0)$), thus the spectrum of $|\Theta_\beta|$ is continuous (due to non-degeneracy). Furthermore, the convergence (modulo the shift in $v$) of the eigenfunctions $\Psi_{\beta,k}$ to the analogous eigenfunctions of the mainstream LQC evolution operator\footnote
{
This follows directly from the observation that both families of eigenfunctions share the same leading order asymptotics (modulo phase shifts), see \eqref{eq:eig-asympt}.
}
allows to conclude that:
\begin{enumerate}[(i)]
	\item the spectrum of $|\Theta_\beta|$ is the entire positive real line, ${\rm Sp}(|\Theta_{\beta}|)=\mathbb{R}^+$
	\item following the reasoning of Appendix \ref{sec:aux-math}, we find the normalization constant
\begin{equation}\label{eq:basis-norm}
	\zeta = \dfrac{4}{\sqrt{|k|}}
\end{equation}
\end{enumerate}
From now on, we will denote the normalized eigenfunctions by $e_{\beta,k}$.

To summarize: throughout this section we have established the existence of self-adjoint extensions of the evolution operator $\Theta_{\rm TR}$; we characterized the family of these extensions and explicitly constructed an orthonormal (in the sense of distributions) basis of a subspace $\Hil_{\beta} \subset \Hil_{\rm gr}$ relevant for the physical model considered. $\Hil_{\beta}$ is spanned by the eigenstates of the corresponding extension $|\Theta_\beta|$ of $|\Theta_{\rm TR}|$.
These structures will be used in the next section to construct the physical Hilbert space and probe the dynamical behavior of the model.

\documentclass[./main.tex]{subfiles}

\section{The dynamical sector}
\label{sec:dynmSector}

In order to complete the Dirac quantization program we need to:
\begin{enumerate}
\item construct the physical Hilbert space
\item construct a sufficiently large family of observables encoding physically relevant properties of the system
\item probe the dynamical behavior of a class of semiclassical states sufficiently rich to provide robust insights
\end{enumerate}
These steps will be performed in the next two subsections, following the methods already introduced in \cite{APS06b,APS06c,PA11}.

\subsection{Physical Hilbert space}

While the construction of a physical Hilbert space for constrained systems is a nontrivial task, systematic methods exist. One of the most convenient is the so-called ``group averaging'' \cite{ALMMT-ga} (which has been applied to mainstream LQC in \cite{APS06b}). 
Its main component is the construction of a rigging map which ``averages'' the kinematical states over a group of transformations generated by constraints.
In the case at hand this map takes the form ($\mathcal{D}_{\rm kin} :=\mathcal{D}\otimes \mathcal{S}(\re)\subset \mathcal{H}_{\rm kin}$)
\begin{equation} \label{eq:rig}
  \eta: \mathcal{D}_{\rm kin}  \to \mathcal{D}_{\rm kin}^\star , 
  \qquad 
  \eta(\psi) = \left( \int_{\re} \rd N e^{iNC_{\beta}} \psi\right)^\dagger ,
  \qquad
  C_{\beta} = -(\id\otimes\partial_{\phi}^2 + \Theta_{\beta}\otimes\id)\ .
\end{equation}
The physical Hilbert space is then defined as $\mathcal{H}_{\rm phy}:= \overline{{\rm Im}[\eta]}$, with an induced physical inner product (cf. \cite{ALMMT-ga})
\begin{equation}\label{phys_innerprod}
(\eta(\psi)|\eta(\psi'))_{\rm phy} := [\eta(\psi)](\psi') = \int_{\re} \rd N (\psi|e^{-iNC_{\beta}}\psi')_{\rm kin} 
\end{equation}
The space of physical states is a union of the positive and negative frequency superselection sectors (corresponding, respectively, 
to the positive and negative part of the spectrum of $i\partial_{\phi}$). The restriction to the positive frequency sector 
(per analogy with Klein-Gordon equation) can be safely performed
by just replacing $C_{\beta}$ in the expressions above with $C_{\beta}^{+} := \id\otimes i\partial_{\phi} + \sqrt{|\Theta_\beta|}\otimes\id$. To characterize the physical states, let us start by expanding $\psi(x,\phi) \in \mathcal D_{\rm kin}$ on the basis $(e_{\beta, k} \otimes \varphi_\sigma)(x,\phi) = e_{\beta, k}(x) e^{i\sigma\phi}$ of eigenstates of $\sqrt{|\Theta_\beta|} \otimes \id$ and $\id \otimes i\partial_\phi$:
\begin{equation} \label{psi-expanded-c}
\psi(x,\phi) = \int \rd k \rd \sigma \; c(k,\sigma) e_{\beta,k}(x) e^{i\sigma\phi}
\end{equation}
Using this, one finds for the physical state
\begin{align} \label{explicit-etapsi}
[\eta(\psi)](x,\phi) & = \left[\int \rd k \rd \sigma \; c(k,\sigma) \int_{\mathbb R} \rd N e^{iN (\omega(k) - \sigma)} e_{\beta,k}(x) e^{i\sigma\phi}\right]^\star = \notag
\\
& = 2\pi \left[\int \rd k \rd \sigma \; c(k,\sigma) \delta(\omega(k) - \sigma) e_{\beta,k}(x) e^{i\sigma\phi}\right]^\star = \notag
\\
& = 2\pi \int \rd k \; c^\star(k,\omega(k)) e_{\beta,k}^\star(x) e^{-i\omega(k)\phi}
\end{align}
where in the first step we observed that $e_{\beta,k}(x) e^{i\sigma\phi}$ is eigenstate of $C_\beta^+$ with eigenvalue $\omega(k) - \sigma$, and in the second we performed the integral over $N$ to obtain $2\pi \delta(\omega(k) - \sigma)$. Equation (\ref{explicit-etapsi}) makes it apparent that we can identify a physical state with a 1-parameter family $\Psi_\phi$ of elements of the gravitational Hilbert space $\mathcal H_{\beta}$: their components on the basis $e_{\beta,k}$ being $f_\phi(k) := 2\pi c(k,\omega(k)) e^{i\omega(k)\phi}$, so we have
\begin{align} \label{phys-state-gr-represented}
\eta(\psi) \ \to \ \Psi_{\phi}(x) := 2\pi \int \rd k \; c(k,\omega(k)) e_{\beta, k}(x) e^{i\omega(k)\phi}, \ \ \ \text{with} \ c(k,\omega(k)) := \big(e_{\beta,k} \otimes \varphi_{\omega(k)} | \psi\big)_{\rm kin}
\end{align}
This identification $\mathcal H_{\rm phy} \to \mathcal H_{\beta}$ preserves the scalar product. Indeed,
\begin{align}
(\eta(\psi)|\eta(\psi'))_{\rm phy} & = [\eta(\psi)](\psi') = \int \rd x \rd\phi \rd N\; \psi^\star(x,\phi) \int \rd k' \rd \sigma' \; c'(k',\sigma') e^{-iN(\omega(k') - \sigma')} e_{\beta,k'}(x) e^{i\sigma'\phi} \notag
\\
& = 2\pi \int \rd k \rd k' \rd \sigma \rd \sigma' \; c^\star(k,\sigma) c'(k',\sigma') \delta(\omega(k') - \sigma') \int \rd \phi \; e^{i\phi (\sigma'-\sigma)} \int \rd x e_{\beta,k}^\star(x) e_{\beta,k'}(x) \notag
\\
& = 2\pi \int \rd k \rd k' \rd \sigma \; c^\star(k,\sigma) c'(k',\omega(k')) \int \rd \phi \; e^{i\phi (\omega(k')-\sigma)} \delta(k-k') \notag
\\
& = (2\pi)^2 \int \rd k \rd \sigma \; c^\star(k,\sigma) c'(k,\omega(k)) \delta(\omega(k)-\sigma) \notag
\\
& = (2\pi)^2 \int \rd k \; c^\star(k,\omega(k)) c'(k,\omega(k))
\end{align}
which coincides with $(\Psi_\phi | \Psi'_\phi)_{\beta} = \int \rd x \Psi_\phi^\star(x) \Psi_\phi'(x) = \sum_{v \in \mathcal L_4} \Psi_\phi^\star(v) \Psi_\phi'(v)$.
\\
\\
Relation (\ref{phys-state-gr-represented}) allows to interpret the structure resulting from group averaging as the deparametrization ``on the quantum level'' of the system with respect to the scalar field, now attaining the role of an internal clock (or a matter time). Under this interpretation, the system is the vacuum one, i.e., only gravitational degrees of freedom are physical: hence, the role of the physical Hilbert space is played by the subspace $\Hil_{\beta}\subset\Hil_{\rm gr}$, and time evolution (in terms of the scalar field) is generated by a true Hamiltonian $\sqrt{|\Theta_{\beta}|}$. The unitary time-evolution operators are then
\begin{equation}
	U_{\beta,\phi,\phi'} : \Hil_{\beta} \to \Hil_{\beta} : \ U_{\beta,\phi,\phi'} := e^{i\sqrt{|\Theta_\beta|}(\phi'-\phi)} , 
	\qquad 
	\Psi_{\phi'}(x) = U_{\beta,\phi,\phi'} \Psi_\phi(x) .
\end{equation}
This interpretation will be used in the next subsection to provide an intuitive construction of physically useful observables.  

\subsection{Observables}\label{sec:obs}

The last component needed to describe the dynamical sector of the theory is a sufficiently rich set of physical observables. Mathematically, these should be Dirac observables, that is, operators $\hat O$ on $\mathcal D_{\rm kin}$ such that $[\hat O, C_\beta^+] = 0$. Indeed, if we are given such an operator, its action can be lifted to the physical Hilbert space $\mathcal H_{\rm phy}$ by the formula
\begin{equation} \label{Oeta-etaO}
\hat O \eta(\psi) := \eta(\hat O^\dag \psi)
\end{equation}
Then, calling $\psi' := \hat O^\dag \psi$, we can find the corresponding $\Psi'_\phi(x)$ according to equation (\ref{phys-state-gr-represented}), and therefore obtain the action of Dirac observable $\hat O$ on $\mathcal H_{\beta}$.
\\
\\
The simplest example of such an operator is the scalar field momentum $\hat P_{\phi} := \id \otimes \hat{p}_\phi : \mathcal{D}_{\rm kin} \to \mathcal{D}_{\rm kin}$ which, as we will now see, plays the role of energy and is a constant of motion. Clearly, it commutes with the constraint, so it is a Dirac observable. Then, its action on physical state $\eta(\psi)$ passes to the action on $\psi$, and so we find
\begin{equation}
\psi'(x,\phi) = (\hat P_\phi^\dag \psi)(x,\phi) = \int \rd k \rd \sigma c(k,\sigma) e_{\beta,k}(x) i \hbar \partial_\phi e^{i \sigma \phi} = \int \rd k \rd \sigma [- \hbar \sigma c(k,\sigma)] e_{\beta,k}(x) e^{i \sigma \phi}
\end{equation}
Comparing this with the form (\ref{psi-expanded-c}), we read off $c'(k,\sigma) = - \hbar \sigma c(k,\sigma)$. Hence, following (\ref{phys-state-gr-represented}), we conclude that the physical state $\hat P_\phi \eta(\psi)$ is represent on $\mathcal H_{\beta}$ by
\begin{equation}
\Psi'_\phi(x) = -2\pi \hbar \int \rd k \; \omega(k) c(k,\omega(k)) e_{\beta,k}(x) e^{i\omega(k) \phi}
\end{equation}
In other words, the action of Dirac observable $\hat P_{\phi}$ is defined on $\mathcal H_{\beta}$ as
\begin{equation}
\hat P_{\phi} \Psi_\phi = -\hbar \sqrt{|\Theta_\beta|} \Psi_\phi
\end{equation}
Since, in light of the discussion above, $\sqrt{|\Theta_\beta|}$ can be thought of as the true Hamiltonian of the system, we see that $\hat P_{\phi}$ is in fact the energy operator. Moreover, in the $k$-representation of $\mathcal H_{\beta}$, the operator $\hat P_\phi$ acts by multiplication. This in particular means that, for the energy Gaussians $c_{\rm Gauss}(k,\omega(k))$ that we will consider for explicit computations later (see equation \eqref{eq:state-Gauss}), the expectation value and variance of $\hat P_{\phi}$ equal
\begin{equation}
\langle \hat P_{\phi} \rangle = \hbar \omega^\star, \ \ \ \ \ \ \ \ \ \ \Delta P_{\phi} = \hbar \sigma/\sqrt{2}
\end{equation}
\\
\\
The scalar field momentum $\hat P_\phi$ is not the only Dirac observable. In fact, given {\it any} self-adjoint operator $\hat L : \mathcal H_{\rm gr} \to \mathcal H_{\rm gr}$, the rigging map \eqref{eq:rig} defines a 1-parameter family of Dirac observables $\hat{L}_{\phi'}: \mathcal D_{\rm kin} \to \mathcal D_{\rm kin}$. These are known as partial observables \cite{Rov90,Rov91,Dit06,Dit07}, and are given as follows (see for example \cite{KLP09}):
\begin{equation}
\hat{L}_{\phi'} = \int_{\re} \rd N e^{-iNC^+_{\beta}} [\hat{L}\otimes\hat{\delta}_{\phi'}] e^{iNC^+_{\beta}}
\end{equation}
where $(\hat{\delta}_{\phi'}g)(\phi)=\delta(\phi-\phi')g(\phi)$ in the scalar field representation. Again, the action of operator $\hat{L}_{\phi'}$ lifts to the physical Hilbert space by \eqref{Oeta-etaO}, and hence it can be defined on $\mathcal H_{\beta}$ by the same procedure. First, we identify the kinematical state $\psi'$ that results from the action of $\hat L_{\phi'}$ on $\psi$:
\begin{align}
\psi'(x,\phi) & = (\hat L_{\phi'}^\dag \psi)(x,\phi) = \int \rd N \big(e^{-iNC_\beta^+} [\hat L \otimes \hat \delta_{\phi'}] e^{iNC_\beta^+} \psi\big)(x,\phi)
\\
& = \int \rd N \int \rd k \rd \sigma [e^{-iNC_\beta^+} e_{\beta,k} \otimes \varphi_\sigma](x,\phi) \big(e_{\beta,k} \otimes \varphi_\sigma | [\hat L \otimes \hat \delta_{\phi'}] e^{iNC_\beta^+} \psi\big)_{\rm kin} \notag
\\
& = \int \rd N \int \rd k \rd \sigma \int \rd k' \rd \sigma' c(k',\sigma') \times \notag
\\
& \quad e^{-iN (\omega(k) - \sigma - \omega(k') + \sigma')} e_{\beta,k}(x) e^{i\sigma\phi} \big(e_{\beta,k}, \hat L e_{\beta,k'}\big)_{\rm gr} \big(\varphi_\sigma | \hat \delta_{\phi'} \varphi_{\sigma'}\big)_{\phi} \notag
\\
& = 2\pi \int \rd k \rd \sigma \int \rd k' c(k',\omega(k') + \sigma - \omega(k)) e_{\beta,k}(x) e^{i\sigma\phi} \big(e_{\beta,k}, \hat L e_{\beta,k'}\big)_{\beta} e^{i\phi' (\omega(k') - \omega(k))}
\end{align}
where in the third step we introduced a resolution of identity in terms of $e_{\beta,k} \otimes \varphi_\sigma$; in the fourth step we expanded $\psi$ on the same basis, and evaluated the operators $e^{iNC_\beta^+}$ using the fact that $e_{\beta,k} \otimes \varphi_\sigma$ is eigenstate of $C_\beta^+$ with eigenvalue $\omega(k) - \sigma$; in the fifth step we observed that $\big(\varphi_\sigma | \hat{\delta}_{\phi'} \varphi_{\sigma'}\big)_{\phi} = \int \rd \phi \varphi_\sigma^\star(\phi) [\hat{\delta}_{\phi'} \varphi_{\sigma'}](\phi) = \int \rd \phi \delta(\phi - \phi') e^{i\phi (\sigma' - \sigma)} = e^{i\phi' (\sigma' - \sigma)}$ and then performed the integral over $N$ obtaining $\delta(\omega(k) - \sigma - \omega(k') + \sigma')$, which we used to consume the integral over $\sigma'$. Comparing this form of $\psi'$ with (\ref{psi-expanded-c}), we read off
\begin{align}
c'(k,\sigma) = \int \rd k' c(k',\omega(k') + \sigma - \omega(k)) \big(e_{\beta,k}, \hat L e_{\beta,k'}\big)_{\beta} e^{i\phi' (\omega(k') - \omega(k))}
\end{align}
Thus, the action of $\hat L_{\phi'}$ is defined on $\mathcal H_{\beta}$:
\begin{align}
[\hat L_{\phi'} \Psi_\phi](x) & = 2\pi \int \rd k c'(k,\omega(k)) e_{\beta,k}(x) e^{i\omega(k) \phi}\notag
\\
& = 2\pi \int \rd k \rd k' c(k',\omega(k')) \big(e_{\beta,k}, \hat L e_{\beta,k'}\big)_{\beta} e^{i\phi' \omega(k')} e^{i (\phi - \phi') \omega(k)} e_{\beta,k}(x) \notag
\\
& = \int \rd k \big(e_{\beta,k}, \hat L \Psi_{\phi'}\big)_{\beta} e^{i (\phi - \phi') \omega(k)} e_{\beta,k}(x) \notag
\\
& = \int \rd k \big(e^{-i (\phi - \phi') \sqrt{|\Theta_\beta|}} e_{\beta,k}, \hat L \Psi_{\phi'}\big)_{\beta} e_{\beta,k}(x) \notag
\\
& = [e^{i (\phi - \phi') \sqrt{|\Theta_\beta|}} \hat L \Psi_{\phi'}](x)
\end{align}
Taking the scalar product with a different state $\Psi'_\phi$, we find the matrix elements of $\hat L_{\phi'}$ on $\mathcal H_{\beta}$:
\begin{align} \label{eq:exp-practical}
\big(\Psi'_\phi | \hat L_{\phi'} \Psi_\phi\big)_{\beta} & = \big(e^{-i (\phi - \phi') \sqrt{|\Theta_\beta|}} \Psi'_\phi | \hat L \Psi_{\phi'}\big)_{\beta} = \big(\Psi'_{\phi'} | \hat L \Psi_{\phi'}\big)_{\beta} = \int \rd x \Psi_{\phi'}'^\star(x)[\hat{L}\Psi_{\phi'}](x)
\end{align}
These matrix elements coincide with the matrix element of $\hat L_{\phi'}$ on physical $\mathcal H_{\rm phy}$:
\begin{align}
\big(\eta(\psi') | & \hat L_{\phi'} \eta(\psi)\big)_{\rm phy} = \big(\eta(\psi') | \eta(\hat L_{\phi'}^\dag \psi)\big)_{\rm phy} = \int_{\mathbb R} \rd N \big(\psi' | e^{-iNC_\beta^+} \hat L_{\phi'}^\dag \psi\big)_{\rm kin} \notag
\\
& = \int_{\mathbb R} \rd N \int_{\mathbb R} \rd M \big(\psi' | e^{-i(N-M)C_\beta^+} [\hat{L} \otimes \hat \delta_{\phi'}] e^{-iMC_\beta^+} | \psi\big)_{\rm kin} \notag
\\
& = \int_{\mathbb R} \rd N \int_{\mathbb R} \rd M \int \rd k \rd \sigma \int \rd k' \rd \sigma' c'^\star(k,\sigma) c(k',\sigma') \times \notag
\\
& \quad \int \rd x \rd \phi \delta(\phi - \phi') \bigg[e^{i(N-M)C_\beta^+} e_{\beta,k}(x) e^{i\sigma\phi}\bigg]^\star [\hat{L} \otimes \id] e^{-iMC_\beta^+} e_{\beta, k'}(x) e^{i\sigma'\phi} \notag
\\
& = \int_{\mathbb R} \rd N \int_{\mathbb R} \rd M \int \rd k \rd \sigma \int \rd k' \rd \sigma' c'^\star(k,\sigma) c(k',\sigma') \times \notag
\\
& \quad e^{-i(N-M)(\omega(k) - \sigma)} e^{-iM(\omega(k') - \sigma')} e^{-i(\sigma - \sigma')\phi'} \int \rd x e_{\beta,k}^\star(x) \hat{L} e_{\beta, k'}(x) \notag
\\
& = (2\pi)^2 \int \rd k \int \rd k' c'^\star(k,\omega(k)) e^{-i\omega(k)\phi'} c(k',\omega(k')) e^{i\omega(k')\phi'} \int \rd x e_{\beta,k}^\star(x) [\hat{L} e_{\beta, k'}](x) \notag
\\
& = \int \rd x \Psi_{\phi'}'^\star(x) [\hat{L} \Psi_{\phi'}](x)
\end{align}
In the fourth step we represented the kinematical scalar product in $(x,\phi)$-variables and expanded $\psi(x,\phi)$ and $\psi'(x,\phi)$ on the basis $e_{\beta,k}(x) e^{i\sigma\phi}$; in the fifth step we used the fact that $e_{\beta, k}(x) e^{i\sigma\phi}$ is eigenstate of $C_\beta^+$ with eigenvalue $\omega(k) - \sigma$, and we consumed the integral over $\phi$; in the sixth step we observed that the integrals over $N$ and $M$ produce $2\pi \delta(\omega(k) - \sigma)$ and $2\pi \delta(\omega(k) - \sigma - \omega(k') + \sigma')$ respectively, and we used them to consume the integrals over $\sigma$ and $\sigma'$; finally, in the last step, we resummed the integeals over $k$ and $k'$, noting that the resulting object is the matrix element of $\hat O$ on wavefunctions of the form (\ref{phys-state-gr-represented}).
\\
\\
The particular ($1$-parameter families of) operators we are interested in will be constructed out of the following gravitational kinematical observables:
\begin{enumerate}[(i)]
	\item The compactified volume
		\begin{equation} \label{eq:obs-theta}
			\hat{\theta}_K := \arctan(\hat{V}/(\alpha K)) ,
		\end{equation}
		where $K$ is a positive real dimensionless constant chosen arbitrarily.
		The compactification is necessary, since the partial observables constructed out of $\hat{V}$ would lead outside of the physical Hilbert space, as it happens in the LQC model with positive cosmological constant \cite{PA11}.
	\item The matter energy density (which, by the constraint, is equal to the gravitational energy density):
		\begin{equation}\label{eq:obs-rho}
			\hat{\rho}_\phi = \frac{1}{2}\hat{V}^{-1}\Theta_{\beta}\hat{V}^{-1} .
		\end{equation}
	\item The Hubble rate 
		\begin{equation}\label{eq:obs-hub}
			\hat{H}_r = \frac{i}{6} [ \hat{V} ,\hat{V}^{-1} \Theta_{\beta}\hat{V}^{-1} ] .
		\end{equation}
\end{enumerate}
These observables together form a sufficiently large set to verify the correctness of the low energy limit of the model, as well as to identify novel properties characteristic of the chosen regularization scheme. The quantum evolution of these observables is analysed in the semiclassical regime and the results are presented and discussed in section \ref{sec:results}. However, before moving to that, we expose the construction and analysis of an effective description of the quantum model, as it is quite useful to evaluate the phenomenological aspects of the quantum theory through an effective model. Indeed, a very interesting feature of the mainstream LQC is that, for all the models whose genuine quantum dynamics was tested, the evolution of the universe was very accurately mimicked by certain classical effective models known under the name of \emph{classical effective LQC}. Since a lot of interesting results of LQC came from classical effective models (as the extrapolation of genuine quantum approach), it would be extremely useful to recover such effective approach for the regularization scheme investigated in this paper. This is the subject of the next section, while the comparison between the quantum and effective models is featured in section \ref{sec:results}.

\documentclass[../main.tex]{subfiles}
\section{Effective dynamics}
\label{sec:effDynm}

As we have seen, the asymptotic analysis of the eigenstates of the evolution operator leads to the conclusion that contributions appear that are due to the dynamics driven by an effective cosmological constant. This is quite surprising (since the ``bare'' theory we started from does not have any cosmological constant), and in stark contrast with standard LQC -- where, if one studies flat FLRW universe without ``bare'' cosmological constant, no ``emergent'' cosmological constant appears.
\\
In this section we shall construct a function $H_{\rm eff}$ on the phase space of cosmology which plays the role of ``effective Hamiltonian'' for the regularization presented in this paper. The name is justified since, as we will see in section \ref{sec:results}, the dynamics it generates well-approximates the quantum evolution of semiclassical states.
\\
Given this function, it is easy to derive the Hamilton's equations of motion which, surprisingly, can be integrated analytically. Once the full solution is known, we will study the asymptotic limit of vanishing energy density of matter, and find that, in the far past (with respect to cosmic time), the universe is essentially a contracting de Sitter solution with emergent cosmological constant (whose value agrees with (\ref{lambda-eff-from-q})). Moreover, as it has been observed in \cite{param-effective}, higher order corrections amount to a rescaling of Newton constant.
\\
Before delving in this analysis, however, it is instructive to consider the case of standard LQC. In this case, if we do not include a bare cosmological constant from the start, then the universe in the far past is a contracting solution of classical Friedmann equations without cosmological constant. The situation changes if a cosmological constant is present from the start.
\subsection{Effective dynamics of LQC with cosmological constant}
The (genuine quantum dynamics) of the flat FRW universe with a positive cosmological constant $\Lambda$ and a massless scalar field $\phi$ has been investigated in detail in \cite{PA11}. It appears that the quantum dynamics of this model is with high accuracy mimicked by the phase space dynamics generated by the effective Hamiltonian constraint of the form
\begin{align} \label{Hlqc-lambda}
C^{\rm LQC,\Lambda}_{\rm eff} = C_{\rm \phi,eff} + C^{\rm LQC,\Lambda}_{\rm gr, eff} = \dfrac{p_\phi^2}{2V} - \frac{3}{8\pi G\Delta\gamma^2} V \sin^2(b) + \dfrac{\Lambda}{8\pi G} V
\end{align}
where $p_\phi$ is the momentum conjugated to $\phi$. Evaluating $\dot{V}$ via Hamilton equation and eliminating the functions of $b$ via the constraint $C^{\rm LQC,\Lambda}_{\rm eff}=0$, one arrives at 
the modified Friedmann equation 
\begin{equation}
H_r^2 := \left(\frac{\dot{a}}{a}\right)^2= \frac{8\pi G}{3} \left[ \frac{\Lambda}{8\pi G}\left(1-\frac{\Lambda}{8\pi G \rho_{c}}\right) + \rho_{\phi} \left(1-\frac{\Lambda}{4\pi G \rho_{c}}\right) - \frac{\rho_{\phi}^2}{\rho_{c}} \right]
\end{equation}
where $\rho_{\phi} := p_{\phi}^2/(2V^2)$ is the energy density of the scalar field and we recall that $\rho_{c} = 3/(8\pi G \Delta \gamma^2)$ is the critical energy density in mainstream LQC when $\Lambda = 0$. In the limit of low energy density of matter, we can neglect the quadratic term in $\rho_\phi$, thus arriving at the effective ``classical'' Friedman equation
\begin{equation}
H_r^2 = \frac{8\pi \bar{G}}{3} \rho_{\phi} + \frac{\bar{\Lambda}}{3}
\end{equation}
with the effective cosmological constant $\bar{\Lambda}$ and gravitational constant $\bar{G}$ given by
\begin{equation}
\bar{\Lambda} = \Lambda\left(1-\frac{\Lambda}{8\pi G \rho_{c}}\right), \qquad
\bar{G} = G\left(1-\frac{\Lambda}{4\pi G \rho_{c}}\right) . 
\end{equation}
In other words, we can say that the asymptotic behavior of the spacetime obeys the classical Friedmann equations provided that we replace the ``bare'' Newton constant and the cosmological constant with ``dressed'' ones. Note that, in particular, if the bare $\Lambda$ is zero, then $\bar G = G$. More generally, solving $\Lambda$ for $\bar\Lambda$ and using the fact that the observed cosmological constant (i.e., $\bar\Lambda$) is extremely small, we find two possibilities:
\begin{align}
\Lambda_1 \approx \bar\Lambda \ \ \ \ \ \text{or} \ \ \ \ \ \Lambda_2 \approx \dfrac{3}{\Delta\gamma^2} - \bar\Lambda
\end{align}
Plugging these in the second equation, we find respectively
\begin{align}
\bar G_1 = \left(1-2\dfrac{\Delta\gamma^2}{3}\bar\Lambda\right) G \approx G \ \ \ \ \ \text{or} \ \ \ \ \ \bar G_2 = -\left(1-2\dfrac{\Delta\gamma^2}{3}\bar\Lambda\right) G \approx -G
\end{align}
We thus conclude that, while in the first case the bare quantities differ from the measured ones by a negligible quantity, in the second case this is not true, and in particular the bare Newton's constant has opposite sign than the measured one!
\subsection{Effective dynamics of the new model (without cosmological constant)}
Let us now go back to the new model, and consider the effective dynamics associated with it. Recall that the Hamiltonian constraint operator reads (from (\ref{eq:Const-Def}) and (\ref{Def_Theta}))
\begin{align}
\hat C_{tot} & := \hat C_{\phi} + \hat C_{E}^{\bar\mu}[1] + \hat C_{L}^{\bar\mu}[1] = \dfrac{1}{2} \hat p_\phi^2 \hat V^{-1} - \dfrac{\hbar^2}{2} \hat V^{-\frac{1}{2}} \Theta_{\rm TR} \hat V^{-\frac{1}{2}} \notag
\\
& = \dfrac{1}{2} \hat p_\phi^2 \hat V^{-1} -\frac{3}{32\pi G \Delta} \left(-s
\hat{\mathcal{N}}^4\hat{V}\hat{\mathcal{N}}^4+ \hat{\mathcal{N}}^2\hat{V}\hat{\mathcal{N}}^2+2(s-1)\hat{V}+\hat{\mathcal{N}}^{-2}\hat{V}\hat{\mathcal{N}}^{-2}- s\hat{\mathcal{N}}^{-4}\hat{V}\hat{\mathcal{N}}^{-4}\right)
\end{align}
Given this quantum Hamiltonian, it is possible to extract an effective one by the replacements of $\hat p_\phi \to p_\phi$, $\hat{\mathcal N} \to \mathcal N = e^{ib/2}$ and $\hat V \to V$. Using the fact that the classical quantities commute, we find
\begin{align}
\hat C_{tot} \to C_{\rm eff}^{\rm TR} & = \dfrac{p_\phi^2}{2V} + \frac{3}{16 \pi G \Delta} V \left[s\cos(4b) - \cos(2b) - (s-1)\right] \notag
\\
& = \dfrac{p_\phi^2}{2V} + \frac{3}{8 \pi G \Delta} V \sin^2(b) (1-4s) \left[1 + \frac{4s}{1-4s}\sin^2(b)\right]
\end{align}
Now, recalling that $s = (1+\gamma^2)/(4\gamma^2)$, we obtain $1-4s = -1/\gamma^2$, and so
\begin{align} \label{H-eff-class}
C_{\rm eff}^{\rm TR} = C_{\rm \phi,eff} + C_{\rm gr,eff}^{\rm TR} = C_{\rm \phi,eff} + C_{\rm gr,eff}^{\rm LQC, \Lambda=0} \left[1 - (1 + \gamma^2) \sin^2(b)\right]
\end{align}
where $C_{\rm \phi,eff} = p_\phi^2/(2V)$ and $C_{\rm gr,eff}^{\rm LQC, \Lambda=0}$ is given by
\begin{align}
C_{\rm gr,eff}^{\rm LQC, \Lambda=0} = -\frac{3}{8\pi G\Delta\gamma^2} V \sin^2(b)
\end{align}
The new regularization of the Lorentzian part -- more in line with the full theory -- has produced a correction with respect to standard LQC) in the gravitational part of the effective constraint proportional to $C_{\rm gr,eff}^{\rm LQC, \Lambda=0} \sin^2(b)$. It is worth noting that the same function, $C_{\rm gr,eff}^{\rm TR}$, can be obtained as the expectation value of LQG Hamiltonian on complexifier coherent states peaked on cosmological data. For details, see \cite{DL17a,DL17b}.

\subsubsection{Energy density of matter}

Recall that the energy density of the scalar field is $\rho_\phi = p_\phi^2/(2V)$. So we can write
\begin{align} \label{C-eff-rho-gen}
C_{\rm eff}^{\rm TR} = V \rho_\phi + C_{\rm gr,eff}^{\rm TR} = V \rho_{\phi} - \frac{3}{8\pi G\Delta\gamma^2} V \sin^2(b) \left[1 - (1 + \gamma^2) \sin^2(b)\right]
\end{align}
Now, solving the constraint $C_{\rm eff}^{\rm TR} = 0$ for the energy density $\rho_\phi$, we get
\begin{align} \label{rho-phi-sinb}
\rho_\phi = \frac{3}{8\pi G\Delta\gamma^2} \sin^2(b) \left[1 - (1 + \gamma^2) \sin^2(b)\right]
\end{align}
This shows that $\rho_\phi$ is bounded. To find the maximum value, let us use the notation $x := \sin(b)^2$. Then $\rho_\phi$ is a polynomial quadratic in $x$, whose maximum is obtained for $x = 1/(2(1+\gamma^2))$. The corresponding value is the {\it critical} energy density of the new model:
\begin{align}
\rho_{c}^{\rm TR} = \frac{3}{32\pi G\Delta \gamma^2(1+\gamma^2)}
\end{align}
The boundedness of $\rho_\phi$ is an indication that the Big Bang singularity is resolved.\footnote
{
While we are here working with a massless scalar field, we notice that (\ref{C-eff-rho-gen}) is true for any other form of perfect fluid, so the boundedness result is general.
}
Also, we observe that the critical energy density of this model is different than the one of standard LQC. Indeed, it is
\begin{align}
\rho_{c}^{\rm TR} = \dfrac{1}{4 (1+\gamma^2)} \rho_{c}
\end{align}
which is smaller than $\rho_{c}$.

\subsubsection{Equations of motion}

The equations of motion of the model can be derived by Hamilton's equation of the effective constraint, which in terms of phase space conjugated variables $(V,b)$ and $(\phi,p_\phi)$ reads
\begin{align} \label{H-eff-tot}
C_{\rm eff}^{\rm TR} = \frac{p_\phi^2}{2V} - \frac{3}{8\pi G\Delta\gamma^2} V \sin^2(b) \left[1 - (1 + \gamma^2) \sin^2(b)\right]
\end{align}
From $\{\phi,p_\phi\} = 1$ we find (denoting by dot the derivative with respect to cosmic time $t$)
\begin{align} \label{matter-dot}
\dot \phi = \{\phi,C_{\rm eff}^{\rm TR}\} = \frac{p_\phi}{V}, \ \ \ \ \ \dot p_\phi = \{p_\phi,C_{\rm eff}^{\rm TR}\} = 0
\end{align}
The second equation, in particular, shows that $p_\phi$ is a constant of motion. Similarly, from $\{b,V\} = 4\pi G \gamma \sqrt\Delta$ we find
\begin{align} \label{v-dot}
\dot V = \{V,C_{\rm eff}^{\rm TR}\} = \frac{3}{2\gamma \sqrt\Delta} V \sin(2b) [1 - 2(1 + \gamma^2) \sin^2(b)]
\end{align}
and
\begin{align} \label{b-dot}
\dot b = \{b,C_{\rm eff}^{\rm TR}\} = -2\pi G \gamma \sqrt\Delta \frac{p_\phi^2}{V^2} - \frac{3}{2\gamma \sqrt\Delta} \sin^2(b) \left[1 - (1 + \gamma^2) \sin^2(b)\right]
\end{align}
Recall that the maximum of $\rho_\phi$ corresponds to $\sin(b)^2 = 1/(2(1+\gamma^2))$. Replacing this in (\ref{v-dot}), we see that $\dot V = 0$. This condition identifies a bounce, for which we thus have
\begin{align} \label{init-b}
b_B = \pm \arcsin\left(\frac{1}{\sqrt{2(1 + \gamma^2)}}\right), \ \ \ \ \ \rho_{\phi,B} = \rho_{c}^{\rm TR}
\end{align}
The first relation can be used to fix $b$ at the bounce (up to a sign), while the second -- recalling that $\rho_{\phi} = p_\phi^2/(2V^2)$ and that $p_\phi$ is a constant of motion -- fixes $V$:
\begin{align} \label{init-v}
V_B = \frac{|p_\phi|}{\sqrt{2\rho_{c}^{\rm TR}}} = |p_\phi| \sqrt{\frac{16\pi G\Delta \gamma^2(1+\gamma^2)}{3}}
\end{align}
These values can be used as initial conditions (at the bounce) and so Hamilton's equations (\ref{v-dot}) and (\ref{b-dot}) can be numerically integrated. The only free parameters (that label the specific solution) are $p_\phi$ and the sign of $b_B$.

\paragraph*{Remark on physical time}

Here we expressed everything with respect to cosmic time $t$. However, the natural choice of physical time in this model is $\phi$. Indeed, from (\ref{matter-dot}) it follows that $\dot\phi$ has definite sign. For example, if we choose the constant $p_\phi$ positive, then $\dot\phi > 0$, and so $\phi$ grows monotonically in $t$. It is therefore a good clock for the whole evolution.\footnote
{
At the technical level, once $V=V(t)$ is computed, equation $\dot\phi = p_\phi/V$ is immediately integrated, giving $\phi = \phi(t)$ up to initial condition which corresponds to the value $\phi_B$. Due to monotonicity, this equation can be inverted, so we have $t = t(\phi)$. The functions $V(t)$ and $b(t)$ can now be expressed in terms of $\phi$.
}
The equations of motion with respect to $\phi$ are
\begin{align} \label{V,b-phi}
\begin{array}{c}
\dfrac{dV}{d\phi} = \dfrac{3}{p_\phi \gamma \sqrt\Delta} V^2 \sin(b) \sqrt{1-\sin^2(b)} [1 - 2(1 + \gamma^2) \sin^2(b)]
\\
\\
\dfrac{db}{d\phi} = -2\pi G \gamma \sqrt\Delta \dfrac{p_\phi}{V} - \dfrac{3}{2p_\phi\gamma \sqrt\Delta} V \sin^2(b) \left[1 - (1 + \gamma^2) \sin^2(b)\right]
\end{array}
\end{align}

\subsubsection{Exact solution of the effective dynamics}\label{SubSec5B3}

While, as said above, we now have everything we need to solve numerically the dynamics, it is actually possible to find the general solution of this model analytically. For this, using the definition
\begin{align} \label{x-def}
x := \sin^2(b)
\end{align}
in equation (\ref{rho-phi-sinb}), we write
\begin{align} \label{rho-x}
\rho_\phi = \frac{3}{8\pi G\Delta\gamma^2} x \left[1 - (1 + \gamma^2) x\right]
\end{align}
This can be inverted, to find
\begin{align} \label{x-sqrt}
x = \dfrac{1 + s \sqrt{1 - \rho_\phi/\rho_{c}^{\rm TR}}}{2 (1 + \gamma^2)}
\end{align}
with $s$ an unspecified sign. We are now going to derive a differential equation for $x$.
\\
\\
Consider $f(x) := (x')^2$, where prime denotes derivative with respect to $\phi$. From (\ref{x-def}) it follows that
\begin{align}
f(x) = [2\sin(b)\cos(b)b']^2 = 4 x (1-x) b'^2
\end{align}
where, from (\ref{V,b-phi}), we have
\begin{align}
b' & = -2\pi G \gamma \sqrt{\Delta} \sqrt{2\rho_\phi} - \dfrac{3}{2 \sqrt{2\rho_\phi} \gamma \sqrt\Delta} x \left[1 - (1 + \gamma^2) x\right] \notag
\\
& = -\sqrt{12\pi G x [1 - (1+\gamma^2)x]}
\end{align}
having used (\ref{rho-x}) in the second step. Hence, we find the equation
\begin{align} \label{x-prime-square}
x'^2 = 48\pi G x^2 (1-x) [1 - (1+\gamma^2)x]
\end{align}
This equation admits a unique\footnote
{
To check that this solution is unique, we observe that equation (\ref{x-prime-square}) can be written as a second order differential equation: if the solution is non-trivial ($x=\text{const}$), we can divide by $x'$, obtaining
\begin{align} \notag
x'' = 24\pi G x [2 - 3x (2 + \gamma^2) + 4x^2 (1 + \gamma^2)]
\end{align}
Now, this equation can be written as a first order differential equation for vector $X := \left(\begin{array}{c}x \\ x'\end{array}\right)$:
$$X' = \left(\begin{array}{c}x' \\ x''\end{array}\right) = \left(\begin{array}{c} X_2 \\ 24\pi G X_1 [2 - 3X_1 (2 + \gamma^2) + 4X_1^2 (1 + \gamma^2)] \end{array}\right)$$
The vector on the rhs of this equation admits continuous partial derivatives in $X_1$ and $X_2$, so the equation satisfies Lipschitz criterion, which in turn means that it admits unique solution.
}
solution of the form
\begin{align} \label{x-new-lqc}
x_{\rm TR}(\phi) = \dfrac{1}{1+\gamma^2 \cosh^2(\sqrt{12\pi G} (\phi-\phi_o)} , 
\end{align}
where the free constant $\phi_o$ reflects the invariance of the equations of motion with respect to the shift in the (matter) time (the freedom of choice of the point of origin of time measurement). Once $x_{\rm TR}(\phi)$ is known, all other interesting quantities can be easily computed: $\rho_\phi$ by (\ref{rho-x}) reads
\begin{align} \label{rho-new-lqc}
\rho_\phi(\phi) = \dfrac{3}{8\pi G\Delta} \left[\dfrac{\sinh(\sqrt{12\pi G} (\phi {-\phi_o}))}{1 + \gamma^2 \cosh^2(\sqrt{12\pi G} (\phi {-\phi_o}))}\right]^2
\end{align}
so the volume is
\begin{align} \label{v-new-lqc}
V(\phi) = \dfrac{|p_\phi|}{\sqrt{2\rho_\phi}} = \sqrt{\dfrac{4\pi G \Delta p_\phi^2}{3}} \dfrac{1 + \gamma^2 \cosh^2(\sqrt{12 \pi G}(\phi {-\phi_o}))}{|\sinh(\sqrt{12 \pi G}(\phi {-\phi_o}))|}
\end{align}
from which we also find the Hubble rate
\begin{align}
H_r & = \dfrac{\dot V}{3V} = p_\phi \dfrac{V'}{3V^2} = -\dfrac{\rho'}{3\sqrt{2 \rho_\phi}} \notag
\\
& = \dfrac{1 + \gamma^2 [1 - \sinh^2(\sqrt{12 \pi G} (\phi - \phi_o))]}{\sqrt\Delta [1 + \gamma^2 \cosh^2(\sqrt{12\pi G}(\phi {-\phi_o}))]^2} \cosh(\sqrt{12 \pi G} (\phi {-\phi_o}))
\end{align}
The derivatives $\rho'$ and $H_r'$ can also be computed analytically (though their explicit form is rather involved, and will therefore be omitted), and we can also compute $dH_r/dt$ by using the fact that, for any function $F$, we have $\dot F = F'\dot\phi = F'\sqrt{2\rho_\phi}$.

\subsubsection{Discussion: coordinate vs physical time}

All our analysis until now was based on the physical time given by the scalar field $\phi$. For completeness, we discuss here the cosmic time $t$. From the equation of motion $\dot \phi = p_\phi/V$, and using the explicit form (\ref{v-new-lqc}), we can integrate {this} equation. {Due to the singularity at $\phi=\phi_o$ the integration has to be performed independently on two domains $\phi>\phi_o$ and $\phi<\phi_o$. The result yields}
\begin{align} \label{t:funct:phi}
t(\phi) = t_o + \dfrac{\gamma^2 \sgn(p_{\phi}(\phi-\phi_o))}{\sqrt{12\pi G}} \left[\cosh\left(\sqrt{12\pi G} (\phi{-\phi_o})\right) - (1 + \gamma^2) \log\left|\coth\left(\sqrt{3\pi G} (\phi{-\phi_o})\right)\right|\right]
\end{align}
This function is plotted in \ref{fig:phi-t}, where we can see that $t(\phi)$ is not invertible. On \emph{each of the two domains} ($\phi>\phi_o$ and $\phi<\phi_o$) separated by the singularity of the equation at $\phi=\phi_o$ the image of $t(\phi)$ covers the whole real line. As a consequence, the cosmic time chart can cover only one of two domains indicated above (later referred to as \emph{aeons}). Due to time reflection symmetry of the equations of motion, we can focus our attention on the aeon $\phi>\phi_o$ (our observations translate to $\phi<\phi_o$ via time reflection $t\to -t$). In this chart, from the point of view of a comoving observer (whose proper time is $t$), the infinite past corresponds to $\phi \to^+ \phi_o$, while the infinite future to $\phi \to \infty$. So, for such observer, the far past consists of a quantum region in which the universe is undergoing a de Sitter contracting phase dominated by emergent cosmological constant, while the far future consists of a classically expanding phase dominated by the matter (scalar field).
\begin{figure}[h!] 
\includegraphics[width=0.5\textwidth]{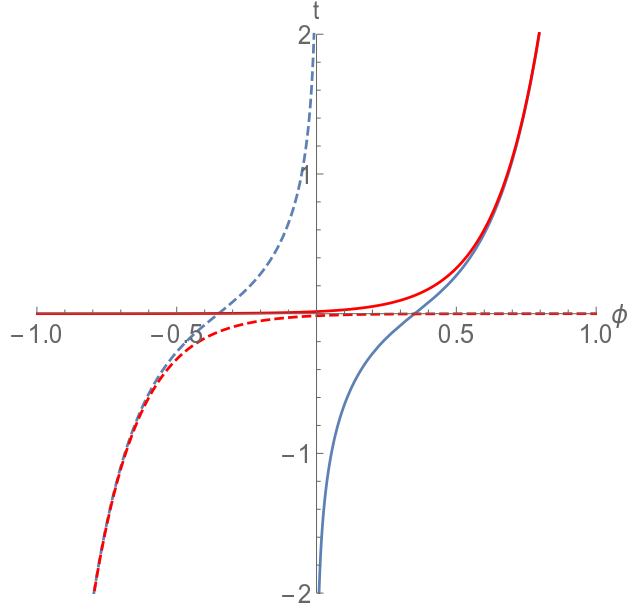}
	\caption{Cosmic time $t$ as a function of $\phi$ in the new model (blue) and GR (red). The red dashed line corresponds to the classical solution obtained by time reversal ($\phi \to -\phi$), whereas the blue dashed one represents $t(\phi)$ for $\phi<\phi_o$ (with $\phi_o$ set to $0$ for the convenience of the presentation). In the new model, $t$ only covers half the $\phi$-chart. This means that, when parametrizing the dynamics with $t$, there exist two solutions (aeons): one covering the $\phi > \phi_o$ region, the other covering the $\phi < \phi_o$ region.}
	\label{fig:phi-t}
\end{figure}
Unfortunately, expression (\ref{t:funct:phi}) cannot be inverted analytically, so we do not have an explicit form for the quantities of interest (such as volume) as functions of cosmic time $t$. Nevertheless, these can still be plotted numerically. As an example, in figure \ref{fig:t} we plot the curvature $R = 2 [\ddot V/V - \dot V^2/(3V^2)]$ and the volume, comparing them with the results of standard LQC. Note, in particular, that in the far past the curvature of the current model reaches a non-zero constant: since this value is still Planckian, it justifies why in the far past the quantum gravity effects are still important (despite the energy density of matter being negligible), and it explains the existence of an emergent cosmological constant.
\\
\\
Finally, as mentioned, earlier the above results translate to the aeon $\phi<\phi_o$ via time reversal transformation. In there, the far past ($\phi \to -\infty$) consists of a classically contracting universe, while the far future ($\phi \to^- \phi_o$) consists of a quantum region in which the universe is undergoing a de Sitter expanding phase dominated by emergent cosmological constant. It is interesting to note that, while with respect to $\phi$ these two solutions are bridged in a finite time (passing through a region of infinite volume at $\phi = \phi_o$), with respect to the cosmic time they are distinct, physically disconnected regions. It is only the use of a matter clock which brings these two aeons together.

\begin{figure}[h!]
	\begin{centering}
		\subfloat[]{\includegraphics[width=0.4\textwidth]{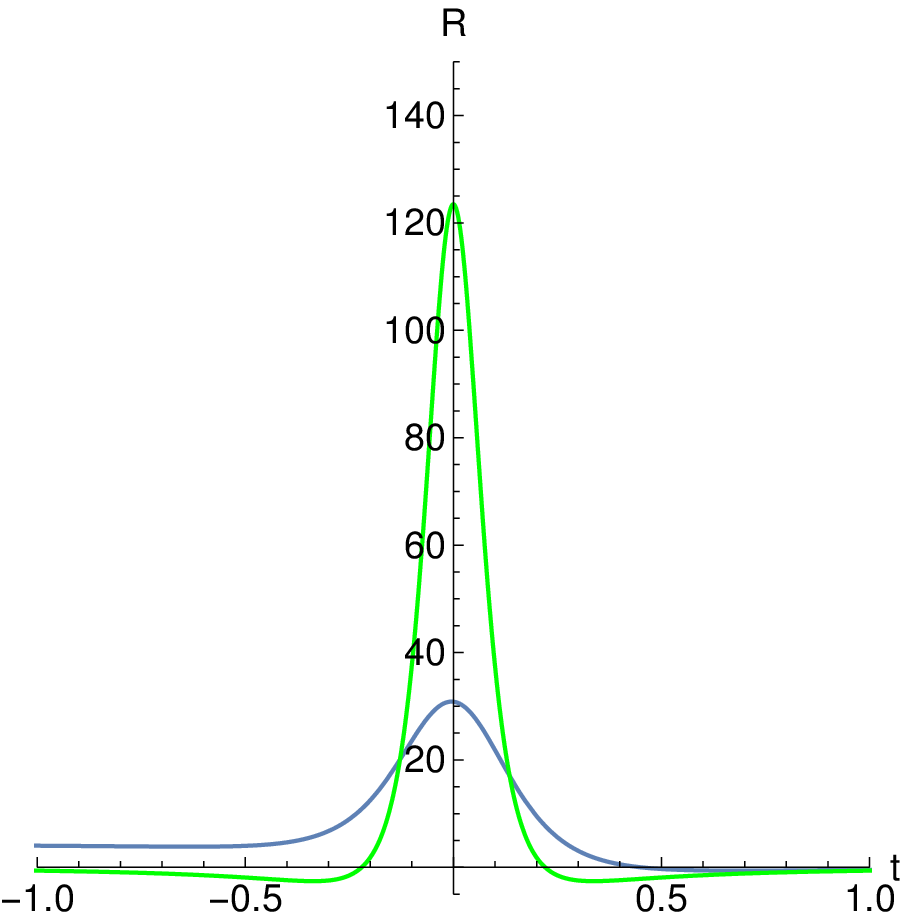} \label{fig:R}}
		\hspace{1cm}
		\subfloat[]{\includegraphics[width=0.4\textwidth]{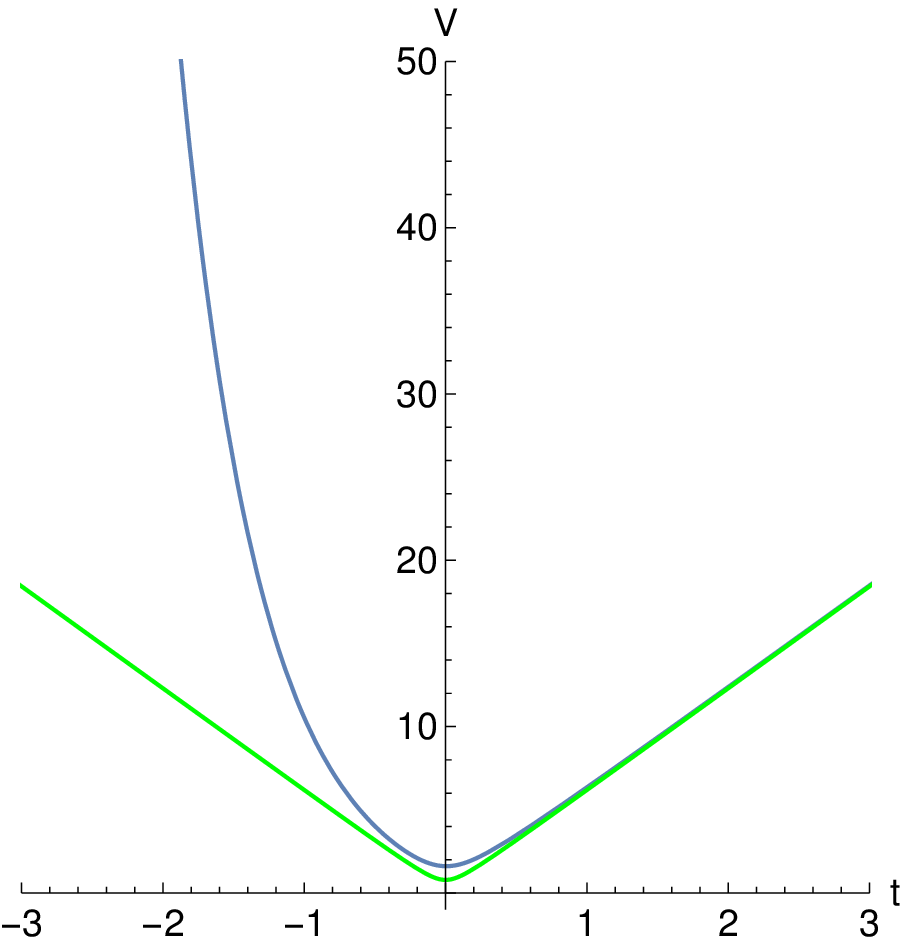} \label{fig:v-t}}
	\end{centering}
	\caption{The evolution of the curvature $R$ and the volume $V$ during the {aeon $\phi \geq \phi_o$} (as functions of cosmological time $t$) for the studied model (blue) is compared against that of mainstream LQC (green). The bounce occurs at $t=0$. Notice that in the new model $R$ reaches $0$ in the future classical FLRW phase, but a finite non vanishing value in the past de Sitter phase (a), consistent with the non symmetric bounce shown in (b).}
	\label{fig:t}
\end{figure}

\documentclass[./main.tex]{subfiles}

\section{Comparative analysis of the effective and quantum models}\label{sec:results}

In this section we {present the details of the methodology and the results of} the analysis of the evolution in the models described in the previous sections. {Since the genuine quantum analysis (based on numerical methods) could be performed only for a finite population of examples, whereas the simplicity of the effective dynamics allows for a systematic probing of the space of solutions, the results regarding quantum trajectories themselves are discussed using the effective dynamics, with the (purely quantum) numerical studies serving as the verification of the accuracy of the effective results. The genuine quantum analysis, however, has to be used for probing the higher order quantum properties, i.e., the behavior of variances. For that reason, we first present the results coming from the} effective dynamics in both standard LQC and in the model we derived with the new Lorentzian term in the Hamiltonian. The results of the analysis are compared together with classical GR from the perspective of the observables of interest in the context of FLRW cosmology, namely the volume, the matter energy density and the Hubble rate. {The genuine quantum analysis} of the evolution in the new model (characterized by the new $\Theta_{\rm TR}$ operator defined in \eqref{final_theta} and whose properties were discussed in section \ref{sec:evol-prop}) {is discussed in the second subsection.} We then conclude with a comparison between the semi-classical evolution obtained in the quantum model of the considered observables, and the effective evolution of their classical counterparts in the aforementioned effective models.

\subsection{Asymptotic analysis of the effective models}

{Given any quantum theory/model built on the nonperturbative level, the first question one needs to ask is whether in an appropriate regime it reproduces the observationally confirmed classical theory (in our case the cosmological sector of GR) in the low energy limit. Provided that the quantum trajectories can be predicted with sufficient level of accuracy by the effective classical dynamics, which is indeed the case here as we show in sec.~\ref{sec:quant-num}, one can address this question by studying} the behavior of the solutions to the equations of motion analyzed in section \ref{SubSec5B3} in the limit $\rho_\phi \ll \rho_c^{\rm TR}$, which is the condition we would expect to determine the semiclassical region. 

{From (\ref{rho-new-lqc}), we see that the above condition will be satisfied in two regimes:} either (i) $\cosh(\sqrt{12\pi G} (\phi{-\phi_o})) \to \pm \infty$, or (ii) $\sinh(\sqrt{12\pi G} (\phi{-\phi_o})) \to 0$. These situations translate respectively into the following conditions on $\phi$:
\begin{itemize}
\item $\phi \to \pm \infty$, corresponding to
\begin{align} \label{xTR-asympt1}
x_{\rm TR}(\phi) \sim \dfrac{4}{\gamma^2} e^{\mp 2\sqrt{12 \pi G} (\phi{-\phi_o})}
\end{align}
\item $\phi \to^{\pm} {\phi_o}$, corresponding to
\begin{align} \label{xTR-asympt2}
x_{\rm TR}(\phi) \sim \dfrac{1}{1+\gamma^2} \left[1 - \dfrac{\gamma^2}{1 + \gamma^2} 6\pi G (\phi{-\phi_o})^2\right]
\end{align}
\end{itemize}
Interestingly, (\ref{xTR-asympt1}) is the same asymptotic behavior found in classical GR. Indeed, in classical cosmology the exact solution for $b$ is\footnote
{
The negative (resp. positive) sign corresponds to a classically expanding (resp. contracting) universe.
}
\begin{align}
b_{\rm class}(\phi) = b_o e^{\mp \sqrt{12\pi G} (\phi{-\phi_o})}
\end{align}
which implies
\begin{align}
x_{\rm class}(\phi) & = \sin^2(b_{\rm class}(\phi)) = \sin^2\left(b_o e^{\mp \sqrt{12\pi G} (\phi{-\phi_o})}\right) \stackrel{\phi \to \pm \infty}{\longrightarrow} \notag
\\
& \to b_o^2 e^{\mp 2 \sqrt{12\pi G} (\phi{-\phi_o})}
\end{align}
So, we conclude that:
\begin{itemize}
\item In the limit $\phi \to +\infty$, the TR-model coincides with a classically expanding universe (with integration constant $b_o$ such that $b_o^2 = 4/\gamma^2$).
\item In the limit $\phi \to -\infty$, the TR-model coincides with a classically contracting universe (with integration constant $b_o$ such that $b_o^2 = 4/\gamma^2$).
\end{itemize}
We can repeat the same procedure (done for the new model) in the context of standard LQC, the only difference being the relation between $x$ and $\rho_\phi$, as well as the form of $b'$: in LQC we have
\begin{align}
x = \dfrac{8\pi G \gamma^2\Delta}{3} \rho_\phi, \ \ \ \ \ b' = - \sqrt{12\pi G x}
\end{align}
Hence, after analogus manipulations, we find the equation
\begin{align}
x'^2 = 48\pi G x^2 (1-x)
\end{align}
whose solutions with initial conditions $x_{\rm LQC}(0) = x_o$ are\footnote
{
This can be rewritten as $x_{\rm LQC}(\phi) = 1 - \tanh^2[\sqrt{12\pi G} (\phi{-\phi_o}) \mp \text{arctanh}(\sqrt{1 - x_o})]$, so it is clear that the sign of $\phi$ is not important: both positive and negative signs correspond to the same one, but shifted by a constant. In particular, if we set $x_o = 1$, we see that both solutions coincide.
}
\begin{align}
x_{\rm LQC}(\phi) = 1 - \tanh^2[\text{arctanh}(\sqrt{1 - x_o}) \mp \sqrt{12\pi G} (\phi{-\phi_o})]
\end{align}
Again, it is easy to check that the asymptotic behavior of $x_{\rm LQC}(\phi)$ in the limit $\phi \to \pm \infty$ coincides with the classical one, and can be made exact by appropriately choosing the integration constant $x_o$:
\begin{align}
x_{\rm LQC}(\phi) = 1 - \tanh^2[\sqrt{12\pi G} (\phi{-\phi_o}) + \ln(\gamma)]
\end{align}
This confirms that the three models -- the TR-model, standard LQC and classical GR -- coincide in the limit $\phi \to \pm \infty$.
But, contrary to LQC, the TR-model presents another ``semiclassical limit'', namely the case $\phi \to^{\pm} {\phi_o}$. In this limit, $x_{\rm TR}$ presents the behavior (\ref{xTR-asympt2}), which can be seen to coincide with the asymptotic behavior of classical GR {\it in presence of a cosmological constant $\Lambda$ and a modified Newton constant $\bar G$}. Indeed, in this case Friedmann equations are
\begin{align} \label{de Sitter-eff}
\left\{\begin{array}{l}
H_r^2 = \dfrac{8\pi \bar G}{3} \left(\rho + \rho_\Lambda\right)
\\
\\
\dfrac{\ddot a}{a} = -\dfrac{16\pi \bar G}{3} \left(\rho - \dfrac{\rho_\Lambda}{2}\right)
\end{array}\right. \ \ \ \ \ \text{with} \ \ \rho_\Lambda := \dfrac{\Lambda}{8\pi \bar G}
\end{align}
whose exact solution for the volume $V = a^3$ is
\begin{align}
V_{\rm dS}(\phi) = \sqrt{\dfrac{4\pi \bar G p_\phi^2}{\Lambda}} \dfrac{1}{|\sinh(\sqrt{12\pi \bar G} (\phi{-\phi_o}))|}
\end{align}
This can be seen to coincide with (\ref{v-new-lqc}) in the $\phi \to {\phi_o}$ limit under the identification\footnote
{For certain values of $\gamma$, $\bar G$ becomes negative, so in light of the first equation of (\ref{de Sitter-eff}) it would seems that such values are forbidden. This is however not true, since that equation only holds for $\rho \ll \rho_c^{\rm TR}$, which means that $\rho_\Lambda$ dominates; but $\rho_\Lambda$ also contains $\bar G$, so the overall sign of the right-hand-side of the first equation in (\ref{de Sitter-eff}) remains positive even if $\bar G$ is not.
}
\begin{align} \label{effective-params1}
\bar G = \dfrac{1-5\gamma^2}{1 + \gamma^2} G, \ \ \ \ \ \ \ \ \ \ \Lambda = 8\pi \bar G \rho_\Lambda = \dfrac{3}{\Delta (1+\gamma^2)^2}
\end{align}
and so
\begin{align} \label{effective-params2}
\rho_\Lambda = \dfrac{3}{8\pi \bar G\Delta (1+\gamma^2)^2}
\end{align}
The above models are compared in figure \ref{V-rho-compare}, where we plot the volume $V$ and energy density $\rho$ for the respective solutions, and in figure \ref{portraits} where we display the $\dot H_r(H_r)$ portrait.
\begin{figure}[h!]
\centering
\subfloat[]{\includegraphics[width=8cm]{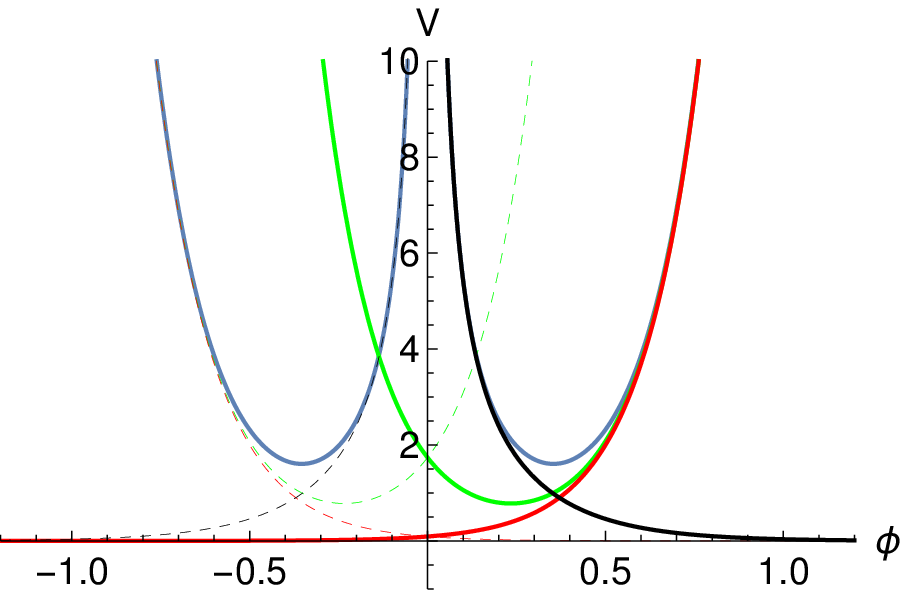}}
\hspace{0.3cm}
\subfloat[]{\includegraphics[width=8cm]{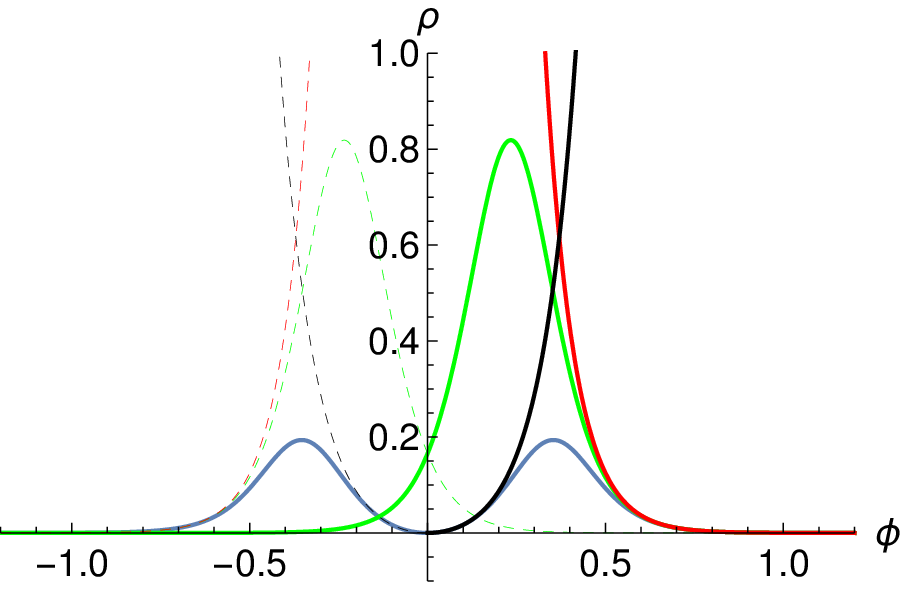}}
\caption{Volume and energy density in the new model (blue), LQC (green), GR (red) and GR with effective $\bar G$ and $\Lambda$ (black). {For presentation convenience $\phi_o$ is set to $0$.} The dashed lines correspond to the solutions obtained by time reversal ($\phi \to -\phi$).}
\label{V-rho-compare}
\end{figure}
\begin{figure}[h!]
\centering
\subfloat[]{\includegraphics[width=6.5cm]{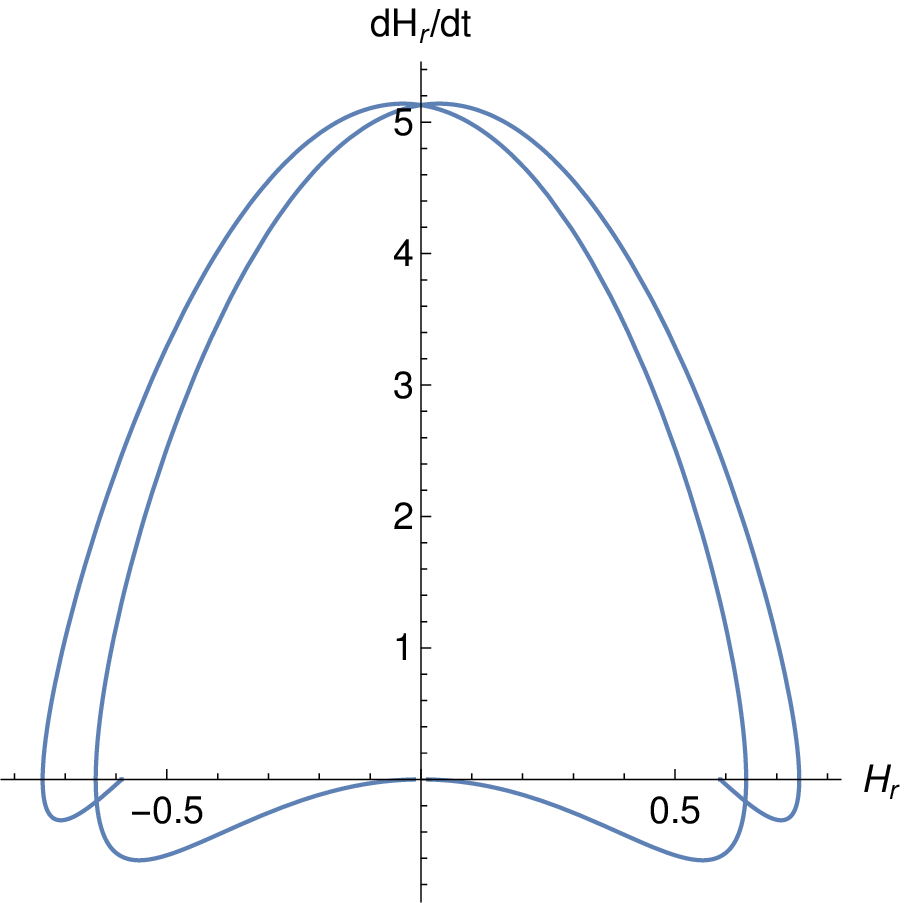}}
\hspace{0.3cm}
\subfloat[]{\includegraphics[width=6.5cm]{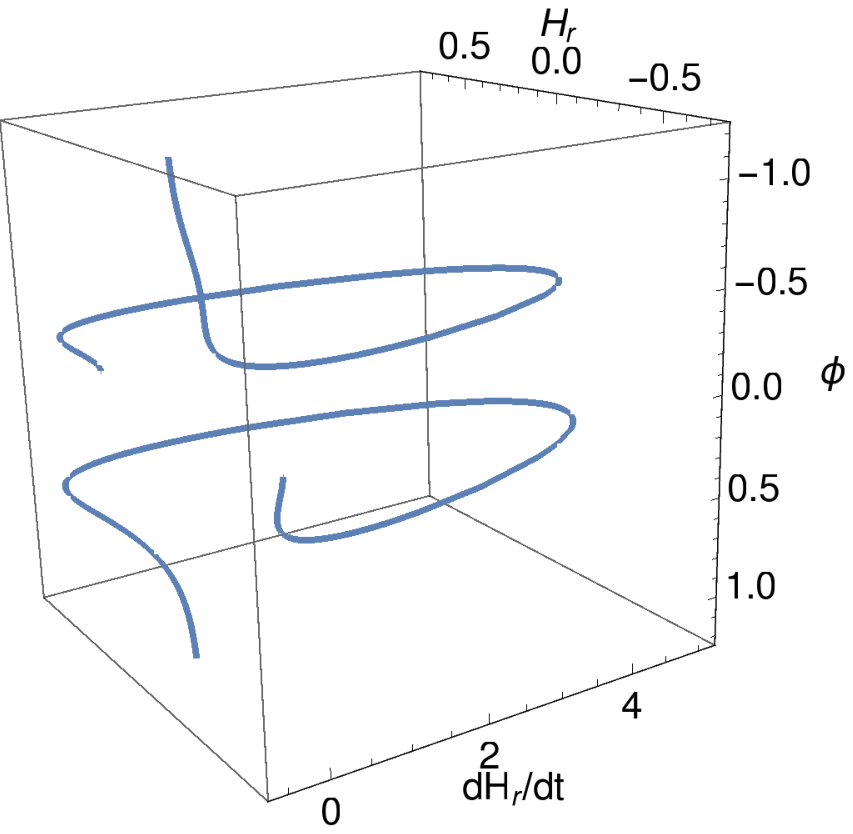}}
\vspace{0.3cm}
\subfloat[]{\includegraphics[width=6.5cm]{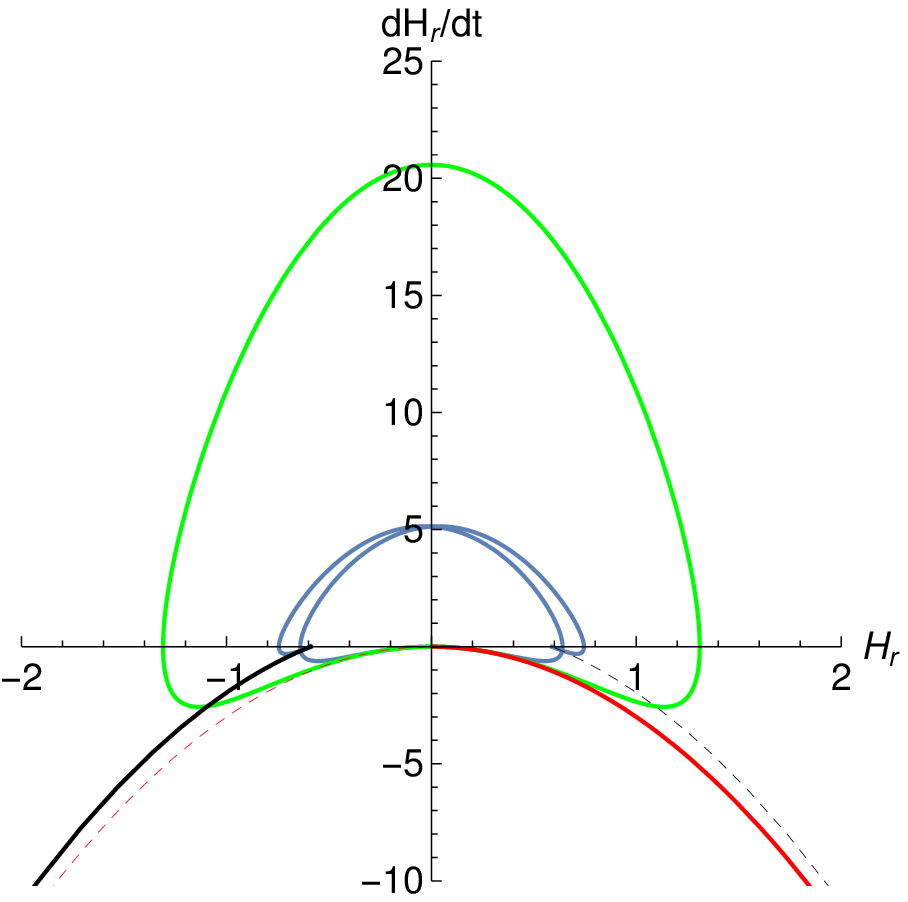}}
\hspace{0.3cm}
\subfloat[]{\includegraphics[width=6.5cm]{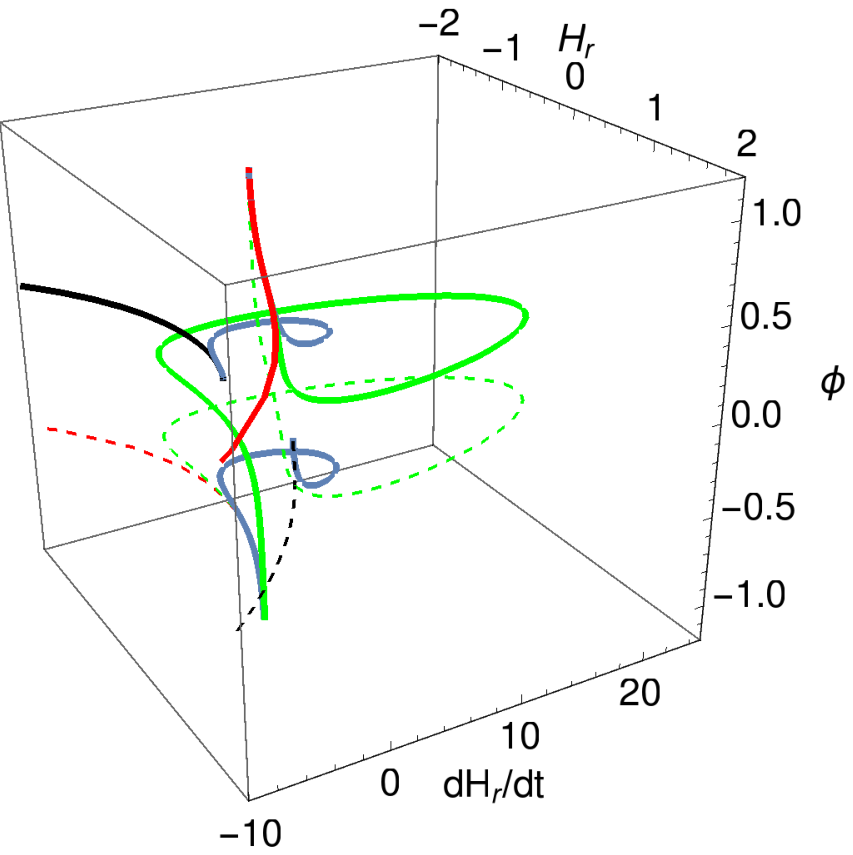}}
\caption{2D and 3D plots of $H_r$ and $\dot H_r$ in the new model (blue) and in comparison with LQC (green) and GR (classical (red) and with effective $\bar G$ and $\Lambda$ (black)). The dashed lines correspond to the solutions obtained by time reversal ($\phi\to -\phi$). The far past ($\phi \to -\infty$) corresponds to the point $(H_r,\dot H_r) = (0,0)$, where both the new model and LQC match the far past of the classical contracting solution. Then, as $\phi$ increases, $H_r<0$ and $\dot H_r < 0$, which denotes a decelerating contraction; here, all model depart, and while the classical universe continues to the big crunch (at negative infinity), the new model and LQC cross the $\dot H_r = 0$ line and enter a phase where gravity becomes ``repulsive'' ($\dot H_r > 0$). This phase ends at $H_r = 0$, where the bounce occurs. After that, the repulsivity of gravity drives a phase of accelerated expansion ($H_r > 0$), which continues until the $\dot H_r = 0$ line is crossed again. At this point, the behavior of LQC and the new model are very different: the former approaches again $(0,0)$, which now corresponds to the far future of the classically expanding solution; the latter approaches $\dot H_r \to^- 0$ at a finite value of $H_r$, which corresponds to the far future of a de Sitter expanding solution. As the 3D plot shows, this super-expansion phase is reached at finite values of $\phi$. In fact, at $\phi = 0$ a discontinuity takes place, the trajectory being mapped to $H_r \to -H_r$. In other words, the universe follows now a de Sitter contracting solution. This solution is soon departed, and a symmetric behaviour takes place, ending at $(0,0)$, where the classically expanding solution is reached.}
\label{portraits}
\end{figure}

\subsection{Numerical analysis of the quantum evolution}
\label{sec:quant-num}

In order to investigate the evolution in the new quantum model with the $\Theta_{\rm TR}$ operator defined in \eqref{final_theta}, we use the families of partial observables defined in the subsection \ref{sec:obs} which allow us to provide a notion of (parametrized by $\phi$) quantum trajectories, defined as the expectation values of the observables (as functions of $\phi$) in suitable states. The steps to obtain these trajectories are detailed in the following.

First we observe that the physical states in the new quantum model have a very simple form in $x$-representation, however in $v$-representation the form of the wave function can be found only numerically. Thus, in probing the dynamics we are forced to focus on particular classes of states, which can be probed in a robust way by a finite set of examples. Among those, the ones of particular interest are the states semiclassical in the low energy sector -- the ones reproducing (in some epoch) the semiclassical universe following the predictions of GR. In mainstream LQC this requirement was satisfied in particular by the (sufficiently sharply peaked) energy\footnote{The name follows from the interpretation of $\sqrt{|\Theta_{\beta}|}$ as vacuum Hamiltonian of the deparametrized system.} Gaussians, which were the class of states used for majority of numerical studies there. Following the previous works we too pick for the investigation the states of spectral profiles (see \eqref{phys-state-gr-represented})
\begin{equation}\label{eq:state-Gauss}
	c(k,\omega(k)) = \frac{1}{\sqrt{2\pi\sigma}} e^{-\frac{(\omega(k)-\omega^\star)^2}{2\sigma^2}} e^{-i\omega(k)\phi_o} =: c_{\rm gauss}(k,\omega(k)),
\end{equation}
where $\sigma\ll\omega^{\star}$ are positive constants and have unit $G^{1/2}$ and $\phi_o$ has unit $G^{-1/2}$.

Second, since (i) $\Theta_{\rm TR}$ has a relatively simple form in $b$-representation, \eqref{eq:theta-brep}, and (ii) thanks to \eqref{eq:x-func} the physical states \eqref{eq:eigenv-real} can be expressed in $b$-representation as integrals, one can be tempted to evaluate quantum trajectories analytically. The problem is, however, that the operator $\hat{V} \propto |\hat{v}|$ cannot be expressed in this representation easily. While it takes a simple form in the auxiliary spaces defined in Appendix \ref{sec:aux-math} (thus one could in principle try to perform the calculations following those of \cite{ACS08}) one then needs to $(i)$ represent the action of operators directly in the $k$-representation and $(ii)$ perform the projections of the physical states onto those auxiliary spaces.
For that reason we decided to evaluate the needed expectation values directly in $v$-representation by numerical means, especially because the methods involved are a straightforward adaptation of those already built for the model of FLRW universe with positive cosmological constant in LQC \cite{PA11}. 

Now, the actual evaluations were performed as follows:
\begin{enumerate}
	\item The form of the wave function in $v$-representation has been evaluated by performing the inverse of the transform \eqref{eq:vb-transform}
		\begin{equation}\label{eq:vb-trans-inv}
			\psi_{\rm gauss}(v,\phi) = \frac{1}{\pi} |v|^{1/2} \int_0^{\pi} {\rd} b \; \Psi_{\phi}^{\rm gauss}(x(b)) e^{-(i/2)vb}, 
		\end{equation}
		where $x(b)$ is given by \eqref{eq:x-func} and
\begin{equation}
			\Psi_{\phi}^{\rm gauss}(x) = 2\pi \int \rd k \; c_{\rm gauss}(k,\omega(k)) e_{\beta, k}(x) e^{i\omega(k)\phi}
		\end{equation}
with $e_{\beta,k}(x)$ the normalized versions of (\ref{eq:eigenv-real}). The integral has been evaluated via an adaptive Romberg method, of which error tolerances have been set in actual simulations to $10^{-6}$. The domain of $b$ has been probed in the uniform grid of $2^{19}\approx 5\cdot 10^5$ intervals.
	\item The expectation values and dispersions (variances) of the observables defined in subsection \ref{sec:obs} are evaluated directly by \eqref{eq:exp-practical}, 
		where we use a standard definition for dispersion
		\begin{equation}
			\Delta^2\hat{O} = \langle\hat{O}^2\rangle - \langle\hat{O}\rangle^2 .
		\end{equation}
		The actions of $\hat \theta_K$, $\hat \rho$, $\hat H_r$ are given by \eqref{eq:obs-theta}, \eqref{eq:obs-rho} and \eqref{eq:obs-hub} respectively, thus straightforward to evaluate. Whereas for $\hat{p}_{\phi} = i\hbar\partial_{\phi}$, the needed derivative $\partial_{\phi}\psi_{\rm gauss}(v,\phi)$ is evaluated analogously to $\psi_{\rm gauss}(v,\phi)$ 
		through the transform \eqref{eq:vb-trans-inv}:
		\begin{equation}
			[\partial_{\phi} \Psi_{\rm gauss}](x(b),\phi) = 2\pi i \int \rd k \; \omega(k)c_{\rm gauss}(k,\omega(k)) e_{\beta,k}(x(b)) e^{i\omega(k)\phi} .
		\end{equation}
\end{enumerate}

In the actual simulations the quantum trajectories have been evaluated for $\omega^\star$ ranging from $500\sqrt{G}$ to $5000\sqrt{G}$ with relative dispersion in $p_{\phi}$ ranging from $0.02$ to $0.1$. 

\begin{figure}[!htb]
	\begin{centering}
		\subfloat[]{\includegraphics[width=0.45\textwidth,height=0.34\textwidth]{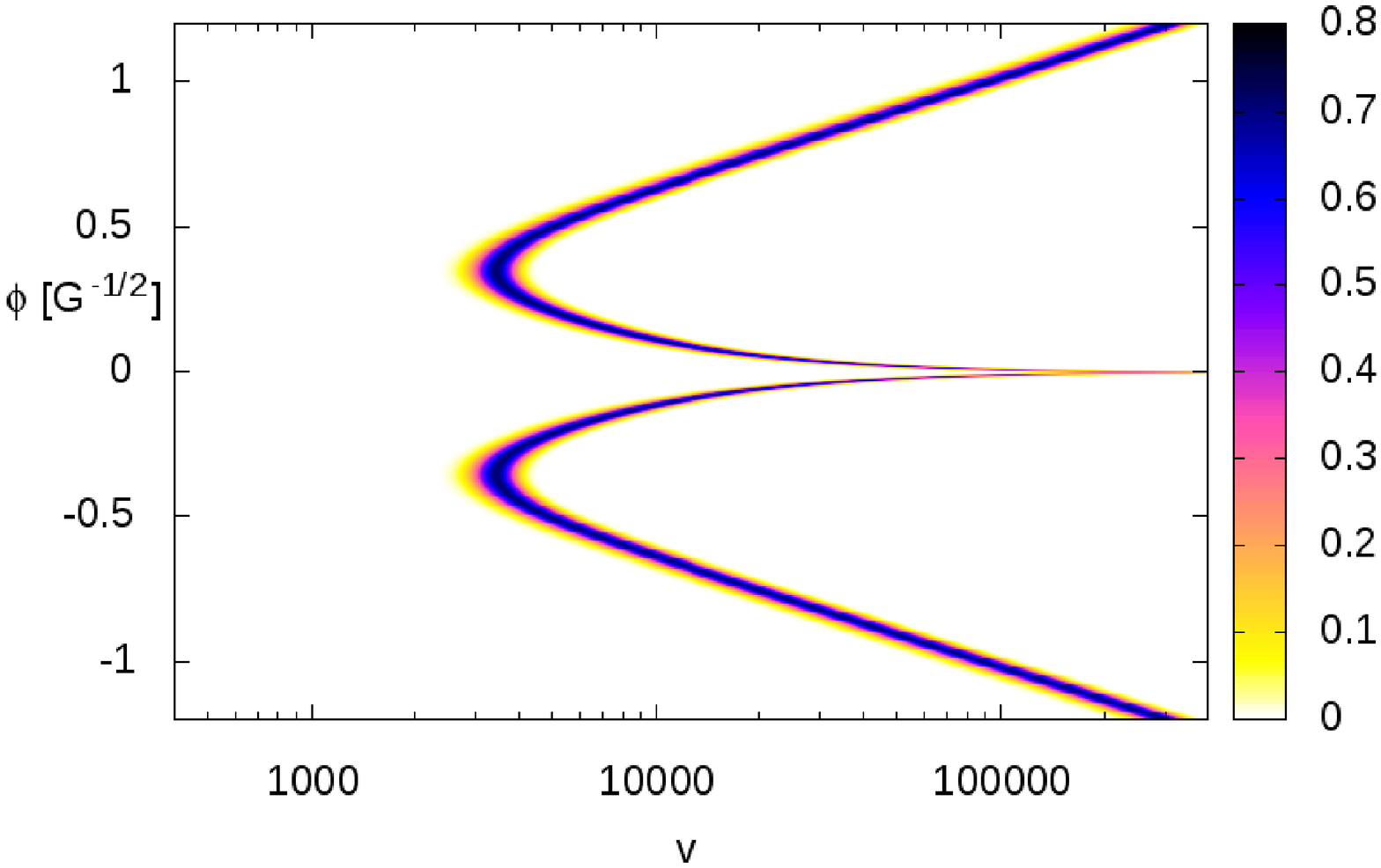}\label{fig:main-map}}
		\hspace{0,5cm}
		\subfloat[]{\includegraphics[width=0.45\textwidth]{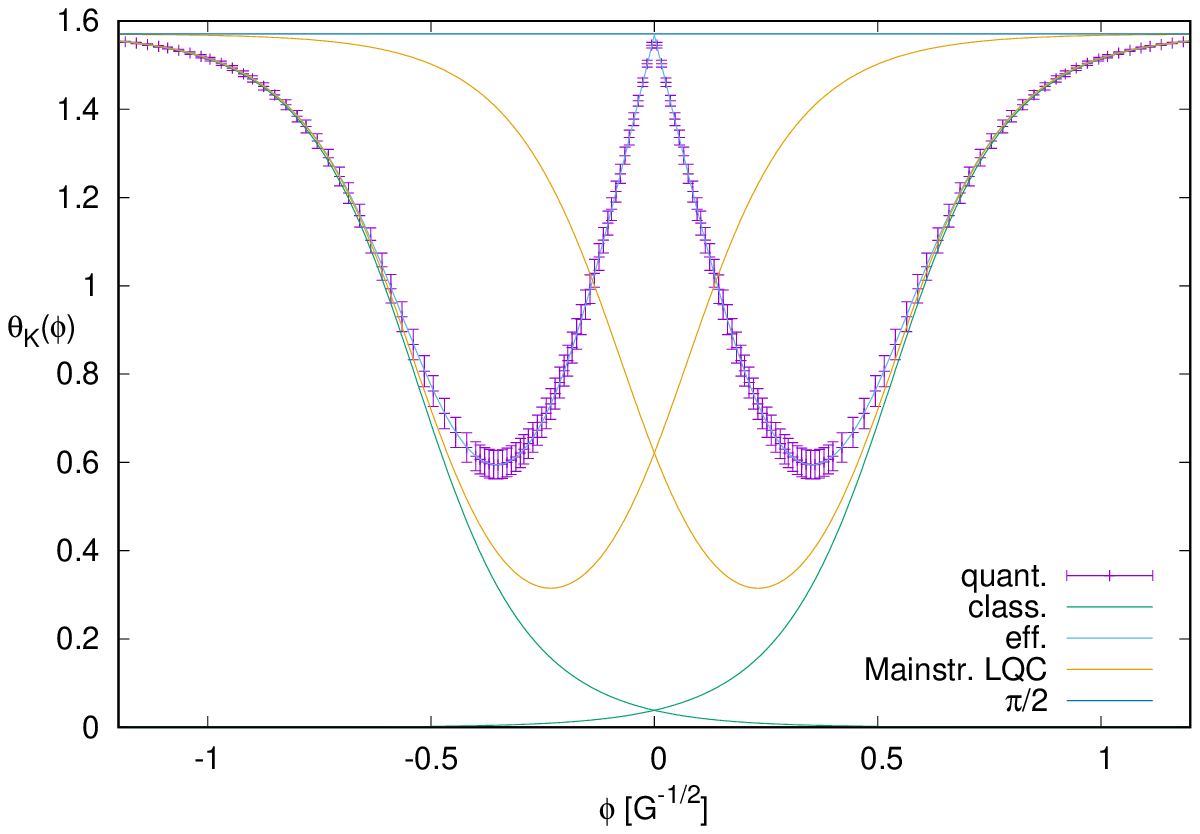}\label{fig:main-theta}} \\
		\subfloat[]{\includegraphics[width=0.45\textwidth]{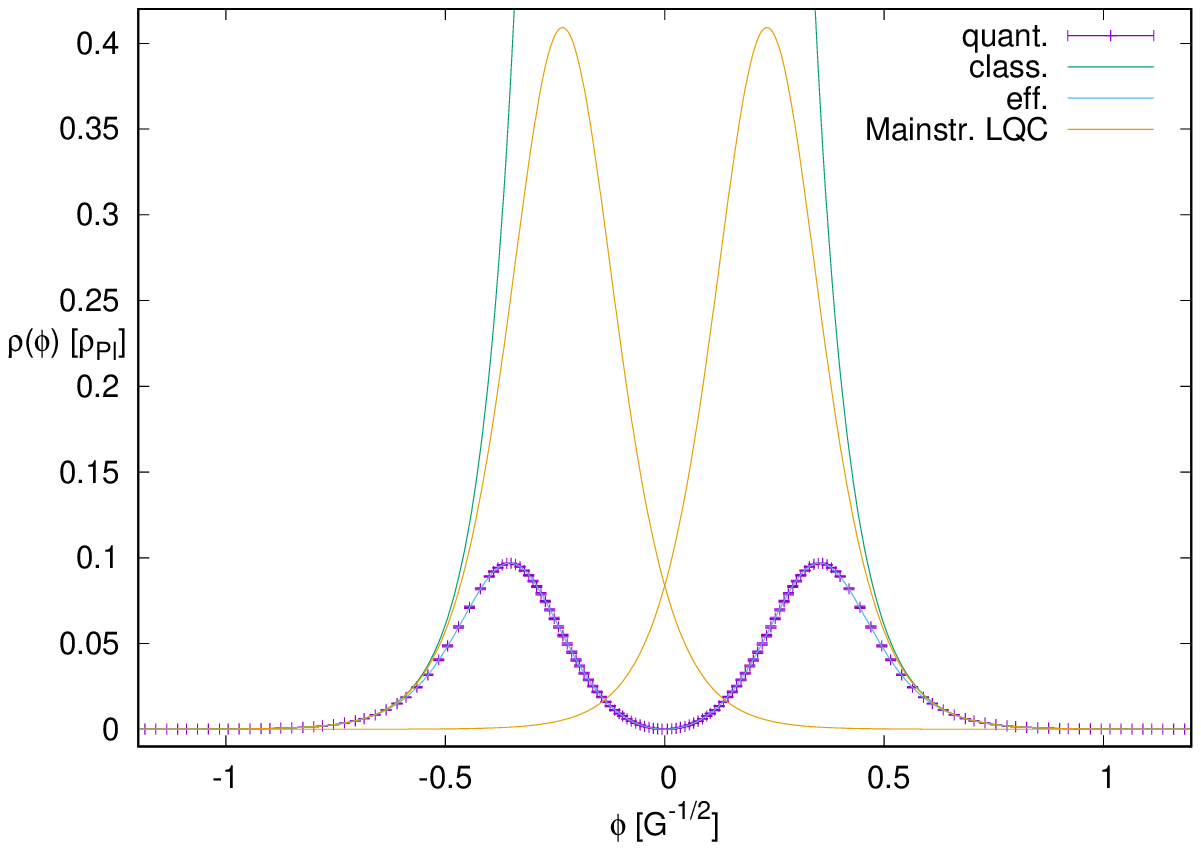}\label{fig:main-rho}}
		\hspace{0,5cm}
		\subfloat[]{\includegraphics[width=0.45\textwidth]{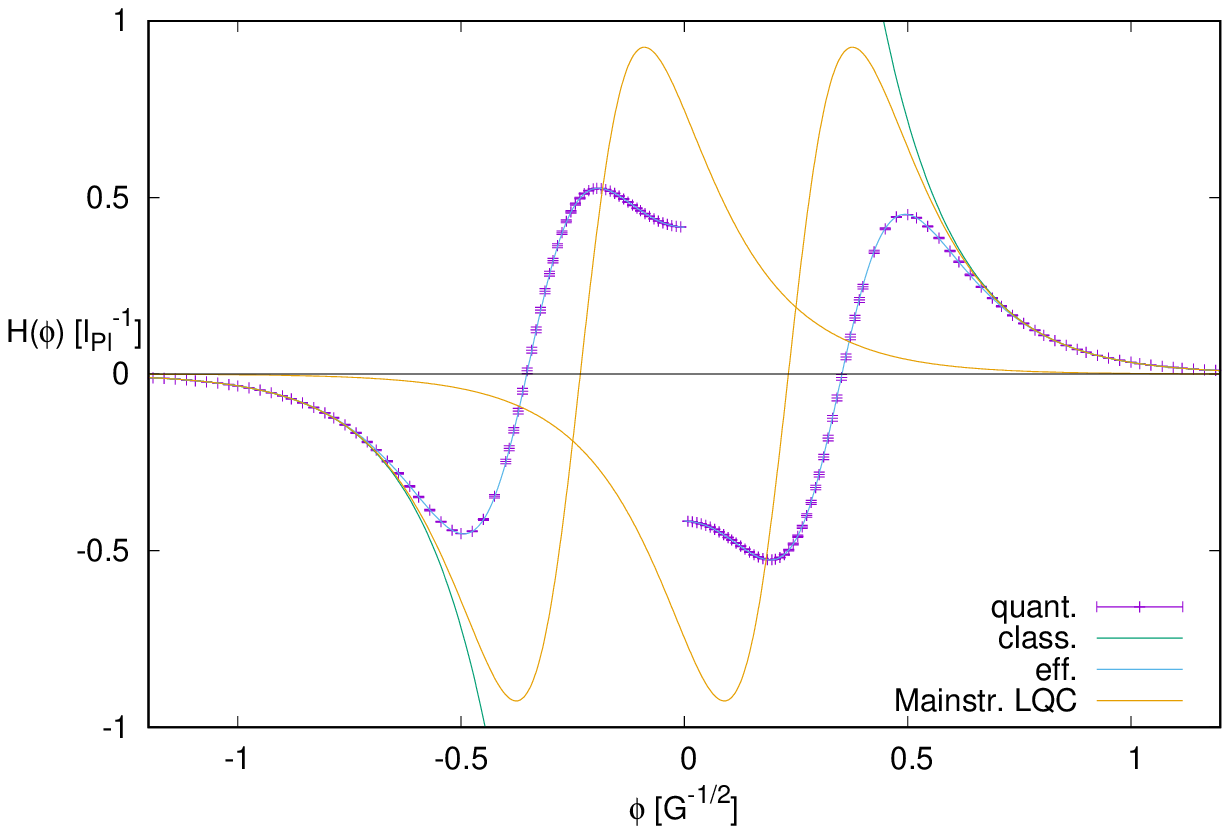}\label{fig:main-hubb}}
	\end{centering}
	\caption{The map of the physical state $(a)$ {on a $v-\phi$} plane [where the volume $V\approx 2.41 |v| \ell_{\rm Pl}^3\ $], and quantum trajectories of the observables: {compactified volume} $\theta_{K=5\cdot 10^3}$ $(b)$, {matter energy density} $\rho_{\phi}$ $(c)$ and {Hubble rate} $H_r$ $(d)$ of the Gaussian state peaked on $p_{\phi} = 5.05\cdot 10^3 G^{1/2}$ with relative spread in $\Delta p_{\phi}$ of about $0.05$. The genuine quantum trajectories of the investigated model (purple error bars) are compared against the predictions of the effective dynamics generated by Hamiltonian \eqref{H-eff-class} (blue lines) and against the classical GR (green lines) and mainstream LQC  effective trajectories (yellow lines), to which the quantum one converges in the asymptotic past/future. While both mainstream LQC trajectories feature a single bounce (each) at (respectively) $\phi\approx\pm 0.25 G^{-1/2}$, for the trajectories obtained with the Hamiltonian we investigate \eqref{final_theta} we observe two bounces at $\phi\approx \pm 0.35 G^{-1/2}$ separated by a a transition point from future to past conformal infinity at $\phi = 0$, where the matter energy density reaches zero and the volume $V$ reaches infinity. The Planck units $\rho_{\rm Pl}$ and $\ell_{\rm Pl}$ are defined respectively as $(G^{2}\hbar)^{-1}$ and $(G\hbar)^{1/2}$. The departure from mainstream LQC lasts only {about $1.2 G^{-1/2}$} in relational time $\phi$, but from each bounce it takes infinite cosmic time to reach the transition at $\phi = 0$.}
	\label{fig:main}
\end{figure}

The results of these numerical simulations are displayed in figure \ref{fig:main}, and are compared with the results in mainstream LQC, and the new effective model discussed earlier. The asymptotic behavior obtained in the new effective model confirms the results found in the quantum theory, namely that in the far past the universe is essentially a contracting de Sitter with an effective cosmological constant, and the effective trajectories mimic to high accuracy the evolution trajectories obtained in the quantum theory. The emerging picture that we observe, in backward evolution, is the following: first an expanding phase following the predictions of GR, beginning with a bounce (resolving the classical singularity) and the transition to a contracting de Sitter phase. This phase is followed by a transition through past scri at $\phi=0$ to an expanding de Sitter phase, which is connected, through another bounce, to a contracting phase approaching the classical solution in the far past. What is remarkable is that the semi-classical states remain sharply peaked throughout the entire evolution. The asymptotic analysis of the eigenstates of the evolution operator leads to the conclusion that in the period between the two bounces the dynamics is driven by an effective cosmological constant. The presence of a cosmological constant is unexpected and quite surprising, since the ``bare'' theory we started from does not have any cosmological constant. This is in stark contrast with standard LQC, where if one studies flat FLRW universe without cosmological constant, there is a single symmetric bounce and no effective cosmological constant. Note that the effective cosmological constant we obtain \eqref{effective-params1} is of quantum gravity origin. It might be surprising that quantum gravity effects are present for large volume and low energy density, however the analysis of the effective dynamics shows that the Ricci scalar curvature remains constant in the region $t\rightarrow -\infty$ (figure \ref{fig:t}), therefore justifying the presence of quantum corrections.

Finally, an interesting aspect of our results in the quantum theory is the existence of a transition from expanding to contracting de Sitter epoch, which in fig.~\ref{fig:main} happens at $\phi=0$. This issue has already been discussed in \cite{KP09}. On one hand, since the de Sitter expanding/contracting Universe with a scalar field is future/past complete, the two sectors $\phi<0$ and $\phi>0$ are geodesically complete, thus from the classical spacetime perspective they constitute separate Universes. On the other hand, the trajectories of locally observable quantities (for example matter energy density) as functions of $\phi$ have a unique analytic extension through that point. Therefore, from the quantum theory perspective (where the time problem forced us to use the matter as a clock) the extension of spacetime past the transition point is natural.

\documentclass[../main.tex]{subfiles}

\section{Conclusions and outlook}
\label{sec:conclusion}

In this article we studied the physical effects of an alternative to the standard regularization of the Hamiltonian constraint in the framework of Loop Quantum Cosmology. We did so on the example of a flat isotropic FRLW universe with massless scalar field as the matter content, focusing the attention on the original proposal of Thiemann introduced for full LQG. The difference with respect to the one used originally \cite{Boj99a,Boj99b,APS06a,APS06b,APS06c} manifests itself in the so called \emph{Lorentzian} part of the constraint (depending on the extrinsic curvature) and leads to a modified evolution operator $\Theta_{\rm TR}$ taking the form as expressed in \eqref{final_theta}. Unlike standard LQC, where in the volume representation the evolution operator is a difference operator of the $2$nd order, in our case $\Theta_{\rm TR}$ is a difference operator of the $4$th order. Nonetheless, in the representation of the volume canonical momentum (denoted as $b$ and classically related to the Hubble rate), both operators are of $2$nd order. 
Also, the structure of the superselection sectors on the (kinematical) Hilbert space induced by the new operator is the same as in the mainstream LQC: $(i)$ division of the wave function supports onto the set of discrete uniform lattices, and $(ii)$ a symmetry with respect to triad orientation change allowing to work with either symmetric or antisymmetric states. In consequence the superselected spaces are separable despite the full kinematical Hilbert space being non-separable.

Unlike the old form of the operator, which was essentially self-adjoint, $\Theta_{\rm TR}$ admits an entire family of self-adjoint extensions parametrized by $U(1)$ group elements, a structure which is very similar to the one featured by model of isotropic universe with massless scalar field and positive cosmological constant in standard LQC. As in there, while the choice of each of the extensions leads to inequivalent unitary evolutions, all of them lead to very similar dynamical predictions. The decomposition of unity for each extension (that is, the eigenbases of extensions of $\Theta_{\rm TR}$) were evaluated numerically in $v$-representation (while having in $b$-representation relatively simple analytic form) and their large $v$ asymptotic behavior was determined analytically \eqref{eq:eig-asympt}.

The construction of the physical Hilbert space for the new model was achieved systematically using the group averaging method as in standard LQC in \cite{ALMMT-ga}. The precise identification of the space and the extended domains of $\Theta_{\rm TR}$ allowed in turn to determine the quantum trajectories corresponding to the physical states defined through expectation values of families of Dirac observables parametrized by the value of the scalar field 
(which plays the role of the internal clock, as in standard LQC). These observables are: the compactified volume \eqref{eq:obs-theta}, the matter energy density \eqref{eq:obs-rho} and the Hubble rate \eqref{eq:obs-hub}. For technical reasons, these quantum trajectories could be evaluated only numerically, thus forcing us to focus on specific classes of states. In particular, in order to compare the predicted dynamics with standard classical cosmology and predictions of mainstream LQC, we have chosen for our studies the states which were semiclassical at some point in (scalar field) time, that is, sharply peaked in the selected observables and corresponded to a large expanding universe. The calculation of the quantum trajectories consisted then in evolving such states backwards in time. We focused our attention on the ``energy Gaussian'' states of spectral profiles specified in \eqref{eq:state-Gauss}. A population of such states, peaked about different values of the scalar field momentum and with various variances, have been probed this way.

In addition to the fully quantum analysis, we constructed an effective description of the model by introducing an effective Hamiltonian \eqref{H-eff-tot} (as function of classical phase space variables), which generates a dynamics approximating very well the genuine quantum evolution. This effective Hamiltonian was constructed in a heuristic way (standard for LQC), i.e., by replacing its component elementary operators by their expectation values. Its form was simple enough that the equations of motion it generates could be solved analytically. It is worth noting that this effective Hamiltonian is in agreement with the one obtained by taking the expectation value of the full LQG Hamiltonian on coherent states peaked about isotropic cosmological spacetimes \cite{DL17a,DL17b}.

Both these approaches gave a consistent dynamical picture of the evolution of a Universe which is semiclassical at late time. That evolution starts with a large contracting Universe following the predictions of GR, until energy density of the matter content reaches the Planckian order. Then, as in standard LQC, the (loop) quantum geometry effects generate an effective repulsive gravity force which modifies the dynamics, leading to a bounce at roughly $1/4$ of LQC critical energy density.
After the bounce the Universe quickly expands, although now (unlike in the old LQC picture), instead of following the classical trajectory, it follows one corresponding to a classical Universe with large (meaning of Planckian order) positive cosmological constant and a modified Newton constant.\footnote{The values of modified cosmological and Newton constants for this model has been found independently in \cite{param-effective}.} In this phase, the volume (as measured by the compactified volume observable) reaches infinity for finite value of the (scalar) clock field. 
At that point, we observe a transition of de Sitter conformal future to conformal past into a contracting de Sitter Universe, similar to that observed in LQC models of the universe with positive cosmological constant. The fine details of the transition depend on the choice of superselection sector. Thus, in order to have a fully deterministic evolution, a specific extension (or, equivalently, the boundary data at conformal infinity) has to be chosen. However, all the extensions provide the same (up to numerically undetectable discrepancies) quantum trajectory.
The now contracting Universe follows an effective trajectory again well agreeing with that of the de Sitter Universe with the same effective $\Lambda$ and $G$ as in the expanding epoch. Once the Universe contracts sufficiently and the matter energy density reaches again Planckian order, we observe the second bounce, after which the Universe enters a classical trajectory describing a large expanding Universe.

Despite the observed consistency with each other, the fully quantum (numerical) approach and the effective approach are not sufficient to establish the complete robustness of the results presented above. This happens because, due to limitations of the numerics (finite computational time), we were able to investigate only a population with a \emph{finite number of examples} of quantum states. The results provided by the effective dynamics (having analytic form) are general, however the method itself relies on the heuristic construction of the effective Hamiltonian and its accuracy has been verified just for a finite number of examples. Fortunately, the key features of the dynamics discussed above can be verified by asymptotic analysis of the physical Hilbert space bases corresponding to each extension. \emph{All} of them share the following properties:
\begin{enumerate}[(i)]
	\item all the asymptotic waveforms have a form of a reflected/standing wave. This implies symmetry of the qualitative picture of the state evolution. In other words, to each contracting phase there is an expanding counterpart, with possibly different details in the features of the Universe. Also, an immediate consequence of this fact is the presence of at least one bounce.
	\item The asymptotic waveforms are combinations of two types of waveforms: the ones appearing in the geometrodynamical quantum description of an isotropic Universe with massless scalar field (see for example \cite{APS06c}) and the ones of the isotropic Universe with massless scalar field and positive cosmological constant (see \cite{Paw16}). This implies the presence of (both expanding and contracting) phases of classical evolution as well as the effective de Sitter epochs.
\end{enumerate}
The properties listed above are features of all the ``energy'' (momentum conjugate to the scalar field used as clock) eigenstates, thus any physical state of sufficiently good semiclassical nature must feature the properties indicated above. It is worth mentioning that the form of the asymptotics allows to identify the value of the effective cosmological constant (although not of the modified Newton constant, for which we would need to determine higher order corrections) in the de Sitter phase, see \eqref{lambda-eff-from-q}.

In the determination of the global evolution picture above, an essential role was played by the choice of parametrization of the evolution by the matter field. It is worth noting that the expanding and contracting de Sitter epochs featured in this picture are, respectively, future and past geodesically complete, which means that the transition point at conformal infinity corresponds to the infinite future/past in standard cosmic time. Therefore, from the point of view of time parametrization natural in classical theory (GR), the epochs before and after the transition can be considered as ``independent'', i.e., separate distinct Universes each of which is geodesically complete. Following this perspective, one could restrict the attention to the evolution of the after-transition branch, and treat the transition point as the ``true'' origin of the Universe, lying in the infinite past. This particular observation is relevant for present and future attempts to study perturbations in this model and its extensions: indeed, in the proper (cosmic) time chart (the one containing the ``present'' large classical expanding Universe epoch), the point of origin of the Universe is a contracting de Sitter region, which allows to select a unique initial state for matter and geometry quantum perturbations (inhomogeneities) known as a Bunch-Davies vacuum \cite{Agu18}.

While the use of proper time is more natural from the perspective of classical GR, we have to remember, that in the quantum description no such notion of time is present. This was the feature that forced us to select the matter clock as the evolution parameter. Thus, from the perspective of the quantum theory, the whole evolution picture where the two geodesically complete ``Universes'' are just epochs of evolution of the same Universe (known as ``aeons'') connected by the de Sitter transition point is the correct one. In this sense, the global evolution resembles the proposal of Cyclic Conformal Cosmology (CCC) \cite{p-cycl}. In comparison to that proposal, however, the picture emerging here differs in a key point: instead of gluing the conformal future infinity of one aeon to the Big Bang singularity of the next one, here we end up with gluing\footnote{In the new model the data are not actually glued, but they evolve through the transition point in a deterministic manner, once a particular self-adjoint extension is selected.} the future conformal infinity of one aeon with the past conformal infinity of the other aeon. Such transition allows for much better understanding of transfer of information from one aeon to the next, since the mathematical results used in CCC were originally developped for future infinity to past infinity transition \cite{Friedrich}. Therefore, the model studied here comes equipped with interesting features of CCC, while not being weighted down by the restrictions imposed by conformal infinity to singularity transition needed there. 

It is worth mentioning, that the transition between the expanding and contracting de Sitter epochs ocurring at the finite (matter) time is not just a result of choosing massless scalar field as the internal clock. This feature would be present also if other non-exotic forms of matter (with the exception of dust) -- for example the radiation \cite{PPW14} -- were used as a clock. Such ``universality'' becomes important once we start trying to answer the question, which choice of time (proper versus matter) -- and in consequence which evolution picture -- we should adopt. This question, while appearing to border on philosophy, can be approached in an operational way: what we perceive as the passage of time are dynamical changes of the configurations of matter fields (the clock's pointer, the electrochemical potentials in neurons); given that, as well as the necessity to use matter clocks in the quantum description of the geometry, suggests that the bigger picture containing both proper time charts might be the more natural one. Such choice will have nontrivial consequences once the inhomogeneities (i.e., perturbations) are included in the model, as now the previously initial perturbations will be generated by the (possibly very rich) history of the Universe before the transition. In principle, this may lead to possible imprints of the existence of the previous aeon, for example through the gravitational wave emissions of black hole mergers as it is hoped for the model of CCC \cite{MN15_rings}.

The results and techniques presented here were provided in the context of the particular model -- flat FRLW universe with massless scalar field -- the simplest one commonly used for testing new ideas in LQC. The inclusion of the studied regularization can however also be performed for more advanced models: with more complicated matter content and extended to the homogeneous anisotropic cosmology with use of more recently available techniques. However, investigating the possible physical significance of all these models would require inclusion of the inhomogeneities, either in terms of perturbations or at the nonperturbative level. For that, the existing techniques need to be better understood and possibly improved. For example the issue of instabilities for some treatments of inhomogeneities, and the discrepancies of predictions between different treatments (see for example \cite{Guillermo}) need to be addressed. 

From a more fundamental point of view, an important consequence of our result is that in LQC different regularizations lead to different physical predictions. Since in the standard quantization schemes the choice of regularization is considered a minor technical detail and the results are often required to not depend on it, our result poses a challenge for the predictive power of the theory. The dependence found here leads to the major task to find a way to single out a physically preferred regularization. This could be achieved by introducing new consistency criteria into the theory: a possible example comes in the form of demanding cylindrical consistency, which has been studied in various contexts for applications to LQG \cite{BD09,Ba14,LLT17}.

\section*{Acknowledgements}
M.\ A.\ acknowledges the support of the project BA 4966/1-1 of the German Research Foundation (DFG).  K.\ L.\ thanks the German  National  Merit  Foundation  for their financial support and the NSF grant PHY-1454832. T.P. acknowledges the support of the Polish Narodowe Centrum Nauki (NCN) grant 2012/05/E/ST2/03308.

\appendix

\documentclass[../main.tex]{subfiles}

\section{Review of LQC}
\label{ReviewA}

In this appendix we review the derivation of the evolution operator in standard LQC, with focus on the regularization choice. We also introduce the new regularization used in this article which is more in line with full LQG.
\subsubsection{Hamiltonian formulation of cosmology}
Let us start by recalling the Hamiltonian formulation of GR in terms of the Ashtekar-Barbero variables\cite{Ash86,Ash87,Ash88,Bar94,Bar95}. In this context, one recasts GR as a gauge theory with internal group $SU(2)$, and identifies the phase space of GR as the one coordinatized by the Ashtekar connection $A^i_a$ and the inverse densitized triad $E^b_j$ ($i,j = 1,2,3$ are $SU(2)$ algebra indices, while $a,b = 1,2,3$ are spatial ones). Explicitly, these are given by
\begin{align}
A^i_a = \Gamma^i_a + \gamma K^i_a, \ \ \ \ \ E^a_i = \sqrt{\det q} \ e^a_i
\end{align}
where $\gamma\in\mathbb{R}-\{0\}$ is a free quantity called the {\it Immirzi parameter}, $e^a_i$ is the (inverse) triad of the metric, and $\Gamma^i_a$ and $K^i_a$ are related respectively to the spin-connection $\Gamma^i_{\ ja}$ and the extrinsic curvature $K_{ab}$ by (we employ the summation convention on repeated indices, and raise/lower $i,j,...$ with the Euclidean metric)
\begin{align}
\Gamma^i_a = -\frac{1}{2} \epsilon^{ijk} \Gamma^j_{\ ka}=-\frac{1}{2}\epsilon^{ijk}e^b_k(2\partial_{[b}^{\;} e^j_{a]}+e^c_je^m_a\partial_be^m_c), \ \ \ \ \ K^i_a = K_{ab} e^b_i
\end{align}
With $(A,E)$ being canonical coordinates, the symplectic form reads
\begin{align} \label{symplectic-structure}
\Omega = \frac{1}{8\pi G \gamma} \int_{\sigma_M} d^3x \ \text dA^i_a(x) \wedge \text dE^a_i(x)
\end{align}
where $\sigma_M$ is some compact cell. $\Omega$ defines the Poisson bracket
\begin{align}
\{A^i_a(x), E^b_j(x')\} = 8\pi G\gamma  \delta^i_j \delta^b_a \delta^{(3)}(x,x')
\end{align}
Due to the symmetries of the theory, one finds that not all the dof's in $(A,E)$ are physical. This fact is encoded in the following constraints:
\begin{itemize}
\item the Gauss constraint, that generates internal $SU(2)$ transformations:
\begin{align}
G_i = \dfrac{1}{16\pi G\gamma} \left[\partial_a E^a_i + \epsilon_{ijk} A^j_a E^a_k\right]
\end{align}
\item the Vector constraint, that generates spatial diffeomorphisms:
\begin{align}\label{diffeo.constraint-full}
C_a = \dfrac{1}{8\pi G\gamma} F^i_{ab} E^b_i
\end{align}
\item the Scalar constraint, that generates time-like diffeomorphisms:
\begin{align}\label{scalar-constraint-full}
C= C_E +C_L
\end{align}
where
\begin{align} \label{C-before-reduction}
C_E = \dfrac{1}{16\pi G} \dfrac{\epsilon_{ijk} E^a_j E^b_k}{\sqrt{\det q}} F^i_{ab},\hspace{50pt}
C_L=
 -(1 + \gamma^2) \dfrac{1}{16\pi G} \dfrac{\epsilon_{ijk} E^a_j E^b_k}{\sqrt{\det q}} \epsilon_{imn} K^m_a K^n_b
\end{align}
\end{itemize}
In these equations, $F^i_{ab}$ is the gauge curvature of connection $A^i_a$, explicitly given by
\begin{align}\label{F_ab}
F^i_{ab} = \partial_a A^i_b - \partial_b A^i_a + \epsilon_{ijk} A^j_a A^k_b
\end{align}
We now apply this framework to the case of flat isotropic cosmology, i.e., the symmetry-reduced metric
\begin{align}
g = -dt^2 + q ,\hspace{30pt} q= a^2(t) \eta
\end{align}
where $a(t)$ is the scale factor and $\eta$ is the Euclidean 3-metric.\\
It is immediate to compute the triads: imposing $q_{ab} = e^i_a e^j_b \delta_{ij}$, we find $e^i_a = a \delta^i_a$, from which it follows that $E^a_i = a^2 \delta^a_i$. On the other hand, since the metric is independent of spatial coordinates, we have $\Gamma^i_{\ ja} = 0$, and so $A^i_a = \gamma K^i_a$. Finally, using the fact that the extrinsic curvature reduces to $K_{ab} = \dot q_{ab}/(2N) = \delta_{ab} a\dot a/N$, we find $A^i_a = \delta^a_i \gamma \dot a/N$. We can therefore summarize this by saying that, for flat isotropic cosmology, Ashtekar variables are
\begin{align} \label{frw-ash}
A^i_a = c \delta^i_a, \ \ \ \ \ E^a_i = p \delta^a_i
\end{align}
with $c = \gamma \dot a/N$ and $p = a^2$. We can think of $(c,p)$ as coordinatizing the subspace of GR phase space representing flat isotropic cosmology. Plugging (\ref{frw-ash}) in (\ref{symplectic-structure}), we find the reduced symplectic structure:
\begin{align}
\Omega = \frac{3}{8\pi G \gamma} \text dc \wedge \text dp \ \int_{\sigma_M} d^3x = \frac{3 V_o}{8\pi G \gamma} \text dc \wedge \text dp, \ \ \ \ \ V_o := \int_{\sigma_M} d^3x
\end{align}
from which it follows that the Poisson bracket on the reduced phase space is
\begin{align}
\{c,p\} = \frac{8\pi G\gamma}{3 V_o}
\end{align}
Finally, plugging (\ref{frw-ash}) in the expression for the gauge curvature, we find $F^i_{ab} = c^2 \epsilon_{iab}$. Using this in the constraints, one sees that the Gauss constraint and the vector constraint vanish identically, while the scalar constraint reduces to
\begin{align}
C = -\dfrac{3}{8\pi G\gamma^2} \sqrt p c^2
\end{align}
This concludes the review of the Hamiltonian formulation of classical cosmology.
\subsubsection{Kinematical Hilbert space of LQC}
We will only give a brief overview of how the kinematical Hilbert space is defined. For further details, we refer to \cite{Boj05,Ash09_Cos,AP11}.
\\
\\
The canonical quantization of full general relativity in terms of its Ashtekar connection leads to the approach called Loop Quantum Gravity (LQG) \cite{Thi07}. As it transpires in LQG, constructing an operator corresponding to the connection $A^i_a(x)$ does not lead to a successful quantization. Instead, the fundamental algebra which will be promoted to quantum operators is the classical {\it holonomy-flux algebra}. The holonomies of the connection $A^i_a$ are constructed as the path-ordered exponentials of $A^i_a$ smeared with respect to some piecewise analytic curves, whose real analytic segments are called {\it edges}, $e$:
\begin{align}
h(e):=\mathcal{P}\exp\left(\int_e A\right), \ \ \ \ \ A = A^i_a dx^a \tau_i
\end{align}
where $\tau_i$ are the generators of the algebra $\mathfrak{su}(2)$, and are chosen to be related to the Pauli matrices by $\tau_i = -i \sigma_i/2$, such that $[\tau_i,\tau_j] = \epsilon_{ijk}\tau_k$. While in LQG one quantizes the holonomies on every edge, for the purposes of LQC it suffices to restrict to certain special edges. The form $A^i_a= c\delta^i_a$ naturally suggest to choose edges oriented along the three axes of coordinates of the fiducial metric $\eta_{ab}$. Since their global position does not matter, we only consider three families of edges parametrized by their coordinate length $\epsilon>0$ and whose tangent are respectively $\dot{e}_{\pm x,\epsilon}=\pm\hat{x}$, $\dot{e}_{\pm y,\epsilon}=\pm\hat{y}$ and $\dot{e}_{\pm z,\epsilon}=\pm \hat{z}$. Hence the holonomies take the explicit form
\begin{align} \label{hol-hom}
h(e_{\pm x,\epsilon}) = e^{\pm c \epsilon \tau_1}, \ \ \ \ \ h(e_{\pm y,\epsilon}) = e^{\pm c \epsilon \tau_2}, \ \ \ \ \ h(e_{\pm z,\epsilon}) = e^{\pm c \epsilon \tau_3}
\end{align}
In LQC, we restrict the algebra further, and only consider edges of one finite length $\epsilon=\mu$. The choice of $\mu$ is a {\it crucial} part of the construction in LQC. Currently, the most widely accepted choice is the so-called {\it $\bar{\mu}$-scheme} (also known as {\it improved dynamics} \cite{APS06c}), which prescribes to keep $\mu$ finite (as opposed to sending it to $0$, as one would do in lattice QFT). The reasoning behind this choice is based on the regularization of gauge curvature $F^i_{ab}$: as we will later see, $F^i_{ab}$ can be approximated in terms of holonomies along a small closed curve; in this case, $\mu^2$ can be thought of as the coordinate area of the surface enclosed by the loop; however, in LQG the area is an operator with discrete spectrum \cite{AL96_Ar}, and so one fixes $\mu$ (which in this scheme is denoted by $\bar\mu$) so that the physical area $p \bar\mu^2$ of the loop coincides with the smallest non-vanishing area eigenvalue, $\Delta$. In other words, we set
\begin{align}
\bar{\mu}:= \frac{\sqrt{\Delta}}{\sqrt{|p|}},\hspace{50pt} \Delta:=2\pi\sqrt{3} \gamma G\hbar\approx 2.61 \ell^2_{\rm Pl}
\end{align}
with $\ell_{\rm Pl}$ the Planck length.
\\
\\
With $\bar{\mu}$ being a small quantity which we want to use as regularization parameter for the physical quantities of interest, it is useful to rescale the connection $c$ by $\bar\mu$. Its canonical momentum is nothing but the spatial volume:
\begin{align}
b := c \bar\mu, \hspace{50pt} V := p^{3/2}
\end{align}
The Poisson algebra between the two reads
\begin{align} \label{b-V-alg}
\{b,V\} = \frac{2\alpha}{\hbar}, \ \ \ \ \ \text{with} \ \alpha = 2\pi G \hbar \gamma \sqrt\Delta
\end{align}
The gravitational Hilbert space $\mathcal{H}_{\rm gr}$ is constructed using the canonical pair $V,b$. Being a real observable, we implement the volume as a multiplication operator on $L_2(\bar{\mathbb{R}}, d\mu_{\rm Bohr}(v))$, which is the space of square integrable functions on the Bohr compactification of the real line \cite{APS06b,Vel07}:
\begin{align}
\hat{V} |v\rangle = \alpha |v| \ |v\rangle, \ \ \ \ \ \langle v | v' \rangle = \delta_{v,v'}
\end{align}
where $|v\rangle$ form an orthonormal basis of eigenstates on $L_2(\bar{\mathbb{R}}, d\mu_{\rm Bohr}(v))$. Given that $\hat{V}$ acts by multiplication, (\ref{b-V-alg}) would suggest to implement $b$ as a derivative with respect to $v$. However, mimicking LQG -- in which the connection $A^i_a$ is not promoted to operator, but $h$ is -- we do not promote $b$ to operator, but rather its exponentiated version, $\mathcal N := e^{i b/2}$. The corresponding quantum operator is therefore acting as a shift:
\begin{align}
\mathcal{\hat{N}} |v\rangle = |v+1\rangle
\end{align}
Note that $L_2(\bar{\mathbb{R}}, d\mu_{\rm Bohr}(v))$ includes square integrable functions with negative $v$. We thus define as kinematical Hilbert space the subspace of symmetric states,
\begin{align}
\mathcal{H}_{\rm gr}:=\{\psi(v)\in L_2(\bar{\mathbb{R}}, d\mu_{\rm Bohr}(v))\; :\; \psi(v)=\psi(-v)\}
\end{align}
by which we encode the fact that $v \to -v$ is a large gauge transformation which does not change the physics of the model \cite{APS06b}.
\\
\\
We will now proceed by promoting all classical quantities of interest to operators on $\mathcal H_{\rm gr}$. Let us start with holonomies (\ref{hol-hom}): using a known property of $SU(2)$ matrices, we can write
\begin{align}
h_a & := h(e_{a,\bar{\mu}}) = \exp(b \tau_a) = \cos(b/2) I + 2 \sin(b/2) \tau_a = \left(\frac{I}{2} - i\tau_a\right) e^{ib/2} + \left(\frac{I}{2} + i\tau_a\right) e^{-ib/2} = \notag
\\
& = \left(\frac{I}{2} - i\tau_a\right) \mathcal N + \left(\frac{I}{2} + i\tau_a\right) \mathcal N^{-1}
\end{align}
where $I$ is the $2 \times 2$ identity matrix. Hence, the quantum version is simply
\begin{align}
\hat{h}_a= \left(\frac{I}{2} - i\tau_a\right)\mathcal{\hat{N}} + \left(\frac{I}{2} + i\tau_a\right) \mathcal{\hat{N}}^{-1}
\end{align}
To extract the physical sector of the Hilbert space, one follows the Dirac program, which consists in promoting the constraints to operators, and then imposing that physical states lie in their kernel. As said, in homogeneous isotropic cosmology the only non-trivial constraint is the scalar one. The matter part of it needs to be treated separately (as it requires quantization of the matter degrees of freedom); now we focus on the geometric part.
\subsubsection{Scalar constraint in LQC}
The implementation of the scalar constraint $C$ as an operator requires a regularization. As already discussed, the need for regularization in full LQG originates from the fact that no quantum operator for the connection $A^i_a(x)$ exists, while its corresponding holonomies naturally lead to the representation theory of the group $SU(2)$. But as the classical scalar constraint $C$ is given in terms of $A^i_a(x)$, it must be rewritten in terms of holonomies before this quantization procedure can be applied. However, it is {\it not} possible to express $C$ exactly as a function of holonomies of finite length. Hence, one must necessarily construct a regularization $C^\epsilon$ of $C$ such that, in the limit $\epsilon \to 0$, the continuum result is restored. The same holds true in the context of cosmology, where $C$ is classically a function of $b$ and $v$:
\begin{align}
C = -\dfrac{3}{8\pi G \gamma^2 \Delta} V b^2
\end{align}
Since there is no operator in LQC corresponding to $b$, we must consider a regularization of $C$ in terms of $\mathcal N$ and $v$. Here, we will recall the regularization commonly used in LQC, and then compare it with a new proposal which is closer to the regularization of the scalar constraint in full LQG.
\\
\\
Let us start by regularizing the Euclidean part of the scalar constraint ($C_E$ in (\ref{C-before-reduction})). Consider first the gauge curvature $F^i_{ab}$, equation (\ref{F_ab}). We define
\begin{align}\label{Holonomie_Loop}
(F^\epsilon)^i_{ab}(x) := -\frac{1}{4\epsilon^2} \sum_{s_a, s_b = \pm 1} s_as_b\text{Tr}\left[ \tau^i \left(h(\Box^\epsilon_{s_aa,s_bb}) - h(\Box^\epsilon_{s_aa,s_bb})^\dagger\right)\right]
\end{align}
where $\Box^\epsilon_{\pm a, \pm b}$ is a small plaquette starting at point $x$ with tangent $\pm x^a$ and ending at the same point with tangent $\pm x^b$. It is not hard to show that
\begin{align}
\lim_{\epsilon \to 0} \ (F^\epsilon)^i_{ab} = F^i_{ab}
\end{align}
Now, using the fact that $h(\Box^\epsilon_{s_aa,s_bb}) = h_{s_aa} h_{s_bb} h_{s_aa}^\dag h_{s_bb}^\dag$ and the explicit expressions (\ref{hol-hom}), one computes $(F^\epsilon)^i_{ab}(x) = \epsilon_{iab} \sin(c\epsilon)^2/\epsilon^2$, which clearly reduces to the classical cosmological $F^i_{ab}$ in the limit $\epsilon\to0$. As already discussed, in LQC one makes the choice $\epsilon=\bar{\mu}$, from which one finds 
\begin{align}\label{F_regul}
\epsilon_{abc} (F^{\bar\mu})^i_{ab} = 2\delta^i_c \dfrac{\sin(b)^2}{\bar\mu^2}
\end{align}
which can be easily written in terms of $\mathcal N$ and hence promoted to an operator in LQC. The other term appearing in $C_E$ besides $F^i_{ab}$ is the non-polynomial expression $\text{sgn}(\det(e)) E^b_kE^c_l/\sqrt{|\det(E)|}$. This can be regularized via the first Thiemann identity \cite{Thi98a}
\begin{align}\label{ThiemannI}
2\pi G\gamma \ \text{sgn}(\det(e)) \epsilon^{jkl}\epsilon_{abc}\frac{E^b_kE^c_l}{\sqrt{|\det(E)|}} = \frac{2}{\bar{\mu}} \text{Tr}\left(\tau^j h_a\{h_a^\dagger,V[\sigma_M]\}\right) + \mathcal O(\bar\mu)
\end{align}
It is immediate to promote the right hand side to an operator in LQC: its action on volume eigenvalue $|v\rangle$ reads
\begin{align}\label{Holonomy_correction}
\text{Tr}\left(\tau^j \hat{h}_a [ \hat{h}_a^\dagger, \hat{V} ]\right) | v\rangle = -\frac{i\alpha}{2}\delta^j_{a}(|v-1|-|v+1|)|v\rangle
\end{align}
Using (\ref{F_regul}) and (\ref{ThiemannI}) in $C_E$, one finds
\begin{align} \label{CE-regularized}
\nonumber C_E[N] & = \dfrac{N}{16\pi G} \dfrac{\epsilon_{ijk} E^a_j E^b_k}{\sqrt{|\det(E)|}} F^i_{ab} = \dfrac{N}{16\pi G} \dfrac{\epsilon_{ijk} E^c_j E^d_k}{\sqrt{|\det(E)|}} \left(\frac{\delta_c^a \delta_d^b - \delta_c^b \delta_d^a}{2}\right) F^i_{ab} =\\
& = \dfrac{N}{16\pi G} \dfrac{\epsilon_{cdf} \epsilon_{ijk} E^c_j E^d_k}{\sqrt{|\det(E)|}}\ \frac{\epsilon^{abf} F^i_{ab}}{2} \longrightarrow \notag
\\
\longrightarrow C^{\bar{\mu}}_E[N] & = \frac{2^4 N}{(16\pi G)^2\gamma\Delta^{3/2}}\sin(b) \sqrt{V} \left(\sum_a \text{Tr}(\tau^a h_a \{h_a^\dagger , V\})\right) \sqrt{V} \sin(b)
\end{align}
where we considered a symmetric ordering on $\mathcal{H}_{\rm gr}$. The action of the corresponding quantum operator is therefore
\begin{align} \label{Euclidean_quant_op}
\hat{C}_E[N]|v\rangle & = \frac{3\times 2 N\alpha^2}{(16\pi G)^2\hbar\gamma\Delta^{3/2}}\times\nonumber\\
&\hspace{15pt}\times (\hat{\mathcal{N}}^2-\hat{\mathcal{N}}^{-2})((|v+1|-|v+3|)|v+2|\hat{\mathcal{N}}^2-(|v-3|-|v-1|)|v-2|\hat{\mathcal{N}}^{-2})|v\rangle \nonumber
\\
& = \dfrac{3N\alpha}{4(16\pi G)\Delta}\left(F(v+2) \hat{\mathcal{N}}^4 - F_0(v) {\rm id} + F(v-2) \hat{\mathcal{N}}^{-4} \right)|v\rangle
\end{align}
where the lapse function has been chosen to be independent of $b$ and $V$ and
\begin{align} \label{F-0-F}
F_0(v):=F(v+2)+F(v-2), \ \ \ \ \ F(v):=-|v|(|v+1|-|v-1|)
\end{align}
This is the LQC quantization of the Euclidean part of the scalar constraint.
\\
\\
Let us now turn to the Lorentzian part, $C_L$ in (\ref{C-before-reduction}). The standard procedure in LQC is based on the observation that, on the flat cosmological sector, the following relations hold:
\begin{align} \label{true-only-in-cosmo}
\gamma K^i_a |_{\rm cos}= A^i_a|_{\rm cos}\hspace{20pt}{\rm and}\hspace{20pt}2\gamma^2 K^i_{[a}K^j_{b]}|_{\rm cos}=\epsilon_{ijk} F^k_{ab}|_{\rm cos}
\end{align}
Using these relations, one finds that in classical cosmology the Lorentzian part is proportional to the Euclidean part, $C_L|_{\rm cos} = -C_E|_{\rm cos} (1+\gamma^2)/\gamma^2$. It can therefore be regularized in the same way. Following this route, one ends up with
\begin{align}
\hat{C}[N]_{\rm LQC}|v\rangle=\frac{-3N\alpha}{4\gamma^2(16\pi G)\Delta}\left(F(v+2) \hat{\mathcal{N}}^4 - F_0(v) {\rm id} + F(v-2)) \hat{\mathcal{N}}^{-4} \right)|v\rangle
\end{align}
which is the quantum operator describing the dynamics in standard LQC \cite{APS06a,APS06b,APS06c,PA11}.

\subsubsection{Scalar constraint with the new (Thiemann) regularization}
As explained in the main text, we now follow the philosophy ``first regularize, then reduce''. This leads us to a regularization which is more in contact with the full theory, where the Lorentzian part is {\it not} proportional to the Euclidean part.\\
\\
We recall the second Thiemann identity, which is true in full GR if the regularization parameter $\epsilon > 0$ is independent of the phase space:
\begin{align}\label{AThiemannII}
 \tau_j K^j_a=\frac{1}{8\pi G\gamma^3}\{\tau_j A^j_a,\{C_E[1],V\}\}=-\frac{1}{8\pi G\gamma^3 \epsilon} h(e_{a,\epsilon})\{h^\dagger(e_{a,\epsilon}),\{C_E^{\epsilon}[1],V\}\}+\mathcal{O}(\epsilon)
\end{align}
This allows to find the new regularization of the Lorentzian part of the scalar constraint:
\begin{align}
\nonumber C_L[N] & = - (1 + \gamma^2) \dfrac{N}{16\pi G} \dfrac{\epsilon_{ijk} E^a_j E^b_k}{\sqrt{|\det E|}} \epsilon_{imn} K_a^m K_b^n = \notag
\\
& = 4 (1 + \gamma^2) \dfrac{N}{16\pi G} \dfrac{\epsilon_{ijk} E^a_j E^b_k}{\sqrt{|\det E|}} \text{Tr}\left(\tau_i \tau_m \tau_n \right) K_a^m K_b^n \longrightarrow \notag
\\
\longrightarrow C_L^{\epsilon}[N] & = -\frac{1+\gamma^2}{\gamma^7(16\pi G)^4}\frac{4^3N }{\epsilon^3}\epsilon^{abc} \text{Tr}\left(h_a\{h_a^\dagger,\{C_E^{\epsilon}[1],V\}\} h_b\{h_b^\dagger,\{C_E^{\epsilon}[1],V\}\} h_c\{h_c^\dagger,V\}\right)\label{ALorentz_reg}
\end{align}
where in the last line we used \eqref{AThiemannII} and
\begin{align}
\tau_j\epsilon^{jkl}\frac{E^b_kE^c_l}{\sqrt{|\det(E)|}}=-\frac{1}{4\pi G\gamma \epsilon}\epsilon^{abc} h_a\{h_a^\dagger,V[\sigma_M]\}+\mathcal{O}(\epsilon)
\end{align}
which is related to \eqref{ThiemannI}. This expression can be evaluated on the cosmological sector by first reducing each argument of the Poisson brackets to the cosmological sector, and then using the Poisson bracket between $c$ and $p$. This yields (we set $V_0=1$ to ease the notation)
\begin{align}\label{40}
C^\epsilon_L[N]|_{\rm cos} & = \frac{1+\gamma^2}{\gamma^7(16\pi G)^4}\frac{4^3N}{\epsilon^2}\left(\frac{6}{16\pi G\epsilon^2}\right)^2\epsilon^{abc} \times \notag
\\
& \times \text{Tr}\left(e^{\epsilon c\tau_a}\{e^{-\epsilon c\tau_a},\sqrt{p}\{\sin(c\epsilon)^2,p^{3/2}\}\}e^{\epsilon c\tau_b}\{e^{-\epsilon c\tau_b},\sqrt{p}\{\sin(c\epsilon)^2,p^{3/2}\}\}\sqrt{p}\tau_c\frac{16\pi G\gamma}{4}\right) = \notag
\\
& = \frac{1+\gamma^2}{\gamma^2 16\pi G}\frac{N}{\epsilon^2}\epsilon^{abc}  \;\text{Tr}\left( \tau_a\tau_b \tau_c\right)\sin(2c\epsilon)^2\sqrt{p} = \notag
\\
& = -\frac{6N}{16\pi G}\frac{1+\gamma^2}{\gamma^2}\sqrt{p}\frac{\sin(2\epsilon c)^2}{4\epsilon^2}
\end{align}
which agrees with the continuum expression for $C_L|_{\rm cos}$ in the continuum limit $\epsilon\to 0$, yet is {\it not} proportional to $C_E^\epsilon$! This realization motivates us to consider a new quantization for the Lorentzian part of the scalar constraint in LQC, based on (\ref{ALorentz_reg}). However, before continuing, one has to take care of how one passes from $\epsilon$ to $\bar{\mu}$, which is phase space dependent. Indeed, Thiemann identity (\ref{AThiemannII}) is only correct for $\epsilon$ independent of the phase space point. Thus, instead of performing the replacement $\epsilon\to\bar{\mu}$ in (\ref{ALorentz_reg}), we make use of the following observation from \cite{YDM09}, which is true {\it only} in cosmology:
\begin{align}
\tau_j K^j_a = -\frac{4}{3\bar{\mu}(16\pi G)\gamma^3} h_a\{h^\dagger_a,\{C_E^{\bar{\mu}}[1],V\}\} + \mathcal{O}(\Delta)
\end{align}
With this identity, following the same steps as in (\ref{ALorentz_reg}) one finds
\begin{align} \label{Acorrect-new-CL}
C^{\bar{\mu}}_L[N] = -\frac{1+\gamma^2}{\gamma^7(4\pi G)^4}\frac{N \epsilon^{abc}}{9\Delta^{3/2}} \text{Tr}\left(h_a\{h_a^\dagger,\{C_E^{\epsilon}[1],V\}\} \sqrt V h_b\{h_b^\dagger,V\} \sqrt V h_c\{h_c^\dagger,\{C_E^{\epsilon}[1],V\}\}\right)
\end{align}
where, as for $C^{\bar\mu}_E$. Reducing this expression to the cosmological case (as we did in (\ref{40})), one finds
\begin{align}
C^{\bar{\mu}}_L[N]|_{\rm cos}=-\frac{3N}{8\pi G}\frac{1+\gamma^2}{\gamma^2}V \frac{\sin(2b)^2}{4\Delta}
\end{align}
which correctly coincides with (\ref{40}) under the replacement $\epsilon\to\bar{\mu}$. This confirms that (\ref{Acorrect-new-CL}) is the correct regularization to use if we want to implement Thiemann identity in the $\bar\mu$-scheme.
\\
\\
The quantization of (\ref{Acorrect-new-CL}) on the Hilbert space of LQC can now be done in the standard way: putting the hats and recalling that $\widehat{\{.,.\}} = [.,.]/(i\hbar)$, we find
\begin{align}
\hat C^{\bar{\mu}}_L[N] = -\frac{(1+\gamma^2)N}{\gamma^7(4\pi G)^4}\frac{\epsilon^{abc} }{9\Delta^{3/2}}\dfrac{1}{(i\hbar)^5} \text{Tr}\left(\hat h_a[\hat h_a^\dagger,[\hat C_E^{\epsilon}[1],\hat V]] \sqrt{\hat V} \hat h_b [\hat h_b^\dagger,\hat V] \sqrt{\hat V} \hat h_c[\hat h_c^\dagger,[\hat C_E^{\epsilon}[1],\hat V]]\right)
\end{align}
To write its action on $|v\rangle$ explicitly, recall the form of $\hat h_a$ in terms of $\hat{\mathcal N}$: using (\ref{Euclidean_quant_op}), we get (no sum over $a$)
\begin{align}
&\text{Tr}\left(\tau_b \hat{h}_a[\hat{h}_a^\dagger [\hat{C}_E[1],\hat{V}]]\right)|v\rangle = -i \frac{\delta_{ab}}{2}\left(\hat{\mathcal{N}}[\hat{C}_E[1],\hat{V}]\hat{\mathcal{N}}^{-1}-\hat{\mathcal{N}}^{-1}[\hat{C}_E[1],\hat{V}]\hat{\mathcal{N}}\right)|v\rangle \nonumber\\
&\hspace{15pt}= i \frac{3\alpha^2}{8(16\pi G)\Delta}\delta_{ab}\left[-(g(v+1)-g(v+3))
\hat{\mathcal{N}}^{4}+(g(v-3)-g(v-1))\mathcal{\hat{N}}^{-4}\right]|v\rangle
\end{align}
with $g(v):=F(v)(|v-2|-|v+2|)$. It follows
\begin{align}
\hat{h}_a[\hat{h}_a^\dagger [\hat{C}_E[1],\hat{V}]]|v\rangle
&=\frac{-i3\alpha^2}{4(16\pi G)\Delta}\tau_a \left[-(g(v+1)-g(v+3))
\hat{\mathcal{N}}^{4}+(g(v-3)-g(v-1))\mathcal{\hat{N}}^{-4}\right]|v\rangle
\end{align}
From this and (\ref{Holonomy_correction}), after some manipulations, we find
\begin{align}
\hat{C}_L^{\bar{\mu}}[N]|v\rangle & = \frac{3N\alpha}{16\pi G \Delta 2^{10}}\frac{1+\gamma^2}{4\gamma^2}\left(G(v-4)\hat{\mathcal{N}}^{-8} - G_0(v){\rm id} + G(v+4)\mathcal{\hat{N}}^8
\right)|v\rangle\label{ALorentzian_quant_op}
\end{align}
where
\begin{align}\label{horriblecoefficents}
\begin{array}{c}
G(v) := -F(v)(g(v-3)-g(v-1))(g(v+1)-g(v+3))
\\
\\
G_0(v) := - F(v-4) (g(v-3)-g(v-1))^2 - F(v+4) (g(v+1)-g(v+3))^2
\end{array}
\end{align}

\documentclass[../main.tex]{subfiles}

\section{Asymptotic analysis}
 \label{app:asympt}

In order to study the semi-classical limit of the model at hand, we need to establish the asymptotic limit of the eigenfunctions of the $\Theta_{\rm TR}$ operator. The eigenvalue equation is
\begin{align}\label{Ein.Theta}
 \Theta_{\rm TR} \Psi(v) = \omega^2 \Psi(v)\ ,
\end{align}
where $\omega^2$ are the corresponding eigenvalues. The $4$th-order system \eqref{Ein.Theta} can be expressed in a $1$st-order form as follows. First one introduces the vector(s)
\begin{align}
 \vec{\Psi}(v):= \left(
 \begin{array}{l}
  \Psi(v+4) 
  \\ 
  \Psi(v)
  \\ 
  \Psi(v-4)
  \\ 
  \Psi(v-8)
  \end{array}
  \right)\ ,
\end{align}
so that equation \eqref{Ein.Theta} takes the form
\begin{align}\label{Lin.Eq}
 \vec{\Psi}(v+4) = \bm A(v) \vec{\Psi}(v)\ ,
\end{align}
where the matrix $A$ is defined as
\begin{align}
 \bm A(v) = \left(
\begin{array}{cccc}
 \frac{f_4(v)}{s f_8(v)} & \frac{2 (s-1) f_0(v)-\frac{4}{3\pi G \gamma^2}\omega^2}{s f_8(v)} & \frac{f_{-4}(v)}{s f_{8}(v)} & -\frac{f_{-8}(v)}{f_8(v)} \\
 1 & 0 & 0 & 0 \\
 0 & 1 & 0 & 0 \\
 0 & 0 & 1 & 0 \\
\end{array}
\right)
\end{align}
where $f_a(v) := \sqrt{|v(v+a)|} |v+a/2|$.
\\
The next step is to express the functions $\Psi$ as linear combinations of appropriately selected asymptotic functions. We denote these functions by $\widetilde \psi_i^\pm$, where $i\equiv F$ stands for the FLRW phase and $i\equiv S$ stands for the de Sitter phase. We then rewrite \eqref{Lin.Eq} as an equation for the coefficients in the linear combination.

Using the results of the asymptotic analysis of LQC with a scalar field \cite{Corichi:2007am, ACS08} and LQC with a cosmological constant \cite{KP09, Paw16}, we select the asymptotic functions $\widetilde \psi_i^\pm$ as follows
\begin{align} \label{asympt-SF}
 \widetilde \psi_F^\pm (v) := \frac{\exp \big(\pm i k \log (v)\big)}{\sqrt{v}}\qquad , \qquad
 \widetilde \psi_S^\pm (v) := \frac{\exp \big(\pm i (\Omega_S v + \kappa/v)\big)}{v}\ .
\end{align}
where $k$, $\Omega_S$ and $\kappa$ are functions of the parameter $s$ and the eigenvalues $\omega$, to be determined. The vector $\vec \chi^\pm$ of coefficients in the linear combination for $\psi^\pm$ can be defined as
\begin{align}
 \vec{\Psi}(v) =: \bm B(v-4) \vec \chi(v)\ ,
\end{align}
where the matrix $\bm B$ is
\begin{align}
 \bm B(v):=\left(
\begin{array}{cccc}
 \widetilde \psi_S^+ (v+8) & \widetilde \psi_S^- (v+8) & \widetilde \psi_F^+ (v+8) & \widetilde \psi_F^- (v+8) \\
 \\
 \widetilde \psi_S^+ (v+4) & \widetilde \psi_S^- (v+4) & \widetilde \psi_F^+ (v+4) & \widetilde \psi_F^- (v+4) \\
 \\
 \widetilde \psi_S^+ (v) & \widetilde \psi_S^- (v) & \widetilde \psi_F^+ (v) & \widetilde \psi_F^- (v) \\
 \\
 \widetilde \psi_S^+ (v-4) & \widetilde \psi_S^- (v-4) & \widetilde \psi_F^+ (v-4) & \widetilde \psi_F^- (v-4) \\
\end{array}
\right)\ .
\end{align}
At this point, equation \eqref{Lin.Eq} becomes
\begin{align}
 \vec \chi(v+4) = \bm B^{-1}(v)\bm A(v) \bm B(v-4)\vec \chi(v) =:\bm M(v)\vec \chi(v) \ .
\end{align}
The matrix $\bm M$ can be computed explicitly. In order to guarantee the existence of the limit $\lim_{v\rightarrow \infty} \vec \chi(v) =: \vec \chi_\infty$ (such that $\vec \chi(v) = \vec \chi_\infty \ + \vec O(v^{-1})$), the matrix $\bm M$ must asymptotically satisfy
\begin{align}\label{Asymp.Cond}
\bm M(v) = {\bm 1} + {\bm O}(v^{-2})\ ,
\end{align}
where $\bm O(v^{-2})$ denotes a matrix whose coefficients asymptotically behave as $O(v^{-2})$. 

The asymptotic condition \eqref{Asymp.Cond} determines the expression of the functions $k$, $\Omega_S$ and $\kappa$:
\begin{align}
k = \frac{\omega}{\sqrt{12\pi G}}\ , \ \ \ \ \ \Omega_S = \frac{1}{4} \arccos\left(\frac{1-2 s}{2 s}\right)\ , \ \ \ \ \ \kappa = \frac{4 s k^2}{\sqrt{4 s-1}} + \frac{2s-3}{2\sqrt{4 s-1}}\ ,
\end{align}
In consequence, we can write
\begin{align}
\Psi(v) = \left(\widetilde \psi_S^+ (v) , \widetilde \psi_S^- (v) , \widetilde \psi_F^+ (v) , \widetilde \psi_F^- (v) \right) \cdot \vec \chi_\infty + O(v^{-2})\ .
\end{align}
Plugging in the explicit expression (\ref{asympt-SF}), we can rewrite the result as
\begin{equation}
	\Psi(v)
	= \dfrac{1}{\sqrt v} N_F(\omega) \cos(k\ln(v)+\sigma_F(\omega)) 
	+ \dfrac{1}{v} N_S(\omega) \cos( \Omega_S v + \kappa(\omega)/v + \sigma_S(\omega)) + O(v^{-2}) , 
\end{equation}
where $N_i$ and $\sigma_i$ are for the moment unknown quantities.

\documentclass[./main.tex]{subfiles}

\section{The Wheeler-DeWitt analog}
\label{app:wdw}


The polymer quantization is of course not the only accessible technique of realizing the program of quantization of geometry. The much older program of geometrodynamics employs in particular the standard Schr\"odinger quantum representation. The application of this program to the cosmological models is known as the Wheeler-DeWitt quantization (see for example \cite{Kiefer}). For the model considered here the comparison of the traditional LQC quantization with its WDW analog has been performed already in \cite{APS06c}. The structure of this analog is critically important for the LQC models themselves as for example the normalization of the Hilbert space basis relies on it extensively. In specific contexts, one can even consider the LQC dynamics as the process of scattering of geometrodynamical (WDW) quantum universe \cite{KP10}. This analog is also a necessary component for defining the Hilbert space structures also in our studies. For that reason we briefly outline its main properties.

\subsection{The structure of the model}

Our point of departure is the classical models of FRW isotropic universe with massless scalar field already specified in sec.~\ref{Section_scalar_field}. Its Wheeler-deWitt quantization is discussed in detail (with use of slightly different variables) in \cite{APS06b} and \cite{ACS08}.
With the family of holonomies $h_a$ being continuous, there is no need for Thiemann regularization and one can start with the original connection and triad variables. The gravitational part of the Hamiltonian constraint reduces then to just a function of the coefficients $v:=V/\alpha$ and $b$ as defined in \eqref{bV_def} and the whole constraint, weighted by lapse $N=2V$, takes the form
\begin{equation}
  C_{tot}[N] = p_{\phi}^2 - 3\pi G\hbar^2v^2b^2 .
\end{equation}  
The standard Schr\"odinger quantization while ignoring the constraints (the kinematical level) yields the Hilbert space
\begin{equation}
  \ub{\Hil}_{\rm kin} = \ub{\Hil}_{\rm gr}\otimes \Hil_{\phi} = L^2(\re,\rd v) \otimes L^2(\re,\rd\phi)   
\end{equation}
and the standard set of canonical operators $(\hat{v},\hat{b})$, $(\hat{\phi},\hat{p}_\phi)$ such that
\begin{equation}
  [\hat{b},\hat{v}] = 2 i , \quad [\hat{\phi},\hat{p}_{\phi}] = i\hbar , 
\end{equation}
defined on the domains of Schwartz spaces within $\ub{\Hil}_{\rm gr}$ and $\Hil_{\phi}$ respectively.

Performing the second stage of the Dirac program is straightforward and gives the (essentially self-adjoint) quantum constraint taking (in convenient symmetric factor ordering) the form
\begin{equation}
  \widehat{C_{tot}[N]} = \id_{\ub{\Hil}_{\rm gr}} \otimes p_{\phi}^2 - 3\pi G\hbar^2 (\sqrt{|\hat{v}|}\hat{b}\sqrt{|\hat{v}|})^2 \otimes \id_{\Hil_{\phi}} 
\end{equation}

In the $v$ representation the constraint has the Klein-Gordon form
\begin{equation}
  \hbar^{-2}\widehat{NC_{tot}} 
  = -\id_{\ub{\Hil}_{\rm gr}} \otimes \partial_{\phi}^2 + 12\pi G (\sqrt{|\hat{v}|}\partial_v\sqrt{|\hat{v}|})^2 \otimes \id_{\Hil_{\phi}}
\end{equation}

A restriction to the positive frequency solutions of the constraint, in addition to the symmetry reduction with respect to the parity symmetry $v\mapsto -v$, then the application of the group averaging procedure via a rigging map defined analogously to \eqref{eq:rig}, gives us the physical Hilbert space 
\begin{equation}
  \ub{\Hil}_{\rm phy} \ni |\Psi\rangle: \Psi_\phi(v) = \int_{\re}\rd k\tilde{\ub{\Psi}}(k) \ub{e}_k(v) e^{i\omega(k)\phi} , 
\end{equation}
where $\omega(k)=\sqrt{12\pi G}|k|$ and the functions $\ub{e}_k$ are the Dirac delta normalized eigenfunctions of the WDW evolution operator 
\begin{equation}
  \ub{\Theta} := - 12\pi G (\sqrt{|\hat{v}|}\partial_v\sqrt{|\hat{v}|})^2 ,
\end{equation}
known to be positive definite and essentially self-adjoint. Its entire spectrum is continuous and consists of positive real line (${\rm Sp}(\ub{\Theta}) = \re^+$). The eigenfunctions $\ub{e}_k$ are of the form
\begin{equation}
  \ub{e}_k(v) = \frac{1}{\sqrt{2\pi|v|}} e^{ik\ln|v|} , \quad 
  [\ub{\Theta}\ub{e}_k](v) = \omega(k)^2 \ub{e}_k(v) , \quad 
  (\ub{e}_k|\ub{e}_{k'}) = \delta(k-k') ,
\end{equation}
and they form an orthonormal basis of $\ub{\Hil}_{\rm gr}$.

\subsection{Physical states in $b$ representation}

Per analogy with \eqref{eq:vb-transform}, we can introduce for the elements of $\ub{\Hil}_{\rm gr}$ the transform between the $v$ and $b$ coordinates
\begin{subequations}\begin{align}
  \psi(b) &= [\ub{\mathcal{F}}\psi](b) 
     = \frac{1}{{2\sqrt{\pi}}} \int_{\re}\frac{\rd v}{\sqrt{|v|}} \psi(v) e^{\frac{ivb}{2}} , 
     \label{eq:wdw-trans-b}\\
  \psi(v) &= [\ub{\mathcal{F}}^{-1}\psi](v) 
    = \frac{\sqrt{|v|}}{{2\sqrt{\pi}}} \int_{\re} \rd b\, \psi(b) e^{-\frac{ivb}{2}} ,
    \label{eq:wdw-trans-v}
\end{align}\end{subequations}
which maps between the real symmetric functions in $v$ and the real symmetric functions in $b$.
Applying this transformation to the expression of the scalar product on $\ub{\Hil}_{\rm gr}$ allows to express it as
\begin{equation}\label{eq:wdw-pre-prod}
  \langle\psi|\chi\rangle = \frac{1}{{4\pi}}\int_{\re} |v|\rd v \int_0^\pi \rd b\, \rd b' \psi^\star(b)\chi(b') e^{i\frac{v}{2}(b-b')} .
\end{equation}
Unfortunately, due to presence of absolute value, the inner product cannot be converted to a local form, that is a single integral over $b$. We sidestep this problem following the analogous treatment in \cite{PA11} (and in part earlier in \cite{ACS08}) by introducing the transformations to auxiliary Hilbert space $\ub{\Hil}^{\pm}$ defined by the projections $P^{\pm}$
\begin{equation}\label{eq:wdw-proj}
	P^\pm:\ub{\Hil}_{\rm gr} \to \ub{\Hil}_{\rm gr}, \ [P^{\pm}(\psi)](v) = \psi(v)\theta(\pm v) , \quad
	\ub{\Hil}^{\pm} := {\rm Im}(P^{\pm}) ,
\end{equation}
where $\theta$ is the Heaviside step function\footnote{Here we apply the convention where $\theta(0)=0$.}. The projection induces scalar products on $\ub\Hil^{\pm}$ given by the restriction of the scalar product on $\ub{\Hil}_{\rm gr}$ to positive/negative $v$ respectively. In turn, upon transformation to the $b$ coordinate, these induced scalar products can be written similarly to \eqref{eq:wdw-pre-prod}, but now they have a local form
\begin{equation}\label{eq:wdw-prod-comp}\begin{split}
	\langle\psi|\chi\rangle_{\pm} 
	&= \frac{1}{{4\pi}} \int_{\re^\pm} (\pm v)\rd v \int_\re \rd b \rd b' \psi^\star(b)\chi(b') e^{i\frac{v}{2}(b-b')}  \\
	&= \mp {2i} \int_\re \rd b \psi^\star(b) \partial_b \chi(b) .
\end{split}\end{equation}
Since the spaces $\ub{\Hil}^{\pm}$ are orthogonal, the scalar product of $\ub{\Hil}_{\rm gr}$ can be rewritten as
\begin{equation}\label{eq:wdw-prod-split}
	\langle\psi|\chi\rangle = \langle P^+\psi|P^+\chi\rangle_+ + \langle P^-\psi|P^-\chi\rangle_- ,
\end{equation}
thus it becomes quite simple to evaluate, provided that the projections of the arguments are known. Unfortunately the form of the operators $P^\pm$ in the $b$-representation is not simple. In order to properly control the inner product we need to find the explicit form of $\ub{{\cal F}}(P^\pm \ub{e}_k)$. In the integral form it is a simple restriction of \eqref{eq:wdw-trans-b}
\begin{equation}\label{Transform}
  \ub{{\cal F}}(P^{\pm}\ub{e}_k)(b) = \frac{1}{2\pi} \int_{\re^{\pm}} \frac{\rd v}{|v|} e^{ik\ln|v|} e^{\frac{ivb}{2}} .
\end{equation}
By extending the integrand function to the complex plane and choosing the integration contours as in fig.~\ref{fig:contour}, these integrals can be converted to (well defined) ones over imaginary semi-axes, which in turn can be expressed in terms of the Gamma special functions
\begin{equation}
  \ub{{\cal F}}(P^{\pm}\ub{e}_k)(b) = \pm\frac{1}{2\pi}e^{\pm\sgn(b)\frac{\pi k}{2}} \Gamma(ik) e^{-ik\ln|b/2|} . 
\end{equation}
This in turn allows to write the full transform of $\ub{e}_k$ as
\begin{equation}\label{eq:wdw-ek-trans}
  \ub{{\cal F}}(\ub{e}_k)(b) = \frac{1}{\pi} \Gamma(ik) \sinh(\sgn(b)\frac{\pi k}{2}) e^{-ik\ln|b/2|} 
   ,
\end{equation}
Using the asymptotic relation $|\Gamma(ik)| \sinh(\pi k/2) \sqrt{|k|} = \sqrt{\pi/2} + O(e^{-k})$, we can further approximate the above transform for large $k$ as
\begin{equation}
 \ub{{\cal F}}(\ub{e}_k)(b) =\frac{1}{\sqrt{2\pi|k|}} e^{-i(k\ln|b|+\ub{\sigma})} + O(e^{-|k|}) .
\end{equation}
where $\ub{\sigma}$ is a $k$-dependent phase shift. 

\begin{figure}[h!]
  \includegraphics[width=0.5\textwidth]{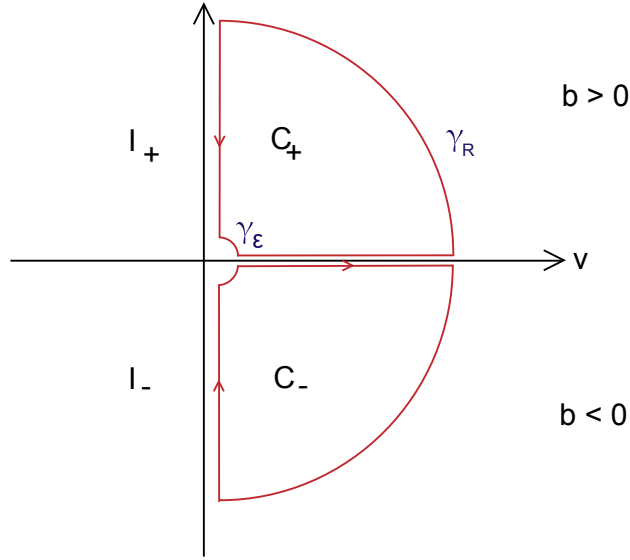}
  \caption{Integration contours for the transform \eqref{Transform}.}
  \label{fig:contour}
\end{figure}

This evaluation will be used in the next subsection to determine the explicit normalization of the Hilbert space basis elements in LQC framework in $b$-representation.

\documentclass[./main.tex]{subfiles}

\section{Gravitational Hilbert space in b-representation}
\label{sec:aux-math}

The $b$-representation is particularly convenient in identifying the spectrum of the evolution operator $\Theta_{\rm TR}$ and the resolution of identity in terms of its eigenstates. However, certain ingredients of the Hilbert space structure are not straightforwardly obtained. An example is the scalar product of $\Hil_{\rm gr}$. This appendix is dedicated to studying the mathematical properties of $\Hil_{\rm gr}$ and of $\Theta_{\rm TR}$, which are necessary for the construction of a basis of the physical Hilbert space. We will focus on the eigenvalue problem for $\Theta_{\rm TR}$ and the normalization of its eigenstates.

Let us start with the scalar product of $\Hil_{\rm gr}$.

\subsection{The scalar product}

Expressing the scalar product of $\Hil_{\rm gr}$ in the $b$-representation can be achieved by the inverse of transformation of \eqref{eq:vb-transform} 
\begin{equation}
	\psi(v) = \frac{\sqrt{|v|}}{\pi} \int_0^\pi \rd b \ \tilde{\psi}(b) e^{-\frac{i}{2}vb}
\end{equation}
This leads to the formula
\begin{equation}\label{eq:pre-prod}
  \langle\psi|\chi\rangle = \frac{1}{\pi^2}\sum_{\lat_4} |v| \int_o^\pi \rd b \rd b' \psi^\star(b)\chi(b') e^{i\frac{v}{2}(b-b')} .
\end{equation}
Like in WDW, due to the presence of the absolute value, the scalar product cannot be converted to a local form, that is a single integral over $b$. We introduce the projections $P^{\pm}$ on $\Hil_{\rm gr}$ analogously to \eqref{eq:wdw-proj}
\begin{equation}
	P^\pm:\Hil_{\rm gr} \to \Hil_{\rm gr}, \ [P^\pm(\psi)](v) = \psi(v)\theta(\pm v) , \quad
	\Hil^{\pm} := {\rm Im}(P^{\pm}) ,
\end{equation}
As in appendix \ref{app:wdw}, the scalar products on $\Hil^{\pm}$ can be written in a form similar to \eqref{eq:pre-prod}:
\begin{equation}\label{eq:prod-comp}\begin{split}
	\langle\psi|\chi\rangle_{\pm} 
	= \frac{1}{\pi^2}\sum_{\lat_4} \pm v \int_o^\pi \rd b \rd b' \psi^\star(b)\chi(b') e^{i\frac{v}{2}(b-b')} = \mp \frac{4i}{\pi} \int_0^\pi \text db \psi^\star(b) \partial_b \chi(b) .
\end{split}\end{equation}
Since the spaces $\Hil^{\pm}$ are orthogonal, the scalar product of $\Hil_{\rm gr}$ can be rewritten as
\begin{equation}\label{eq:prod-split}
	\langle\psi|\chi\rangle = \langle P^+\psi|P^+\chi\rangle_+ + \langle P^-\psi|P^-\chi\rangle_- ,
\end{equation}
thus it becomes quite simple to evaluate, provided that the projections of the arguments are known. Unfortunately, the form of operators $P^\pm$ in 
$b$-representation is not simple, thus evaluating the scalar product this way is not convenient. On the other hand the relations \eqref{eq:prod-comp}, \eqref{eq:prod-split} are useful in probing various properties of elements of the physical Hilbert space (being a subspace of $\Hil_{\rm gr}$). In particular we will apply them to analyze the weak solutions to the eigenvalue problem \eqref{eq:eig-prob}.

We conclude this section by expressing the auxiliary scalar products in terms of the coordinate $x$:
\begin{equation} 
  \langle\psi|\chi\rangle_{\pm} 
  = \mp \frac{4i}{\pi} \int_\re \rd x \psi^\star(x) \partial_x \chi(x) 
\end{equation}

\subsection{The eigenvalue problem for the evolution operator}

For given $\psi\in\Hil_{\rm gr}$ we will denote its components with respect to the projections defined above as $\psi^\pm := P^{\pm}(\psi)$. Furthermore, 
in order to express the weak eigenvalue problem, we switch to the coordinate $x$ defined in \eqref{eq:x-func}. Noting that the action of $\Theta_{\rm TR}$ preserves the sub-spaces $\Hil^{\pm}$,\footnote
{
While $\Theta_{\rm TR}$ involves shifts both in the positive and negative directions, by explicit computation one can check that shifts across $v=0$ are multiplied by $0$. Hence, a function with support on the positive sub-lattice will remain on the positive sub-lattice upon repeated action of $\Theta_{\rm TR}$.
}
we can rewrite the eigenvalue equation (for the eigenvector $\Psi_\lambda$ with corresponding eigenvalue $\lambda$) as
\begin{equation}\label{eq:eig-app}\begin{split} 
	\forall \chi\in\dom \ \ \ \ \ 0 &= (\Psi_\lambda|\Theta_{\rm TR}^\dagger-\lambda^{\star}\id|\chi\rangle \\ 
	&= (\Psi^+_\lambda|\Theta_{\rm TR}^\dagger-\lambda^{\star}\id|\chi^+\rangle_+ + (\Psi^-_\lambda|\Theta_{\rm TR}^\dagger-\lambda^{\star}\id|\chi^-\rangle_-
\end{split}\end{equation}
(where $\dom$ is the domain already defined in \eqref{eq:dom-def}, of which elements are necessarily smooth in $b$). This equation requires that both the components on the right hand side vanish independently (since it must hold for all $\chi$, it holds in particular for $\chi$ such that $\chi^- = 0$). For each of these components we split the domain of $x$ into three intervals where $x(b)$ is regular: $\mathcal I_i \in \{ (-\infty,-\pi/2), (-\pi/2,\pi/2), (\pi/2,\infty) \}$, so that $\Theta_{\rm TR}$ can be easily expressed in terms of $x$: for all $\chi \in \dom\cap\Hil^\pm$, we have
\begin{align}
(\Psi^{\pm}_\lambda & |\Theta_{\rm TR}^\dagger-\lambda^{\star}\id|\chi^{\pm}\rangle_{\pm} = \mp \frac{4i}{\pi} \left[\int_{-\infty}^{-\pi/2} dx + \int_{-\pi/2}^{\pi/2} dx + \int_{\pi/2}^{\infty} dx\right] [\Psi_\lambda^\pm]^\star(x) \partial_x \left[(\Theta_{\rm TR}^\dag - \lambda^\star \id)\chi^\pm\right](x) = \notag
\\
& = \mp \frac{4i}{\pi} \int_{-\infty}^{-\pi/2} dx \; [\Psi_\lambda^\pm]^\star(x) (-12\pi G \partial_x^2 - \lambda^\star \id) \partial_x \chi^\pm(x) \notag
\\
& \mp \frac{4i}{\pi} \int_{-\pi/2}^{\pi/2} dx \; [\Psi_\lambda^\pm]^\star(x) (12\pi G \partial_x^2 - \lambda^\star \id) \partial_x \chi^\pm(x) \notag
\\
& \mp \frac{4i}{\pi} \int_{\pi/2}^{\infty} dx \; [\Psi_\lambda^\pm]^\star(x) (-12\pi G \partial_x^2 - \lambda^\star \id) \partial_x \chi^\pm(x) = \notag
\\
& = \mp \frac{4i}{\pi} \int_{-\infty}^{-\pi/2} (-12\pi G [\Psi_\lambda^\pm]''^\star - \lambda^\star [\Psi_\lambda^\pm]^\star) [\chi^\pm]' \pm 48i G \lim_{x \to^- -\pi/2} \left[[\Psi_\lambda^\pm]^\star [\chi^\pm]'' - [\Psi_\lambda^\pm]'^\star [\chi^\pm]'\right](x) \notag
\\
& \mp \frac{4i}{\pi} \int_{-\pi/2}^{\pi/2} (12\pi G [\Psi_\lambda^\pm]''^\star - \lambda^\star [\Psi_\lambda^\pm]^\star) [\chi^\pm]' \pm 48i G \left(\lim_{x \to^+ -\pi/2} - \lim_{x \to^- \pi/2}\right) \left[[\Psi_\lambda^\pm]^\star [\chi^\pm]'' - [\Psi_\lambda^\pm]'^\star [\chi^\pm]'\right](x) \notag
\\
& \mp \frac{4i}{\pi} \int_{\pi/2}^{\infty} (-12\pi G [\Psi_\lambda^\pm]''^\star - \lambda^\star [\Psi_\lambda^\pm]^\star) [\chi^\pm]' \mp 48i G \lim_{x \to^+ \pi/2} \left[[\Psi_\lambda^\pm]^\star [\chi^\pm]'' - [\Psi_\lambda^\pm]'^\star [\chi^\pm]'\right](x) = \notag
\\
& = \mp \frac{4i}{\pi} \int_{-\infty}^{\infty} dx \; \left([\Theta_{\rm TR} \Psi_\lambda^\pm]^\star(x) - \lambda^\star [\Psi_\lambda^\pm]^\star(x)\right) \partial_x \chi^\pm(x) \notag
\\
& \mp 48i G \left[\lim_{x \to^+ \pi/2} + \lim_{x \to^- \pi/2} - \lim_{x \to^+ -\pi/2} - \lim_{x \to^- -\pi/2}\right] [\Psi_\lambda^\pm]^\star(x) [\chi^\pm]''(x) = \notag
\end{align}
\begin{align}
& = \mp \frac{4i}{\pi} \int_{-\infty}^{\infty} dx \; \left([\Theta_{\rm TR} \Psi_\lambda^\pm](x) - \lambda [\Psi_\lambda^\pm](x)\right)^\star \partial_x \chi^\pm(x) \notag\hspace{175pt}
\\
&\pm \frac{4i}{\pi} \left[\lim_{x \to^+ \pi/2} - \lim_{x \to^- \pi/2} + \lim_{x \to^+ -\pi/2} - \lim_{x \to^- -\pi/2}\right] [\Psi_\lambda^\pm]^\star(x) [\Theta_{\rm TR} \chi^\pm](x)
\end{align}
where in the third step we integrated by part twice using $fg'' = f''g + (fg' - f'g)'$ and disregarded the boundary contributions at infinity, while in the fourth step we observed that $\partial_x \chi^\pm(\pm \pi/2) = 0$ (due to smoothness of $\chi^\pm$ in $b$, since $\partial_x \chi^\pm = (\partial b/\partial x) \partial_b \chi^\pm$ and $(\partial b/\partial x)_{x = \pm \pi/2} = 0$). From this equation, we see that $(\Psi^{\pm}_\lambda |\Theta_{\rm TR}^\dagger-\lambda^{\star}\id|\chi^{\pm}\rangle_{\pm} = 0$ for every $\chi \in \dom\cap\Hil^\pm$ if and only if $\Psi^{\pm}_\lambda$ satisfies
\begin{equation}\label{eq:eig-int}
	\Theta_{\rm TR}\Psi^{\pm}_\lambda = -12\pi G\sgn(|x|-\pi/2)\partial_x^2 \Psi^{\pm}_\lambda = \lambda\Psi^{\pm}_{\lambda}
\end{equation}
and it is continuous (but not necessarily differentiable) at $x = \pm \pi/2$. It is then easy to see (recalling that $\Psi_\lambda(-x) = \Psi_\lambda(x)$) that the general solution is given by
\begin{equation}\label{eq:basis-sol-app}
\Psi_{\beta,k}(x) = \zeta \begin{cases} 
                          \cos(k|x|+\varphi({\beta,} k)) , & |x|>\pi/2 , \\
                          \frac{\cos(k\pi/2+\varphi({\beta,} k))}{\cosh(k\pi/2)}\cosh(kx) , & |x|\leq \pi/2 ,
                        \end{cases}
\end{equation}
where $\zeta$ and $\varphi$ are free constants.

Having this form at our disposal, we can systematically find the eigenstates of $\Theta_{\rm TR}$ relevant for the construction of physical states. This has been done in Sec.~\ref{self-adj-ext} of the paper. What remains is fixing the normalization constant $|\zeta|$. The asymptotic form of the eigenfunctions \eqref{eq:eig-asympt} implies that they are not explicitly normalizable, thus $|\zeta|$ cannot be determined in a straightforward way or by purely numerical means. We will focus on this problem in the next subsection. 

\subsection{Normalization of the eigenstates}

Consider the set of generalized eigenstates $\Psi_k$ of $\Theta_{\rm TR}$ corresponding to the eigenvalue $\omega^2 = 12\pi G k^2$. Applying the asymptotic decomposition \eqref{eq:eig-asympt} to these states, we can write the inner product between two such states as a distribution
\begin{equation}\label{eq:app-wdw-limit}
	(\Psi_k,\Psi_{k'}) = \sum_{v\in\lat_4} N_F(k)N_F(k') \left[ |v|^{-1}\cos(k\ln|v|+\sigma_F(k))\cos(k'\ln|v|+\sigma_F(k')) + \bar{O}(|v|^{-3/2}) \right], 
\end{equation}
where $\bar{O}(\cdot)$ denotes the \emph{bounded} remnant of the rate of decay specified in the argument. 
The trigonometric components are combinations of the basis elements of the Wheeler-DeWitt analog of the model under study (see App.~\ref{app:wdw}) and define the so called Wheeler-DeWitt limit of LQC (see \cite{APS06c} and \cite{KP10} for details).
Introducing an auxiliary variable $\eta:=\ln |v|$, we can further approximate the above sum by an integral. Indeed, as the consecutive steps lengths in $\eta$ between points of the summation decay exponentially, and due to the boundedness of $\cos(k\eta+\sigma_F(k))$ and its derivatives, we have an estimate
\begin{equation}
	(\Psi_k,\Psi_{k'}) = \frac{1}{2} N_F(k)N_F(k') \int_0^{\infty} \rd\eta \cos(k\eta+\sigma_F(k))\cos(k'\eta+\sigma_F(k')) 
	+ \bar{O}(\eta^{-2}) .
\end{equation}
By expressing the cosines in terms of exponentials, using the identity
\begin{equation}
	\int_0^\infty \rd x e^{ikx} = \pi\delta(k) + \frac{i}{k}  ,
\end{equation}
and taking into account that $k,k'>0$, we arrive at the following form of the scalar product
\begin{equation}\label{eq:ek-pre-norm}
  (\Psi_k,\Psi_{k'}) = N_F(k)N_F(k')\frac{\pi}{8}\delta(k-k') + f(k,k') , 
\end{equation}
where $f$ is possibly singular at $k=k'$. This function, however, must vanish due to the orthogonality of the eigenspaces for $k\neq k'$; thus, the orthonormality condition allows us to determine the asymptotic normalization constant $N_F(k)$ as\footnote%
{
To do this, we observe that $\Psi_k$ is normalized to Dirac delta as the principal component of \eqref{eq:ek-pre-norm}. Thus, in the sense of distributions we impose $\int dk \varphi(k) (\Psi_k,\Psi_{k'}) = \varphi(k')$ for all test functions $\varphi$. Since $\int dk f(k,k') \varphi(k) = 0$ for all $\varphi$ (due to the fact that $f$ is non-vanishing only in a set of measure $0$ and it is not proportional to a Dirac delta), we conclude that $\varphi(k') = \int dk \varphi(k) \delta(k-k') N_F(k)N_F(k') \pi/8 = \varphi(k') N_F(k')^2 \pi/8$, which can be solved for $N_F$.
}
\begin{equation}
	N_F = \frac{4}{\sqrt{2\pi}} .
\end{equation}

In order to determine the constant $|\zeta|$ in the expression of the eigenstates $\Psi_k$ given in \eqref{eq:basis-sol-app}, we employ the following observations:
\begin{enumerate}
  \item In the limit $b\to 0,\pi$, the function $x(b)$ approaches (up to a constant) the logarithmic function, that is
    \begin{equation}
      \lim_{b\to 0^+} [x(b)-\ln|b|] = \ln\left(\frac{1+\gamma^2}{2}\right) - \frac{\pi}{2} , \qquad 
      \lim_{b\to \pi^-} [x(b)+\ln|\pi-b|] = - \ln\left(\frac{1+\gamma^2}{2}\right) + \frac{\pi}{2} .
    \end{equation}
    The function $\ln|b|$ is in Wheeler-DeWitt model the analog of $x(b)$ in the model we are studying.
  \item As a weak solution to the eigenvalue problem, for $|x|>\pi/2$ each of the projections 
    ${\cal F}(P^{\pm}e_k)$ need to be linear combinations of $e^{\pm ik x(b)}$.
\end{enumerate}
These two observations allow to relate $|\zeta|$ to the norms of the WDW limits specified in \eqref{eq:app-wdw-limit} and expressed in the $b$-representation using \eqref{eq:wdw-ek-trans} as follows. Defining the quantity
\begin{equation}
  \ub{f}_k := N_F |v|^{-1/2} \cos(k\ln|v|+\sigma_F(k)) , 
\end{equation}
we have
\begin{equation}\begin{split}
  [\ub{{\cal F}}(\ub{f}_k)](b) = [\ub{{\cal F}}(P^+\ub{f}_k+P^-\ub{f}_k)](b) 
  = \frac{8i}{\sqrt{2\pi}}\sgn(b)|\Gamma(ik)|\sinh(\frac{\pi}{2}k)\sin(k\ln|b|+\tilde{\sigma}(k)) , 
\end{split}\end{equation}
where $\tilde{\sigma}$ is some $k$-dependent phase shift. Having the convergence of ${\cal F}(\Psi_k)$ to $\ub{{\cal F}}(\ub{f}_k)$ in the limit $b\rightarrow 0$ we can determine the absolute value of the multiplicative constant $\zeta$ in \eqref{eq:basis-sol-app}. Thus we have
\begin{equation}
|\zeta| = \frac{4\sqrt{2}}{\sqrt{\pi}} |\Gamma(ik)|\sinh(\frac{\pi}{2}k) = \frac{4}{\sqrt{|k|}} + O(e^{-|k|})
\end{equation}

\end{document}